\documentclass[a4paper,11pt]{article}
\pdfoutput=1
\usepackage{heppub}
\usepackage{amsthm}
\usepackage{amsmath}
\usepackage{bbm}
\usepackage{graphicx,accents}
\usepackage{slashed}
\usepackage{amssymb}
\usepackage{footmisc}
\usepackage{enumitem}
\usepackage{xcolor}
\usepackage{braket}
\usepackage{bm,subcaption,comment}
\usepackage{xfrac}

\newcommand{\bs}{\boldsymbol}

\title{Adiabatic Hydrodynamization and the Emergence of Attractors: a Unified Description of Hydrodynamization in Kinetic Theory}

\author[1,2]{Krishna Rajagopal,}
\author[1]{Bruno Scheihing-Hitschfeld}
\author[1]{and Rachel Steinhorst}

\affiliation[1]{Center for Theoretical Physics, Massachusetts Institute of Technology, Cambridge, MA  02139, USA}

\affiliation[2]{Theoretical Physics Department, CERN, 1211 Geneva 23, Switzerland}

\abstract{``Attractor" solutions for the pre-hydrodynamic, far-from-equilibrium, evolution of the matter produced in relativistic heavy ion collisions have emerged as crucial descriptors of the rapid hydrodynamization of quark-gluon plasma (QGP). Adiabatic Hydrodynamization (AH) has been proposed as a framework with which to describe, explain, and predict attractor behavior that draws upon an analogy to the adiabatic approximation in quantum mechanics. In this work, we systematize the description of pre-hydrodynamic attractors in kinetic theory by showing how to use the AH framework to identify these long-lived solutions to which varied initial conditions rapidly evolve, demonstrating the robustness of this framework. In a simplified QCD kinetic theory in the small-angle scattering limit, we use AH to explain both the early- and late-time scaling behavior of a longitudinally expanding gluon gas in a unified framework. In this context, we show that AH provides a unified description of, and  intuition for, all the stages of what in QCD would be bottom-up thermalization, starting from a pre-hydrodynamic attractor and ending with hydrodynamization. We additionally discuss the connection between the notions of scaling behavior and adiabaticity and the crucial role of time-dependent coordinate redefinitions in identifying the degrees of freedom of kinetic theories that give rise to attractor solutions. The tools we present open a path to the intuitive explanation of how attractor behavior arises and how the attractor evolves in all stages of the hydrodynamization of QGP in heavy ion collisions.}

\emailAdd{krishna@mit.edu}
\emailAdd{bscheihi@mit.edu}
\emailAdd{rstein99@mit.edu}

\preprint{MIT-CTP/5724}

\begin{document}

\maketitle

\section{Introduction}
\label{sec:Intro}

The discovery that quark-gluon plasma (QGP) is a strongly coupled fluid~\cite{PHENIX:2004vcz,BRAHMS:2004adc,PHOBOS:2004zne,STAR:2005gfr,Gyulassy:2004zy} has opened a window to study the many-body physics of hot liquid QCD matter under reproducible conditions using particle colliders, particularly via heavy-ion collisions (HICs). While strongly correlated many-body systems are ubiquitous across physics, QGP is special because it is the most perfect liquid ever discovered, its constituents are the elementary quarks and gluons of the standard model of particle physics and --- because the theory that describes these constituents, QCD, is asymptotically free --- we know that the dynamics at the earliest moments of the very high energy collisions in which QGP is later formed is weakly coupled. Therefore, by studying how strongly coupled liquid QGP forms (hydrodynamizes) from initially weakly coupled quarks and gluons, how a droplet of this liquid evolves, expands and cools in HICs, we deepen our understanding of the fundamental building blocks of matter
in conditions as extreme as those present in the first microseconds after the Big Bang, as well as learning how a complex strongly correlated phase of matter can emerge from the fundamental laws that govern matter at the shortest distance scales.

Among the plethora of QGP-related phenomena that have been studied in the past two decades via heavy-ion collisions (for reviews see Refs.~\cite{Muller:2006ee,Jacak:2012dx,Muller:2012zq,Heinz:2013th,Shuryak:2014zxa,Akiba:2015jwa,Romatschke:2017ejr,Busza:2018rrf,Nagle:2018nvi,Schenke:2021mxx,Harris:2023tti}), the discovery that the time it takes for the highly excited, far-from-equilibrium, initially weakly coupled quarks and gluons liberated just after the collision to become near-hydrodynamic is around 1~${\rm fm}/c$ was one of the first and one of the most striking~\cite{Heinz:2001xi,Heinz:2002un,Kolb:2003dz,Heinz:2004pj}. This discovery emerged via comparison of early RHIC measurements of the anisotropic flow of off-center heavy ion collisions with early hydrodynamic calculations and the realization that if the onset of hydrodynamic behavior were much later than 1~fm$/c$, anisotropies as large as those observed were not possible. Later estimates~\cite{Shen:2010uy,Shen:2012vn} applying the same logic with more fully developed calculations and analyses of data indicated that the hydrodynamization time in RHIC collisions could be as short as 0.4~fm$/c$ to 0.6~fm$/c$, with the hydrodynamic fluid formed at that time having a temperature $\sim 500$~MeV to $\sim 350-380$~MeV.
These estimates of the hydrodynamization timescale were seen as surprisingly rapid, simply because they are comparable to the time it takes light to cross a proton and also because early attempts to use the theory of how thermalization proceeds at weak coupling in QCD kinetic theory~\cite{Baier:2000sb} seemed to yield longer timescales.
This became less puzzling when
calculations done in strongly coupled theories with holographic duals~\cite{Chesler:2008hg,Chesler:2009cy,Chesler:2010bi,Heller:2011ju,Heller:2012je,Heller:2012km,vanderSchee:2012qj,Heller:2013oxa,Casalderrey-Solana:2011dxg,Chesler:2015lsa,Chesler:2015fpa,Heller:2016gbp} showed that in the strong coupling limit hydrodynamization should be expected within $\sim 1/T_{\rm hyd}$ after the collision, where $T_{\rm hyd}$ is the temperature of the hydrodynamic fluid that forms. This implicit criterion corresponds to a hydrodynamization time in HICs that is comparable to the fastest estimates that had previously been inferred from RHIC data.  
That said, all of these developments posed a pressing challenge: can we find
a microscopic understanding of this process in terms of the kinetic theory of the fundamental degrees of freedom of QCD itself, noting that at the earliest moments of a high energy collision their dynamics must be weakly coupled.
This
has been a subject of intense study over the past decade; for reviews, see Refs.~\cite{Schlichting:2019abc,Berges:2020fwq}. The successes of hydrodynamic modeling in describing HIC data only reinforce the need for a qualitative and quantitative understanding of the processes that connects the initial state (two highly Lorentz-contracted atomic nuclei) with a hydrodynamic droplet of strongly coupled QGP in local thermodynamic equilibrium.

Due to the famous intractability of calculations of dynamics in QCD as a quantum field theory, the understanding of this process has relied on several different effective descriptions, such as M\"uller-Israel-Stewart-type extensions of hydrodynamics~\cite{Israel:1979wp}, QCD effective kinetic theory (EKT)~\cite{Baier:2000sb,Arnold:2002zm,Kurkela:2015qoa,Kurkela:2018vqr,Kurkela:2018wud}, classical-statistical simulations~\cite{Aarts:2000wi,Berges:2007re,Berges:2008mr}, or holographic modelling~\cite{Chesler:2008hg,Chesler:2009cy,Chesler:2010bi,Heller:2011ju,Heller:2012je,Heller:2012km,vanderSchee:2012qj,Heller:2013oxa,Casalderrey-Solana:2011dxg,Chesler:2015lsa,Chesler:2015fpa,Heller:2016gbp,Chesler:2015bba,Chesler:2016ceu,Grozdanov:2016zjj,Folkestad:2019lam}, depending on the initial state and which stage of the thermalization process is to be described (i.e., the energy scale of interest) and also depending on where strongly coupled dynamics is employed. Indeed, the most successful approaches involve a multi-stage process where different effective descriptions are matched to maximize their effectiveness (e.g., glasma $\to$ kinetic theory $\to$ hydrodynamics as described in~\cite{Kurkela:2015qoa,Kurkela:2018vqr,Kurkela:2018wud}). 
Furthermore, since the estimates made in Ref.~\cite{Kurkela:2015qoa}, it has been understood that numerical simulations of the kinetic theory description of hydrodynamization, whose stages (``bottom-up thermalization'') were first described analytically at very weak coupling in Ref.~\cite{Baier:2000sb}, that employ a reasonable intermediate value of the QCD coupling $g\sim 2$, which corresponds to $\alpha\sim 0.3$ and $g^2 N_c \sim 12$, yield
hydrodynamization time scales $\lesssim 1~{\rm fm}/c$. That is, QCD kinetic theory, when applied boldly at reasonable as opposed to very weak coupling, describes hydrodynamization that is comparable to what was first inferred via comparisons between experimental data and hydrodynamic calculations and also comparable to what had previously been inferred via employing similar intermediate values of the coupling in results obtained via holographic calculations in strongly coupled gauge theories. This is pleasing, but if anything it makes the goal of understanding the physical intuition behind hydrodynamization at weak coupling more pressing.

Because of the diversity of these descriptions, it is not obvious that one will be able to identify unique underlying principles behind the process of QCD hydrodynamization across all of the energy scales through which QCD matter transits in a HIC, despite the fact that the final state (local thermal equilibrium) is universal in the sense that it has lost all memory of which initial state it originated from. Furthermore, one may worry that the early stages of hydrodynamization could have maximal sensitivity to what the initial state is, and therefore that a separate calculation would need to be carried out in full for each posited initial state in order to have predictive power for observables that are sensitive to the pre-hydrodynamic stages of a HIC.
As it turns out, the situation is not as grim. Indeed, much progress in this direction has been enabled by studies of far-from-equilibrium phenomena that are themselves universal in the sense referred to above and, in particular, of so-called ``attractor'' solutions in either a kinetic theory or a holographic description of pre-hydrodynamic physics. These are (families of) solutions to which generic initial conditions converge in a finite time, meaning that they are dynamically selected in the phase space of the theory. The presence of attractor solutions makes it natural for the system to lose memory of (and have little sensitivity to) its initial state before hydrodynamization. Such attractor solutions have been sought and found in most effective descriptions of out-of-equilibrium QCD matter; see, e.g., Ref.~\cite{Kurkela:2019set}.

What attractor solutions crucially encode is \textit{how} the sensitivity to the initial conditions of a HIC is lost, \textit{what} information is carried through hydrodynamization, and perhaps an answer to \textit{why} the hydrodynamization time of QGP is $\sim 1 \, {\rm fm}/c$ for generic initial conditions, while also preserving quantitative information that may give predictive power over HIC observables that are particularly sensitive to the pre-hydrodynamic stage. Even quantities that are usually thought of as independent from the initial hydrodynamization stage in a HIC, such as heavy quark diffusion and the quenching of, and transverse momentum broadening of, high energy partons in a jet shower  
may receive modifications in this initial period~\cite{Avramescu:2023qvv,Boguslavski:2023fdm,Boguslavski:2023alu,Boguslavski:2023waw,Boguslavski:2023jvg,Pandey:2023dzz} that affect how HIC data is to be interpreted, modifications that may be described by studying the corresponding pre-hydrodynamic attractor~\cite{Boguslavski:2023jvg}. As such, it is desirable to have a framework in which to organize and describe the emergence of attractor solutions systematically. This is the task that concerns us in the present paper. Concretely, we will discuss how the Adiabatic Hydrodynamization (AH) picture~\cite{Brewer:2019oha} provides such a framework using the kinetic theory description of QCD in the small-angle scattering approximation~\cite{Mueller:1999pi,Blaizot:2013lga} as a proof of concept.

A key observation was made in recent work~\cite{Brewer:2022vkq}, which we shall refer to as BSY (after its authors), regarding the role of time-dependent coordinate redefinitions (in particular, rescalings) in identifying the attractor solution and explaining its rapid emergence relative to the other time scales in the system. Inspired by the fact that the same self-similar scaling had been observed in classical-statistical simulations~\cite{Berges:2013eia,Berges:2013fga}, in small-angle scattering kinetic theory~\cite{Tanji:2017suk} and in QCD EKT~\cite{Mazeliauskas:2018yef}, BSY proposed that the reduction of dynamically relevant degrees of freedom was most naturally understood in the (time-dependent) frame in which the typical momentum scales of the particle distribution functions are approximately constant. BSY then demonstrated this explicitly by analytically solving for the instantaneous eigenstates and eigenvalues of the generator of time evolution of the theory (which, out of familiarity with the quantum mechanics nomenclature, was referred to as a Hamiltonian) in a simplified version of the small-angle scattering collision kernel, applicable in the earliest stages of the hydrodynamization process of a weakly coupled, boost-invariant gluon gas. 

In this work, we extend the results of BSY in two distinct ways that demonstrate the effectiveness of the AH framework to identify and describe attractors in kinetic theories:
\begin{enumerate}
    \item We show that the close connection between time-dependent scalings, adiabaticity, and universality persists even in situations where one does not have explicit analytic control over the eigenvalues and eigenstates of the time evolution operator.
    \item We demonstrate, within the small-angle scattering approximation and with fewer additional approximations than BSY had to make~\cite{Brewer:2022vkq}, that the AH scenario describes the evolution of the gluon distribution function all the way from the times described in the work of BSY until local thermal equilibrium is reached and the system hydrodynamizes. The process of relaxation to a thermal distribution takes place in stages, firstly along the longitudinal direction, driven by expansion along the boost-invariant direction, where the distribution relaxes to a set of slow modes characterized by a unique profile in this direction, and secondly from this set of slow modes to a thermal distribution where a unique slow mode is singled out. We show that AH provides a unified description of both stages.
\end{enumerate}

More generally, our work provides a systematic method to study the emergence of universal behavior in kinetic theories and identify attractor solutions. Building on the previous papers on this subject~\cite{Brewer:2019oha,Brewer:2022vkq}, we posit that adiabaticity, in the sense we will define in the next Section, provides a robust underlying principle that allows one to identify the slow (low-energy) degrees of freedom of the theory. While we do not aim to prove that it is always possible to find a description where the evolution of the system is adiabatic, the power of the AH framework is that when an attractor solution exists it provides a natural method by which to single it out from the rest.

For simplicity's sake, the collision kernel we work with throughout this paper omits qualitatively and quantitatively important terms of the full QCD EKT collision kernel; this simplification makes the hydrodynamization time of the system unrealistically long. Our goal in this work is therefore not the explanation of the (rapid) timescale for hydrodynamization in QCD. Rather, we seek (and have found) a formalism
that provides a common physical description of, and intuition for, the processes occuring through all the stages of hydrodynamization in the kinetic theory with a simplified collision kernel that we employ, with the goal of employing this formalism and applying this intuition in QCD EKT in future work.
We also omit spatial gradients, which means that we are neglecting the transverse expansion of the 
droplet of QGP throughout the hydrodynamization process. Relaxing this assumption is also a worthy goal for future work, but we do not anticipate that doing so will make a qualitiative difference to the processes of hydrodynamization in the collisions of nuclei whose transverse extent is much larger than the hydrodynamization timescale.
There is no fundamental barrier to generalizing the tools developed here to a kinetic theory which includes the full QCD EKT collision kernel and transverse expansion. We expect that by applying the AH framework to QCD EKT using the systematic procedure we develop here, we will be able to employ the intuitive understanding of the rapid reduction of dynamically relevant degrees of freedom (i.e. hydrodynamization) that we find in this paper in a context where the rapid hydrodynamization expected in a more complete description of QCD is realized. In this way, our work paves the way for a satisfying physical description of and intuition for how and why hydrodynamization occurs in HICs.

This work is organized as follows: In Section~\ref{sec:AH}, we review the AH framework, explaining its usefulness and purpose; we also describe the concrete kinetic theory setup that we shall work with throughout this paper and the role of conserved quantities in this picture. In Section~\ref{sec:scaling}, we show that the introduction of time-dependent rescalings of the momentum coordinates allows one to find adiabatic descriptions of the hydrodynamization process of a weakly coupled gluon gas. We consider the examples of a static, non-expanding case as well as of a boost-invariant longitudinally expanding gluon gas that exhibits two separate scaling regimes: one at early times (discussed in the BSY paper) and one at late times, which is identified with reaching the hydrodynamic regime. While these rescalings are sufficient to describe each regime in terms of an adiabatically evolving state, they do not provide a way to smoothly connect early (pre-hydrodynamic) and late (hydrodynamic) times and attractors. Finally, Section~\ref{sec:non-scaling} provides the tools needed to make this connection, generalizing the relation between scaling and adiabaticity to more general time-dependent variables, and demonstrating point 2 in the preceding discussion via the explicit construction of a unified description of the pre-hydrodynamic and hydrodynamizing attractors. We present our concluding remarks and outlook in Section~\ref{sec:outlook}.

\section{Adiabatic Hydrodynamization} \label{sec:AH}

In this Section, we discuss how the Adiabatic 
Hydrodynamization (AH) framework constitutes a systematic approach to characterize aspects of the out-of-equilibrium dynamics of many-body theories that are common to many systems. Even though beginning in Section~\ref{sec:kin} and then throughout what follows we will employ kinetic theory to describe the dynamics of interest, we expect that the AH framework has applications that go beyond kinetic theory and we therefore introduce its logic with considerable generality.

As laid out in the original work on this approach~\cite{Brewer:2019oha}, the AH scenario can be attained if the dynamics of the system is described by an evolution equation with the form
\begin{equation} \label{eq:general-evol}
    \frac{\partial \ket{\psi}}{\partial t} = - H[\psi; t] \ket{\psi} \, ,
\end{equation}
where $\ket{\psi}$ is a state vector containing all of the many variables needed to describe the state of the many-body system at time $t$. In the case of a strongly coupled many-body theory with a holographic description, the state vector would be specified by the metric and other fields along some hypersurface in a spacetime with one additional dimension. 
More relevant for us in this paper, in a kinetic theory the state vector can encode a distribution function via
$f({\bs x},{\bs p},t) = \braket{{\bs x},{\bs p} | \psi(t)}$. In our discussion in the rest of this paper, we will often refer to $f$ and $\ket{\psi}$ as the state of the system interchangeably. 
The form of Eq.~(\ref{eq:general-evol})  will enable us to carry over some of the intuition developed for analog problems in quantum mechanics, and in particular will enable us to use the adiabatic approximation as an organizing principle.
Nevertheless, there are important conceptual differences between Eq.~(\ref{eq:general-evol}) and the standard formulation of quantum mechanics itself, all of which are related to the time evolution operator $H$:
\begin{enumerate} 
    \item The prefactor in front of the $H$ on the RHS of~\eqref{eq:general-evol} is real not imaginary, and by convention is chosen to be $(-1)$. 
    \item In general, $H$ will be a non-Hermitian operator.
    \item $H$ can depend on the state of the system $\ket{\psi}$, and therefore the evolution of the system is, in general, nonlinear.
\end{enumerate}
The sign convention in the first point is so that if the real part of the eigenvalue spectrum of $H$ is bounded from below, as will be the case in examples of physical interest, we may be able to organize the directions in the vector space of states by the ``speed'' at which they evolve. This will later allow us to single out ``slow modes,'' i.e., solutions that are long-lived compared to all the others. 
While the first point just described is only a convention that aids us in organizing our description of the dynamics, the second and third points are necessary ingredients to describe interacting many-body theories through a kinetic equation with collisions. We shall discuss these two points in turn.

Non-hermiticity of the time-evolution operator means that the left and right eigenstates of $H$ at time $t$ (henceforth the instantaneous eigenstates) will not be naively related by adjoint conjugation. That is to say, if $\ket{n}_R$ is a right eigenstate of $H$, i.e. $H \ket{n}_R = \epsilon_n \ket{n}_R$, it does not follow that $(\ket{n}_R)^\dagger H = \bra{n}_R H = \bra{n}_R \epsilon_n$. However, there does exist a set of states $\{\bra{n}_L\}_n$, which together with $\{\ket{n}_R\}_n$ and the eigenvalues $\epsilon_n$ satisfy
\begin{align}
    H \ket{n}_R = \epsilon_n \ket{n}_R \, , & & \bra{n}_L H = \bra{n}_L \epsilon_n \, , & & \bra{m}_L \ket{n}_R = \delta_{mn} \, , \label{eq:mutual-orthogonality}
\end{align}
where we take ${\rm Re}\{\epsilon_m\} \leq {\rm Re}\{\epsilon_n\}$ if $m < n$, and $n,m \in \{0,1,2,\ldots\}$.
We can then decompose the state of the system as a linear superposition of instantaneous eigenstates
\begin{equation} \label{eq:state-decomposition}
    \ket{\psi(t)} = \sum_n a_n(t) \ket{n(t)}_R = \sum_n \tilde{a}_n(t) e^{- \int^t dt' \epsilon_n(t') } \ket{n(t)}_R \, 
\end{equation}
where the coefficients $a_n(t)$ that specify the superposition by telling us the occupation of each of the instantaneous eigenstates at time $t$ evolve as
\begin{equation}
    \partial_t \ln \frac{a_n}{a_m} = \partial_t \ln \frac{\tilde{a}_n}{\tilde{a}_m} - (\epsilon_n(t)-\epsilon_m(t)).
    \label{eq:2.4}
\end{equation}
This means that if there is a ``low-energy'' state $m$ (or a set thereof) such that ${\rm Re} \{\epsilon_n\} > {\rm Re} \{\epsilon_m\} $ for all $n > m$ (that is, ${\rm Re} \{\epsilon_n\} > {\rm Re} \{\epsilon_m\} $ for all $m$ in said set of low energy states and all $n$ not in said set), and if the condition
\begin{equation} \label{eq:adiabatic-approx}
    \partial_t \ln \left| \frac{\tilde{a}_n}{\tilde{a}_m} \right|   <  {\rm Re} \{\epsilon_n(t)\}-{\rm Re} \{\epsilon_m(t)\},
\end{equation}
is satisfied for all low energy $m$'s and all high energy $n$'s, then via (\ref{eq:2.4}) we have $\partial_t \left| a_n/a_m \right|<0$ and can therefore conclude that the occupation of states with larger relative values of ${\rm Re}\{\epsilon_n\}$ (``high-energy" states) will decay faster than that of ``low-energy'' states. 
In such a system, the long-lived solutions necessarily correspond to states in which only these low-energy states of the ``effective Hamiltonian'' $H$ are occupied. After an early transient period during which the occupation of the high-energy states decays away, the subsequent evolution of the system follows that of the evolving low-energy states, and is referred to as adiabatic.  
The condition (\ref{eq:adiabatic-approx}) is thus the condition that must be satisfied in order for the evolution to become adiabatic, provided the inequality never comes arbitrarily close to being violated.  Furthermore, if all the $\tilde{a}_n$'s and $\tilde{a}_m$'s are constant in time, 
then the adiabatic condition (\ref{eq:adiabatic-approx}) is strictly satisfied as its left-hand side vanishes and
the system is perfectly (and in a sense trivially) described by adiabatic evolution. The BSY paper~\cite{Brewer:2022vkq} 
provides us with an example of this circumstance. In such a case, the
so-called ``attractor'' solutions of the theory are  described described exactly by the ground state(s) of $H$~\cite{Brewer:2022vkq}. In the situation considered in the BSY paper this was achieved by finding a set of time-dependent coordinates in which the instantaneous eigenstates of $H$ were time-independent which, as per Eq.~\eqref{eq:evol-coeffs} below, fulfills the condition (\ref{eq:adiabatic-approx}). 
If instead the system is such that 
$\partial_t \tilde{a}_n \neq 0$ but the
adiabatic approximation~\eqref{eq:adiabatic-approx} is nevertheless satisfied, 
then the ground state(s) of $H$ will describe the attractor(s) up to corrections controlled by the severity of the deviation from $\partial_t \tilde{a}_n = 0$. In practical situations, the attractor will be realized up to transients stemming from a general initial condition, which means that the adiabatic approximation will be useful to identify the long-lived solutions.

By substituting the eigenstate decomposition \eqref{eq:state-decomposition} into the evolution equation \eqref{eq:general-evol}, one finds that 
\begin{equation} \label{eq:evol-coeffs}
    \frac{\partial_t a_n}{a_n} = - \epsilon_n - \sum_{n'} \frac{a_{n'}}{a_n} \bra{n}_L\partial_t \ket{n'}_R \,.
\end{equation}
Let us therefore introduce an adiabaticity criterion that we will find more useful in practice:
\begin{equation}
    \delta_A^{(n,m)} \equiv \left| \frac{\bra{ n}_L\partial_t \ket{m}_R}{\epsilon_n-\epsilon_m} \right| \ll 1 \, , \quad \forall n, m \ \ {\rm s.t.} \ \epsilon_n \neq \epsilon_m \, .
\label{eq:adiabaticitynm}
\end{equation} 
We shall show below that if this criterion is satisfied then the adiabatic criterion that we first introduced in the form \eqref{eq:adiabatic-approx} 
is satisfied.
For our purposes, though, with the goal of identifying attractors in mind we first note that since we expect that the excited states decay faster than the ground state we can conclude immediately that if the condition \eqref{eq:adiabaticitynm} holds for all $m$ in the set of low energy states and all $n$ outside that set then this will ensure that the high-energy states are not driven to large occupation numbers $a_n$ on account of their coupling to the set of instantaneous ground states in the evolution equation~\eqref{eq:evol-coeffs}. 

In fact, we can reproduce~\eqref{eq:adiabatic-approx} from~\eqref{eq:adiabaticitynm} by writing
\begin{align}
    \partial_t \ln \frac{a_n}{a_m}  &= \frac{\partial_t a_n}{a_n} - \frac{\partial_t a_m}{a_m} \nonumber \\
    &= -\epsilon_n + \epsilon_m - \sum_{n'} \frac{a_{n'}}{a_n} \bra{n}_L\partial_t \ket{n'}_R + \sum_{n'} \frac{a_{n'}}{a_m} \bra{m}_L\partial_t \ket{n'}_R  \, , \label{eq:relating-adiab-criteria}
\end{align}
from which it is clear that if the coefficients $a_n$ are all ${\cal O}(1)$ numbers then if the criterion~\eqref{eq:adiabaticitynm} holds for all $n,m$ this implies that $\partial_t \ln |a_n/a_m| \approx - {\rm Re} \{ \epsilon_n - \epsilon_m \}$ (with both of these quantities  $< 0$ if $n > m$, as hypothesized). 
Using \eqref{eq:2.4}, this in turn implies that the adiabatic criterion in the form \eqref{eq:adiabatic-approx} that we first stated is satisfied.
In order to convince ourselves that a strict inequality in~\eqref{eq:adiabatic-approx} is preserved under time evolution, we must check the case in which $|a_n| \ll |a_m|$ if $n$ labels a high-energy state and $m$ labels a low-energy state. In this case, the last sum in~\eqref{eq:relating-adiab-criteria} is either small or ${\cal O}(1)$, 
meaning that the sign of the whole expression will not depend on it provided the gap $\epsilon_n - \epsilon_m$ is sufficiently large. However, $a_{n'}/a_{n}$ in the first sum in \eqref{eq:relating-adiab-criteria} can be large when $n'$ labels a low energy state because $n$ is a high-energy state. In this case it becomes imperative that~\eqref{eq:adiabaticitynm} holds when $m$ is a low-energy state and $n$ is a high-energy state, so that the fact that $a_{n'}/a_{n}$ is large does not prevent the excited states from decaying. This is exactly the statement that long-lived attractor solutions are completely captured by the set of low-energy states. If~\eqref{eq:adiabaticitynm} didn't hold when $m$ is a low-energy state and $n$ is a high-energy state, then excited states would be sourced by the low energy $n'$ states in the first sum of~\eqref{eq:relating-adiab-criteria} and the late time solution would not be dominated by low-energy states alone. 

In keeping with the particular importance of slow variation of low-energy states, we can focus on the case in which the system begins in a state that is close to its ground state (or close to a superposition of its low energy states) which allows us to sharpen \eqref{eq:adiabaticitynm}, because the only aspect of this criterion that then matters is  
\begin{equation} \label{eq:adiabaticityn0}
    \delta_A^{(n)} \equiv \left| \frac{\bra{n}_L\partial_y\ket{0_m}_R}{\epsilon_n-\epsilon_0} \right| \ll 1 
\end{equation}
where $\ket{0_m}_R$ is any one of the low-energy states and where $n$ labels any one of the higher energy states.
If the criterion \eqref{eq:adiabaticityn0} is satisfied, 
the subsequent evolution will rapidly converge to the adiabatically evolving ground state(s).
We see that, in addition to rendering it in a form with practical utility, phrasing the adiabatic criterion 
in the form \eqref{eq:adiabaticityn0} naturally encapsulates the essence of the adiabatic approach that is at the core of all the analyses that follow later in this paper.

Returning to the three listed differences between our adiabatic framework and adiabaticity in quantum mechanics, in practice, the main obstacle to making progress is point 3. This is because the instantaneous eigenstates and eigenvalues of $H$ will also depend on the state of the system, and so solving for them in order to write down the decomposition~\eqref{eq:state-decomposition} is, in general, highly nontrivial for an arbitrary state $\ket{\psi}$.
Nonetheless, this does not obstruct the previous reasoning: given a state $\ket{\psi}$ at time $t$, one can calculate the instantaneous eigenstates $\ket{n[\psi,t]}_R$ and decompose the state on this basis. If after doing this one finds that $\ket{\psi}$ is approximately equal to a superposition of the low-energy instantaneous eigenstates, then the system will evolve slowly, remaining dominated by these low-energy states. If the adiabatic criterion is met, then even if the system is initialized with a non-zero occupancy for the excited states, these will decay and only leave the slow modes driving the system.

If one is able to overcome the practical barriers to framing the system's evolution in these terms, as we will outline later in this paper, one gains both intuitive understanding of the physics as well as predictive power. Specifically, one will have achieved:
\begin{itemize}
    \item A systematic organization of the theory in terms of long- and short-lived modes.
    \item A characterization of the dynamically preferred solutions of the theory, i.e., of out-of-equilibrium attractors.
    \item Predictive power for complex systems in terms of a small number of degrees of freedom. Concretely, once the ``relevant'' degrees of freedom have been identified, one can truncate the evolution equation~\eqref{eq:general-evol} to the relevant subspace and solve it only for that small set of degrees of freedom.
\end{itemize}

The key ingredient, which is not guaranteed to be present in a physical system, is that we are able to describe its evolution in a framework where the adiabatic criterion is met. To achieve this, the rest of this paper deals with the problem of how to set up a vector space describing a distribution function where the instantaneous eigenstates of the time-evolution operator stemming from the underlying kinetic theory evolve adiabatically. As we will see, one ingredient that can solve this problem almost entirely is to first characterize the evolution of the relevant dimensionful scales of the system, treat those as ``background'' quantities for the evolution of the system, and then find the slow modes of the remaining degrees of freedom. This was first realized by BSY, where the fact that the distribution function had a self-similar evolution was exploited to achieve exact adiabaticity after a time-dependent scaling of the momenta was performed. In Section~\ref{sec:scaling}, we review and extend their results towards a complete treatment of QCD kinetic theory in the small-angle scattering approximation by looking at the different limits where scaling phenomena appear. Later, in Section~\ref{sec:non-scaling} we will see how the reduction of degrees of freedom appears even in solutions where there is no self-similar scaling solution valid for all times, thus explicitly showing how Adiabatic Hydrodynamization can take place in kinetic theories derived directly from QCD.

In the remainder of this Section, we will introduce the concrete kinetic theory setup that we shall employ in the rest of the paper, and discuss how conserved quantities provide another extra ingredient that will aid us in organizing the theory.

\subsection{Kinetic theory setup} \label{sec:kin}

To study the hydrodynamization process of a Yang-Mills plasma, we consider a kinetic theory description of the gluon phase space density encoded in a distribution function $f({\bf x}, {\bf p},t)$, where $(t, {\bf x})$ are Minkowski coordinates and ${\bf p}$ labels the 3-momentum of the gluons. The gluon distribution function evolves according to a kinetic equation, also referred to as a Boltzmann equation:
\begin{equation} \label{eq:Boltzmann}
    \frac{\partial f}{\partial t} + \frac{\bf p}{p} \cdot \nabla_{\bf x} f = - \mathcal{C}[f] \, ,
\end{equation}
which is specified by the collision kernel $\mathcal{C}[f]$. For weakly coupled QCD, the collision kernel was obtained in~\cite{Arnold:2002zm} and the resulting theory is called QCD Effective Kinetic Theory (EKT).

In the context of heavy-ion collisions, it is appropriate to work in coordinates that incorporate some information from the geometry of the collisions. Taking $z$ to be the coordinate along the beam axis, it is convenient to go to Milne coordinates $(\tau,{\bf x}_\perp, \eta)$, specified by
\begin{align}
    t = \tau \cosh \eta \, , & & z = \tau \sinh \eta \, , & & {\bf x}_\perp = {\bf x}_\perp \, ,
\end{align}
where the natural momentum variables at each point $(\tau,{\bf x}_\perp, \eta)$ are
\begin{align}
    p_\tau = p \cosh \eta - p_z \sinh \eta \, , & & p_\eta = p_z \cosh \eta - p \sinh \eta \, , & & {\bf p}_\perp = {\bf p}_\perp \, .
\end{align}
The collision is assumed to take place at $\tau = 0 \implies (t, z) = (0, 0)$.

In terms of these new coordinates, the kinetic equation is given by
\begin{equation} \label{eq:Boltzmann-Milne}
    \frac{\partial f}{\partial \tau} + \frac{{\bf p}_{\perp}}{p_\perp} \cdot \nabla_{{\bf x}_\perp} f + \frac{p_\eta}{\tau p_\perp} \frac{\partial f}{\partial \eta} - \frac{p_\eta}{\tau} \frac{\partial f}{\partial p_\eta} = - \mathcal{C}[f] \, .
\end{equation}
In a realistic description of a heavy-ion collision, one should explicitly take into account the dependence of $f$ on all of these variables. In practice, however, we will use two simplifying assumptions:
\begin{enumerate}
    \item We will assume boost invariance of $f$, i.e., $\partial f/\partial \eta = 0$. In the CM frame of an $AA$ collision, this will be true at mid-rapidity $\eta = 0$, and approximately true for some range of $\eta$ around $\eta=0$ that becomes larger in collisions with higher energy. Therefore, for observables that focus on the mid-rapidity region in collisions at top RHIC energies and at the LHC, this assumption is not too costly. Furthermore, because at $\eta = 0$ we have $p_\eta = p_z$, we will use $p_z$ throughout this paper in the place of $p_\eta$.
    \item We will assume translation invariance in the transverse plane, $\nabla_{{\bf x}_\perp} f = 0$. This is certainly a major simplification that blatantly ignores the fact that the droplets of QGP produced in a HIC have finite transverse size. However, by simple inspection of the kinetic equation, it can be a good approximation at early times $\tau$ because of the $1/\tau$ prefactor in the $\partial f/\partial p_\eta$ term, provided the transverse gradients are not large. In particular, as long as the transverse extent of the droplet is much larger than the hydrodynamization time, neglecting the build up of radial flow during the pre-hydrodynamic epoch is a reasonable starting point for its analysis. 
\end{enumerate}
Clearly, the one of these two assumptions that would be most interesting to relax for HIC phenomenology would be translation invariance in the transverse plane; neglecting the lumpiness (in the transverse plane) of the matter produced initially in a HIC -- for example that arising because nuclei are made of nucleons -- cannot be justified by the argument above. However, we leave the investigation of its consequences to future work, as they are beyond the scope of this paper. 
We note, also, that both of these assumptions have been widely used in studies of QCD EKT applied to HICs~\cite{Keegan:2016cpi,Kurkela:2018vqr,Kurkela:2018wud,Mazeliauskas:2018yef}. In fact, the pioneering paper on the ``bottom-up'' thermalization scenario~\cite{Baier:2000sb}, which describes the stages via which thermalization would occur in a HIC within the framework of perturbative QCD and upon assuming weak coupling employs precisely these assumptions.

The only ingredient we have yet to discuss is the collision kernel $\mathcal{C}[f]$. Ideally, we would simply use the full collision kernel of QCD EKT. However, as doing so will suffice for our present purposes, namely to establish how hydrodynamization can be understood in terms of the instantaneous ground state of an effective Hamiltonian as prescribed by the AH scenario~\cite{Brewer:2019oha}, we will make further simplifications compared to the full collision kernel. Concretely, we will study the small-angle scattering approximation~\cite{Mueller:1999pi,Serreau:2001xq,Blaizot:2013lga}, where the collision kernel takes a Fokker-Planck form
\begin{equation}
	\label{eq:small-angle-kernel}
  \mathcal{C}[f] = -\lambda_0 \ell_{\rm Cb} [f] \left[ I_a[f] \nabla_{\bf p}^2 f + I_b[f] \nabla_{\bf p} \cdot \left( \frac{\bf p}{p} (1+f) f \right) \right]\, ,
\end{equation}
where $\lambda_0 = 4\pi \alpha_s^2 N_c^2$. (Note that $\lambda_0=\lambda_{\rm 't~Hooft}^2/(4\pi)$, where the 't Hooft coupling is given by $\lambda_{\rm 't~Hooft}=4\pi\alpha_s N_c$.) The functionals $I_a$, $I_b$ are given by
\begin{align}
  I_a[f] = \int_{\bf p} f (1+f), \, 
  \qquad
  I_b[f] = \int_{\bf p}\, \frac{2}{p} f\, , 
  \label{eq:IaIb}
\end{align}
where here and throughout the paper we use the shorthand notation $\int_{\bf p} \equiv \int \frac{d^3 p}{(2\pi)^3}$. 
The integral $I_{a}$ represents the effective density of scatterers, including a Bose enhancement factor $(1+f)$ due to the bosonic nature of gluons.  $I_{b}$ is related to the Debye mass $m_{D}$, which is the typical momentum exchange per collision, through
\begin{equation}
\label{mD}
  m_D^2 = 2 N_c g^2_{s} I_b\, .
\end{equation}
Furthermore, one can check that a Bose-Einstein distribution $f = (\exp(-p/T_{\rm eff}) - 1)^{-1}$ with $T_{\rm eff} \equiv I_a/I_b$ satisfies $\mathcal{C}[f] = 0$, meaning that this is therefore the equilibrium distribution for this collision kernel. For this reason, we shall refer to $T_{\rm eff} = I_a/I_b$ as the effective temperature of the distribution function, even when $f$ is not the equilibrium distribution. 
The remaining ingredient in (\ref{eq:small-angle-kernel}) is the Coulomb logarithm $\ell_{\rm Cb}$. This is a (perturbatively divergent) integral over the small scattering angle~\cite{Mueller:1999pi} 
\begin{equation}
\label{eq:lcb-def}
  \ell_{\rm Cb} [f] =\ln \left( \frac{p_{\rm UV}}{p_{\rm IR}} \right)\, ,
\end{equation}
where $p_{\rm UV}$ and $p_{\rm IR}$ are UV and IR cutoffs, respectively. Following Refs.~\cite{Mueller:1999fp,Brewer:2022vkq}, we set $p_{\rm IR} = m_D$ and $p_{\rm UV} = \sqrt{\langle p^2 \rangle},$\footnote{The authors of Ref.~\cite{Brewer:2022vkq} (BSY) chose $p_{\rm UV} = \sqrt{\langle p_\perp^2 \rangle}$ because in the regime considered in that work $\langle p^2 \rangle \sim \langle p_\perp^2 \rangle \gg \langle p_z^2 \rangle$} where the brackets $\langle \cdot \rangle$ denote the average over the distribution function $f$:
\begin{equation}
    \langle X \rangle \equiv \frac{\int_{\bf p} X f }{\int_{\bf p} f} \, .
\end{equation}

Throughout our exposition we will consider this collision kernel in various regimes, starting from the simplest cases and building up to a nearly complete treatment of it. We will at times study it in regimes where the approximations that are used to derive it (weak coupling, small-angle scatterings) are not justified from first principles, in particular in Sections~\ref{sec:hydro} and~\ref{sec:connectstages} where we consider couplings as large as $g_s = 2.6$. Numerical details of the implementation are available in Appendix~\ref{app:numerics}. 
Our treatment is ``nearly complete'' in that we also make one further simplifying assumption throughout this work, dropping
the term proportional to $I_b f^2$ in the collision kernel~\eqref{eq:small-angle-kernel}. We do so for simplicity, and because BSY (and we) have checked that this is a good approximation in the early pre-hydrodynamic attractor and because it is manifestly a good approximation as hydrodynamization is achieved in our analysis, since $f$ is then small. We provide further discussion of this approximation in Appendix~\ref{app:f2}. This approximation has a further consequence in that our description does not exhibit the Bose-Einstein condensation dynamics that 
can arise in other treatments based upon the small-angle scattering collision kernel~\eqref{eq:small-angle-kernel} with only number-conserving $2\rightarrow 2$ collisions~\cite{Blaizot:2011xf,Blaizot:2014jna,Blaizot:2016iir,BarreraCabodevila:2022jhi}. We discuss this further in Appendix~\ref{app:f2}.

As we have already discussed in Section~\ref{sec:Intro}, our goal in this work is to find a common physical description of, and intuition for, the processes occuring during all the stages of hydrodynamization in the kinetic theory with the simplified collision kernel that we have specified here. Extending our analysis to apply it to the computationally more challenging QCD EKT collision kernel is a priority for future work, but we anticipate that the physical intuition that we shall glean from the analyses to come below, in particular from Sect.~\ref{sec:connectstages} where we put all the pieces together, will carry over.

\subsection{Conserved quantities in Adiabatic Hydrodynamization}
\label{sec:ConservationLaws}

The fact that hydrodynamics is an effective theory for the dynamical evolution of densities of the conserved quantities of a given system suggests that in the process of hydrodynamization such quantities should also play a central role. Indeed, as we will see momentarily, they can provide crucial information to organize the analysis of the pre-hydrodynamic evolution more efficiently. In the present context, if a quantity $X[f]$ is conserved by the kinetic equation, and the AH scenario is realized, i.e.~if the system approaches hydrodynamics via first being driven into and then following an evolving instantaneous ground state (or set of slow modes which can collectively be considered as ``ground states''), then the conserved quantity must be encoded in such ground state(s).

In the AH framework, the conserved quantities that are most useful are those that can be written in terms projections of the physical state $\ket{\psi}$, i.e., those quantities $K$ that can be written as
\begin{equation}
	K = q_K(\tau) \braket{P_K | \psi(\tau) } \, ,
\end{equation}
where $\bra{P_K(\tau)}$ is a projector that, once applied on the state yields the conserved quantity $K$, and $q_K(\tau)$ is a time-dependent prefactor that does not depend on the state (i.e., its time dependence is explicit).

For a spatially homogeneous, non-expanding gluon gas (i.e., without assuming boost invariance in Milne coordinates), in Minkowski coordinates as in~\eqref{eq:Boltzmann}, and described by the collision kernel~\eqref{eq:small-angle-kernel}, there are two conserved quantities
\begin{align}
	K_N = \braket{P_N | \psi} \equiv \int_{\bf p} f({\bf p}, t) \, , & & K_E = \braket{P_E | \psi} = \int_{\bf p} p \, f({\bf p}, t) \, ,
\end{align}
which are proportional to the number density and energy density of the gluon gas, respectively. 

In the case of a boost-invariant, transversely homogeneous, expanding gluon gas, neither of the previous quantities is conserved. However, the rescaled number density $K_{\tilde{N}} = \tau K_N$ is conserved. Explicitly,
\begin{equation}
	K_{\tilde{N}} = \tau \braket{P_{N} | \psi} \equiv \tau \int_{\bf p} f({\bf p}, \tau) \, ,
\end{equation}
is conserved. Note that the fact that there can be a time-dependent prefactor $q_{\tilde{N}} = \tau$ is crucial for this definition.

As we shall now explain, identifying such a conserved quantity is extremely helpful for implementing the construction that we have described abstractly above,  in particular when choosing a set of left and right basis states to solve the Boltzmann equation~\eqref{eq:Boltzmann-Milne} by writing it in the effective Hamiltonian form~\eqref{eq:general-evol}.
Because of the crucial role that rescalings take in this step, the precise way in which this rewriting is done will be one of the central matters of the following Sections.  Even without an explicit version of~\eqref{eq:Boltzmann-Milne} in the form of~\eqref{eq:general-evol} in hand, one can formally see right away how taking account of conserved quantities will be useful in this regard. If one chooses the left basis $\{\bra{\psi_n}_L\}_{n=1}^{N_{\rm basis}}$ such that the first basis state $\bra{\psi_1}_L$ is exactly the projector $\langle P_N |$, then because acting with $\partial/\partial \tau$ on the conserved quantity $\tau \braket{P_{N} | \psi(\tau) } $ yields zero, from (\ref{eq:general-evol}) we have 
\begin{align}
	\bra{P_{N}} H = \frac{1}{\tau} \bra{P_N} \, ,
\end{align}
i.e., by construction, the first basis state will be a left eigenvector of the Hamiltonian,\footnote{In this discussion, $H$ is the Hamiltonian that generates time evolution with respect to the coordinate $\tau$. It will later turn out to be convenient to introduce a new time coordinate $y = \ln (\tau/\tau_I)$ such that the eigenvalue of the corresponding Hamiltonian acting on this left state is constant.} which we will denote by $\bra{\phi_1}_L$.

Furthermore, this means that there will be a right eigenvector $\ket{\phi_1}_R$ of the Hamiltonian with the same eigenvalue, and, moreover, that all other right eigenvectors $\ket{\phi_n}_R$ (with $n \neq 1$) of the Hamiltonian will be orthogonal to this projector, i.e., $\bra{\phi_1}_L \ket{\phi_n}_R = 0$ $\forall n \neq 1$. As such, all the number density of the system is exclusively contained within the corresponding right eigenvector $\ket{\phi_1}_R$, and the value of the number density is determined by the expansion coefficient $a_1$ in the decomposition~\eqref{eq:state-decomposition}. It follows that if a system described by a positive-definite distribution function $f$ (i.e.~a system with nonzero number density) has an attractor solution described by the AH scenario, then $\ket{\phi_1}_R$ must be (one of) the ground state(s) determining the attractor  and driving the evolution of the system. 

In addition to this, because we know beforehand that the systems we are studying can (and will eventually) hydrodynamize, we also know what the late-time ground state of the system should be: $\ket{\phi_1}_R$ should describe either a Boltzmann distribution if the Bose-enhancement terms are dropped, such that $f \propto \exp(-p/T)$, or a Bose-Einstein distribution $f \propto (\exp(p/T)-1)^{-1}$ if the effects of bosonic quantum statistics are not neglected. It is thus appropriate to choose the first basis state (at least at late times) such that it describes a distribution in local thermal equilibrium, because if the system does hydrodynamize, this will guarantee a quick convergence of the sum in the basis state decomposition of the ground state, and consequently a good quantitative description. Having fixed the first left and right basis states, and provided with an inner product (which we take to be simply determined by $\int_{\bf p}$ or a rescaling thereof, as we discuss in the next Section) and a family of functions to generate the basis, the Gram-Schmidt method uniquely determines the form of all the other (mutually orthogonal) left and right basis states.\footnote{One may wonder what would change in this discussion if we had two conserved quantities, as in the case of a spatially homogeneous, non-expanding gluon gas. In practice, as it turns out, it is impossible to choose a basis where the left basis contains the projectors associated with both number and energy density, and the right basis contains a state describing a positive definite distribution function while satisfying the mutual orthogonality conditions~\eqref{eq:mutual-orthogonality}. This means that in practice when we apply the construction described here we must do so upon choosing only one of the two conserved quantities, and we shall use number conservation.}

Having laid out the AH framework, in the next two Sections we shall make everything we have described above fully explicit (including in particular fully specifying $H$) in several kinetic theory calculations with increasing completeness. First, though, in the next Section we shall discuss the one remaining ingredient that one needs in order to achieve an adiabatic description of out-of-equilibrium kinetic theories. Namely, how should we address
the fact that 
the variables that describe the state of the system (the eigenstates/values of $H$ and the
coefficients that specify the occupation of these eigenstates) and consequently
the typical scales that characterize the state of the system (such as the typical longitudinal momentum of gluons $\sqrt{\langle p_z^2 \rangle}$ or the typical transverse momentum $\sqrt{\langle p_\perp^2 \rangle}$) 
may be rapidly evolving even if the dynamics have driven the system to an attractor solution. Note that the projections of the state $\langle \psi^{(L)}_n | \psi \rangle$ onto a given basis $\{|\psi^{(R)}_n \rangle \}_n$ (satisfying  orthogonality conditions analogous to those for the eigenstate basis) are in practice moments of the distribution function, of which $\langle p_z^2 \rangle$, $\langle p_\perp^2 \rangle$ are examples. This means that in order to make it possible to describe said attractor solutions as adiabatically evolving (sets of) ground state(s), one needs to choose a set of basis states and/or rescale the momentum and time coordinates  in such a way as to recast the description of the evolution of the attractor solution in a way that makes it adiabatic.
There is a large class of collision kernels and kinematic regimes, including all of those described in Section~\ref{sec:scaling}, for which this issue is most appropriately dealt with by introducing ``scaling variables'' that map the physical momentum coordinates to rescaled momentum coordinates, thus giving one the freedom to choose different frames (this use of the term ``frame'' originates from BSY, in a loose analogy with the reference frame concept in relativity where one is free to choose coordinates at will) to study the system. Following BSY, we will refer to a frame that makes the evolution adiabatic, in the sense described above and made explicit in the next Section, as an ``adiabatic frame.''

\section{Scaling and adiabaticity}
\label{sec:scaling}

Scaling phenomena are a hallmark of universality in physical systems.
They also play a prominent role in the behavior of out-of-equilibrium systems across a whole range of phenomena including HICs~\cite{Baier:2000sb,Berges:2013eia,Kurkela:2014tea,Mukherjee:2016kyu,Mukherjee:2017kxv,Mazeliauskas:2018yef} and cold atoms~\cite{Mikheev:2018adp,Prufer:2018hto,Erne:2018gmz}, as well as more general theories~\cite{PineiroOrioli:2015cpb}, the Kibble-Zurek phenomenon of defect formation after out-of-equilibrium phase transitions in cosmology or condensed matter phyics~\cite{Kibble:1976sj,Zurek:1985qw,Chuang:1991zz,Zurek:1996sj,Chandran:2012cjk}, and even turbulence~\cite{frisch1995turbulence}. In this Section, we will detail how identifying time-dependent scaling in a theory allows us to analyze its underlying adiabatic evolution.

A lot can be gained quantitatively by treating the evolution of the typical scales of the system on a different footing than the rest of the system's evolution. Specifically, if one knows that the evolution of the expectation value of a quantity $x$ is given by some calculable function of time $\langle x\rangle (t)$, then the rest of the features of the distribution of $x$ are easier to analyze as a function of $x/\langle x \rangle$, where the dominant time dependence of the expectation values have been scaled out. Provided that the system can undergo scaling, this is the natural way in which to compare long- and short-lived modes.

To put this on a concrete footing, we now specialize to scaling phenomena in kinetic theory, and demonstrate how the intuitive picture we just described can be realized in terms of the AH framework. For the purposes of this discussion, we will consider a distribution function $f$ that is homogeneous 
in the spatial Milne coordinates $(\eta, {\bs x}_\perp)$ and symmetric under rotations in the ${\bs p}_\perp$ plane.
Such a distribution function can be said to be ``scaling" if it takes the form
\begin{equation} \label{eq:scalingform}
    f(\bm{p},\tau) = A(\tau) w\left(\frac{p_\perp}{B(\tau)},\frac{p_z}{C(\tau)}\right)
\end{equation}
for some time-dependent $A,B,$ and $C$, such that the rescaled distribution function $w = w(\zeta,\xi)$ is independent of time. If the dynamics of the system causes $f$ to generically fall into such a scaling form with a universal underlying shape $w$, this $w$ can be viewed as an attractor. This is true in, for example, the bottom-up thermalization picture~\cite{Baier:2000sb}, in which the early-time dynamics of the system is described by the scaling form (\ref{eq:scalingform}) with 
\begin{equation}
    A_\text{BMSS}(\tau) \propto \tau^{-\frac{2}{3}}, \;\;\; B_\text{BMSS}(\tau)  \propto \tau^0, \;\;\; C_\text{BMSS}(\tau)  \propto \tau^{-\frac{1}{3}} \, , \label{eq:BMSS-scalings}
\end{equation}
where BMSS refers to the authors of~\cite{Baier:2000sb}. In this specific case, the ``scaling exponents"
\begin{equation}
    \alpha \equiv \frac{\tau}{A} \frac{\partial A}{\partial \tau}, \;\;\; \beta \equiv -\frac{\tau}{B} \frac{\partial B}{\partial \tau}, \;\;\; \gamma \equiv -\frac{\tau}{C} \frac{\partial C}{\partial \tau}
    \label{eq:scaling-exponents}
\end{equation}
are time-independent with
\begin{align}
    \alpha_\text{BMSS}= -\frac23 \, , \;\;\; \beta_\text{BMSS}=0 \, , \;\;\;  \gamma_\text{BMSS}=\frac13 \, .
    \label{eq:BMSS-exponents}
\end{align} 
We see from \eqref{eq:scalingform} that the physical interpretation of the scaling exponents defined via \eqref{eq:scaling-exponents} is that they describe the instantaneous logarithmic rate of change of the typical occupancy and momentum scales in the distribution function: 
\begin{equation}
    \alpha =\frac{\partial \ln \langle f\rangle}{\partial \ln \tau}, \;\;\; \beta = -\frac{1}{2}\frac{\partial\ln\langle p_\perp^2\rangle}{\partial\ln\tau}, \;\;\; \gamma =-\frac{1}{2}\frac{\partial\ln\langle p_z^2\rangle}{\partial\ln\tau}\,.
    \label{eq:scaling-exponents-2}
\end{equation}
The scaling phenomenon introduced by BMSS has been verified in numerous numerical simulations, both in classical-statistical Yang-Mills simulations~\cite{Berges:2013eia,Berges:2013fga}, as well as more recently in QCD EKT~\cite{Mazeliauskas:2018yef}, where it was observed that these exponents could be well-defined even away from their fixed point values. That is to say, the distribution function collapsed onto the form in Eq.~\eqref{eq:scalingform} even before Eq.~\eqref{eq:BMSS-exponents} came to be fulfilled.

The values of the scaling exponents $\alpha$, $\beta$ and $\gamma$ in early-time BMSS dynamics were correctly predicted via analytical arguments~\cite{Baier:2000sb} in early work, but only recently was the analytical description refined to include the evolution of the exponents as the fixed point is approached~\cite{Brewer:2022vkq}. (See also Ref.~\cite{Mikheev:2022fdl} for a stability analysis around the fixed point.) Our discussion in earlier Sections motivates interpreting the fact that the scaling exponents can be extended to times before the fixed point is attained by postulating that the scaling form~\eqref{eq:scalingform} is the \textit{effective} ground state of the system, undergoing adiabatic time evolution, and that this form (with the same functional form $w$ as for the fixed point) is approached because of the emergence of an energy gap between this ground state and
all the other shapes that the distribution function can attain. This is exactly what was conjectured and proven by BSY~\cite{Brewer:2022vkq}.

The BMSS scaling regime, as described above, is not the only scaling regime to (or through) which a weakly coupled gluon plasma can evolve. BSY also observed~\cite{Brewer:2022vkq} that the small angle scattering collision kernel admits another fixed point for the scaling exponents when the system undergoes longitudinal expansion, with a distribution function of the same form as~\eqref{eq:scalingform}. At this fixed point, called the \textit{dilute} fixed point, the scaling exponents are time-independent with
\begin{align}
    \alpha_\text{dilute}= -1 \, , \;\;\; \beta_\text{dilute} = 0 \, , \;\;\;  \gamma_\text{dilute} = 0 \, .
\label{eq:dilute-exponents}
\end{align}
Corrections to the values of the exponents as they approach the fixed point, both in the dilute and BMSS cases, can be calculated explicitly as shown by BSY, and were shown to evolve logarithmically in proper time $\tau$. A similar correction has been observed in the approach to fixed points in non-expanding QCD EKT~\cite{Heller:2023mah}.

Another example that is important to keep in mind is that in the absence of collisions, i.e., if one sets $C[f] = 0$, the particles in the kinetic theory stream freely. In this case, the evolution of the distribution function (assuming boost invariant longitudinal expansion and homogeneity in the transverse plane) \textit{always} takes the scaling form as in~\eqref{eq:scalingform}, with $w$ determined by the initial condition. The scaling exponents in this case are given by
\begin{align}
    \alpha_\text{fs}= 0 \, , \;\;\; \beta_\text{fs} = 0 \, , \;\;\;  \gamma_\text{fs} = 1 \, ,
    \label{eq:fs-exponents}
\end{align}
where ``fs'' denotes free streaming. 
We note that the dilute scaling regime we just described, with scaling exponents~\eqref{eq:dilute-exponents}, is distinct from free streaming,  in the sense that dilute scaling evolution only arises in the presence of collisions as it involves a balance between longitudinal expansion and collisions. In the analytic calculations of BSY, the term that balances the longitudinal expansion is the $I_a \nabla^2 f$ term in the collision kernel.

All the scaling regimes described above are pre-hydrodynamic. Hydrodynamization must end with the system evolving to a
hydrodynamic state in which the kinetic theory is in local thermal equilibrium, with the distribution function taking the form of a thermal distribution (i.e., Fermi-Dirac, Boltzmann, or Bose-Einstein). 
It has been known since Bjorken~\cite{Bjorken:1982qr} that the hydrodynamic evolution of a boost-invariant longitudinally expanding fluid is itself 
a scaling regime. The kinetic theory description of this regime takes the form
\begin{equation} \label{eq:scalingform-late}
    f(\bm{p},\tau) = w\left(\frac{p}{D(\tau)}\right) \, ,
\end{equation}
where $w(\chi)$ is the appropriate thermal distribution, and $D(\tau)$ plays the role of the local temperature, which evolves as $D(\tau) \propto \tau^{-1/3}$ due to the boost-invariant longitudinal expansion of the system. This distribution function can equivalently be written in the form of~\eqref{eq:scalingform}, with the scaling exponents \eqref{eq:scaling-exponents} then taking values
\begin{align}
    \alpha_\text{thermal}= 0 \, , \;\;\; \beta_\text{thermal} = \frac13 \, , \;\;\;  \gamma_\text{thermal} = \frac13 \, .
    \label{eq:hydro-exponents}
\end{align}

We will encounter all of these scaling regimes --- free-streaming, BMSS, dilute, and hydrodynamic --- later on in this paper.
As we shall reproduce and further elucidate in Sect.~\ref{sec:pre-hydro}, BSY succeeded in using the AH formalism to describe the pre-hydrodynamic evolution of a longitudinally expanding gluon gas with a small-angle scattering collision kernel that begins with free-streaming, is rapidly attracted toward the BMSS fixed point, and subsequently evolves toward the dilute fixed point. In Sect.~\ref{sec:hydro} we introduce a distinct variation of the AH formalism that is designed to describe the approach to hydrodynamic scaling and see explicitly how AH provides an intuitive understanding of hydrodynamization itself.  Finally, in Sect.~\ref{sec:non-scaling} we shall set up the AH framework that provides a unified description of the early evolution governed by a pre-hydrodynamic attractor (that 
takes the kinetic theory distribution from free-streaming to BMSS to dilute) followed by the evolution governed by a hydrodynamizing attractor that takes the distribution from the dilute scaling regime to the hydrodynamic regime. AH provides a unified intuition for this complex dynamics via elucidating how the approach to each new attractor involves the opening up of new gaps in the instaneous spectrum of an evolving effective Hamiltonian, followed by the adiabatic evolution of the remaining ground state(s). We shall see in Sects.~\ref{sec:hydro} and \ref{sec:non-scaling} that hydrodynamization itself occurs when only one isolated instantaneous  
ground state remains, with that state evolving adiabatically toward the hydrodynamic regime where the scaling exponents take the values \eqref{eq:hydro-exponents}.

First, though, in Sect.~\ref{sec:scaling-H} we describe the conceptual underpinnings of the findings of BSY and then, in Sect.~\ref{sec:isotropic},  demonstrate their applicability in a setting that is on the one hand simple but that nevertheless goes beyond the regime described by BSY, in so doing setting the stage for a complete description of the hydrodynamization process of gluons in the small-angle scattering approximation using the AH framework.

\subsection{Effective Hamiltonians for scaling distributions} \label{sec:scaling-H}

Consider a time before the distribution function $f$ takes a scaling form, which, for definiteness, we take to be of the form (\ref{eq:scalingform}) we just discussed. It is always possible to write $f$ in the form
\begin{equation}
    f(\bm{p},\tau) = A(\tau) w\left(\frac{p_\perp}{B(\tau)},\frac{p_z}{C(\tau)},\tau\right),
\end{equation}
and any choice of $A,B,$ and $C$ would comprise a valid choice of ``frame" for $f$, as all of the time dependence could be moved into the time evolution of $w(\cdot,\cdot, \tau)$. However, as $f$ approaches its scaling form, we expect that there is an optimal ``frame", namely an optimal choice of $A,B,$ and $C$ necessary to match Eq.~\eqref{eq:scalingform} as $f$ enters the scaling regime, which is to say necessary to ensure that $w(\zeta,\xi, \tau)$ approaches the functional form of $w(\zeta,\xi)$ in Eq.~\eqref{eq:scalingform}. Equivalently, we could change from one frame to another (and in so doing seek the optimal frame) by making 
time-dependent rescalings of $p_\perp$, $p_z$ and $\tau$.

Let us examine the behavior of a general rescaled distribution function $w$ to try to understand why it should be attracted to the scaling form $w(\zeta,\xi)$ in Eq.~\eqref{eq:scalingform}. Recalling Eq.~\eqref{eq:general-evol}, we shall do so by recasting the Boltzmann equation describing $f$ into an evolution equation for $w$ of the form
\begin{equation} \label{eq:hdef}
    \partial_y w = -H_{\rm eff} w
\end{equation}
where $y \equiv \ln \left( \frac{\tau}{\tau_0} \right)$, and where we shall refer to the operator $H_{\rm eff}$ as the ``effective Hamiltonian''. The effective Hamiltonian will in general be non-Hermitian, non-linear, and depend on our choice of rescalings $A,B,$ and $C$. Furthermore, the variables that define the space of states will be $\zeta$ and $\xi$, with $\zeta \equiv p_\perp/B$ and $\xi \equiv p_z/C$. 

This effective Hamiltonian is determined explicitly by the form of the collision kernel $C[f]$ and the choice of scaling variables. For a kinetic theory undergoing Bjorken flow, in the coordinate basis it reads
\begin{equation}
    H_{\rm eff} = \frac{\partial_y A}{A} - \frac{\partial_y B}{B} \zeta \partial_\zeta - \left( 1 + \frac{\partial_y C}{C} \right) \xi \partial_\xi - \tau \tilde{\mathcal{C}} \left[ f = A w(\zeta, \xi, \tau)|_{\substack{p_\perp \, = \, \zeta B \\ p_z \, = \, \xi C }} \right] \,\, ,
\end{equation}
where $\tilde{\mathcal{C}}[f]$ is a linear operator defined such that $\mathcal{C}[f] = \tilde{\mathcal{C}}[f] f$, i.e., such that its action on $f$ reproduces the collision kernel. In the case of the small-angle scattering approximation where $C[f]$ is given by Eq.~\eqref{eq:small-angle-kernel}, the effective Hamiltonian takes the explicit form
\begin{align}
    H_{\rm eff} =\alpha + \beta \zeta \partial_\zeta + (\gamma-1) \xi \partial_\xi & - \tau \lambda_0 \ell_{\rm Cb} I_a \left[ \frac{1}{B^2} \left( \frac{1}{\zeta} \partial_\zeta + \partial_\zeta^2 \right) + \frac{1}{C^2} \partial_\xi^2 \right] \label{eq:Heff-smallanglescatt-BC}  \\
& - \tau \lambda_0 \ell_{\rm Cb} I_b \left[ \frac{2(1+Aw)}{p} + \frac{(1+2Aw)}{p} \left( \zeta \partial_\zeta +  \xi \partial_\xi \right) \right] \, , \nonumber
\end{align}
where here $w = w(\zeta, \xi, y)$ because we have not yet shown explicitly that the adiabatic approximation is satisfied and the system rapidly collapses onto an adiabatic evolution in which $w$ depends only on its first two arguments.  The quantities $\ell_{\rm Cb}$, $I_a$, $I_b$ that appear in Eq.~(\ref{eq:Heff-smallanglescatt-BC}) are given by the expressions~\eqref{eq:IaIb} and~\eqref{eq:lcb-def} from Section~\ref{sec:kin}. Furthermore, we have introduced the scaling exponents  $\beta = - \partial_y B/B$, $\gamma = - \partial_y C/C$, $\alpha = \partial_y A/A$ as in~\eqref{eq:scaling-exponents}. In practice, we will choose $\alpha$ so that the real part of the lowest eigenvalue of $H_{\rm eff}$ is 0.

Finding the eigenstates and eigenvalues of $H_{\rm eff}$ can be challenging. Nonetheless, once a frame has been specified by choosing $A,$ $B$ and $C$ we can in principle write down the set of right eigenstates $\ket{n}_R$ of $H_{\rm eff}$ and decompose any state -- i.e.~any distribution function $f$ specified by the function $w$ -- at a given time $y$ in this basis:
\begin{equation}
    w(\zeta,\xi,y) = \sum_n a_n(y) \braket{\zeta,\xi|n(y)}_R \, ,
\label{eq:w-decomposition}
\end{equation}
where the dependence of 
$\ket{n}_R$ on $y$ contains both explicit and implicit dependencies, in the sense that it depends on $y$ through $B(y), C(y)$, and $w(\zeta,\xi,y)$. Furthermore, constructing the basis states $\ket{n}_R$ used in the decomposition \eqref{eq:w-decomposition} requires knowing the state, meaning that \eqref{eq:w-decomposition} is an implicit equation even at one time $y$.
However, this does not impede the applicability or implementation of the AH picture.

Now consider the notion of adiabaticity quantified by Eq.~\eqref{eq:adiabaticityn0} which was discussed in Sec.~\ref{sec:AH}. Inspecting this adiabaticity condition, we can see a clear connection between scaling solutions and adiabatic ground states: if we are able to choose $A, B$ and $C$ such that the ground state $\ket{0}_R$ takes the scaling form, then  $\delta_A^{(n)} = 0$ is automatically satisfied exactly because $\partial_y \ket{0}_R = 0$. This means that if the system is in the ground state of $H_{\rm eff}$ then the evolution will be adiabatic. 
However, it is not guaranteed that any given scaling form for $w$ will be the ground state, but if this scaling form behaves like an attractor, it is reasonable to hypothesize that this interpretation will hold, and this hypothesis was confirmed analytically by BSY~\cite{Brewer:2022vkq} in the kinetic theory that they analyzed. Concretely, BSY found that if the term in the collision kernel proportional to $I_b$ is neglected, then if $B$ and $C$ are chosen according to
\begin{align}
    \beta = - \frac{\partial_y B}{B} = - \frac{\tau \lambda_0 \ell_{\rm Cb} I_a }{B^2} \, , & & \gamma =-\frac{\partial_y C}{C} = 1 - \frac{\tau \lambda_0 \ell_{\rm Cb} I_a }{C^2} \, ,
\label{eq:bsyscalings} \end{align}
and $A(y)$ is fixed such that $\alpha = \gamma + 2\beta - 1$, the ground state of $H_{\rm eff}$ is time-independent and the spectrum of states is discrete. Consequently, there is an energy gap between the ground state and all the other states, which means that the occupation of all states other than the ground state decays away exponentially as we have discussed and the ground state quickly comes to dominate.  This is the key physical intuition needed to understand how the sensitivity to the initial conditions in a HIC can be lost at very early times, long before hydrodynamization.

In the remainder of this paper, we shall demonstrate the robustness of the adiabatic description of pre-hydrodynamic evolution and subsequent hydrodynamization in kinetic theory via a set of examples in all of which (unlike in the example treated by BSY) the eigenvalue solutions are not known analytically and in which (also unlike in BSY) the LHS of the adiabatic criterion \eqref{eq:adiabatic-approx} is not identically zero. We will begin in Sec.~\ref{sec:isotropic} by discussing the example of a dilute, non-expanding, gluon gas. This is a case where the appropriate rescaling will be different from that we have just discussed because there is no reason to treat $p_z$ differently from $p_\perp$. In our analysis of this example, we shall discuss how to select the rescaling so that the evolution is as adiabatic as possible. After working out this comparatively simple example, in Sec.~\ref{sec:expanding} we will return to the longitudinally expanding gluon gas and show that the AH framework works in this case even without the simplifying assumptions made by BSY. In Sec.~\ref{sec:non-scaling}, we extend our AH description further, all the way until the distribution becomes hydrodynamic. This will require us to extend the AH framework to describe attractors that are not scaling solutions.

\subsection{Example 1: Dilute, non-expanding, gas of weakly coupled gluons} \label{sec:isotropic}

As a simple initial demonstration, we will consider an isotropic, non-expanding, dilute gas of gluons in the weak coupling regime. 
With no expansion terms and in the dilute limit $f \ll 1$, the Boltzmann equation~\eqref{eq:Boltzmann-Milne} with the small-angle scattering collision kernel~\eqref{eq:small-angle-kernel} reduces to
\begin{equation} \label{eq:isotropickineq}
    \frac{\partial f}{\partial \tau} = \lambda_0 \ell_{\rm Cb}[f] \big( I_a[f] \nabla_p^2 f + I_b[f] \nabla_p \cdot (\hat{p} f ) \big) \, .
\end{equation}
A consequence of this simplification is that the equilibrium distribution of this kinetic theory is a Boltzmann distribution because 
$f \propto \exp(-p/T_{\rm eff})$ with $T_{\rm eff} = I_a / I_b$ makes the RHS of Eq.~\eqref{eq:isotropickineq} vanish.

We can further assume that the gluon distribution function $f$ is isotropic, and write the rescaled distribution function
\begin{equation}
    f(p,\tau) = A(\tau) w\left( \frac{p}{D(\tau)}, \tau \right) = A(\tau) w(\chi,\tau)
\end{equation}
as a function of the rescaled momentum $\chi = p/D(t)$. We then write the evolution equation in the form
\begin{equation}
    \partial_\tau w = - H w,
\end{equation}
(a slight variation of Eq.~\eqref{eq:hdef} for this simplified case), and find that the effective Hamiltonian operator $H$ is 
\begin{equation} \label{eq:isotropich}
    H = \alpha + \delta \chi \partial_\chi - \lambda_0 \ell_{\rm Cb}[f] \frac{I_a[f]}{ D^2} \left(\frac{2}{\chi} \partial_\chi + \partial_\chi^2\right) -  \lambda_0 \ell_{\rm Cb}[f] \frac{I_b[f]}{D} \left( \frac{2}{\chi} + \partial_\chi \right) \, ,
\end{equation}
where we have introduced $\delta \equiv - \partial_\tau D / D$.

To study the dynamics of this system numerically, it is necessary to describe the full distribution function in terms of a finite number of dynamical variables. In the AH framework, the natural variables are the basis state coefficients (that specify the occupation of each basis state) determined by the series expansion of the distribution function in a given basis. In principle, any basis of integrable functions suffices for this purpose because the integral of the distribution function over all of momentum space is finite (and corresponds to the spatial number density of gluons). However, in an actual calculation we keep only some finite number of basis states and, in order to be least sensitive to the effects of this truncation, it is best to choose a basis that is well adapted to describe the physical phenomena of interest.  We shall see how to do this in the present simple case here, and then subsequently in each of the more complete examples we introduce in later Sections.

As we have seen in Sec.~\ref{sec:AH}, because of the fact that the time evolution operator of the theory is non-Hermitian, a better description of the eigenstates is achieved if we choose different left and right bases. And, as discussed in Sec.~\ref{sec:ConservationLaws}, because the collision kernel conserves particle number the constant function will always be a left eigenstate in all of our examples, and so it behooves us to include it as one of the left basis states.
Furthermore, because in the particular example that we are considering here the late-time solution of Eq.~\eqref{eq:isotropickineq} is a Boltzmann distribution,\footnote{Not a Bose-Einstein distribution because (motivated by the more realistic examples to come, not by this simple example) we have dropped the $I_b f^2$ term in the collision kernel \eqref{eq:small-angle-kernel}. See Sec.~\ref{sec:AH} and Appendix~\ref{app:f2} for further discussion.} we will use the basis
\begin{equation} \label{eq:boltzmannbasis}
    \psi_i^{(R)} = N_i \,{\rm L}_i^2(\chi) e^{-\chi}, \;\; \psi_i^{(L)} = N_i \,{\rm L}_i^2(\chi)
\end{equation}
where ${\rm L}_i^2(\chi)$ are associated Laguerre polynomials, and $N_i$ is a normalization constant chosen such that the basis satisfies
\begin{equation}
    \int d^3\chi \psi_i^{(L)}  \psi_j^{(R)} = \delta_{ij} \, .
\end{equation}
As we will see below, provided the scaling parameter $D$ is chosen appropriately, this basis provides a quickly convergent expansion of the dynamics near thermalization.

Once the basis is set up, we can calculate matrix elements
\begin{equation} \label{eq:hprojection}
    H_{ij} = \int d^3\chi \psi_i^{(L)}  H \psi_j^{(R)}.
\end{equation}
If we truncate at some finite number of basis states, it is straightforward to calculate an approximate instantaneous ground state by solving the eigenvalue problem of the truncated Hamiltonian, for a given state of the system. Furthermore, for the effective Hamiltonian given in 
Eq.~\eqref{eq:isotropich} and the basis in 
Eq.~\eqref{eq:boltzmannbasis}, we notice that the matrix element associated with the first left basis state (which, as discussed in Sect.~\ref{sec:AH}, projects onto the particle number) is given by
\begin{equation} \label{eq:hfirstrow}
    H_{1j} = \delta_{1j} (\alpha - 3\delta) \, ,
\end{equation}
confirming that the constant function is indeed a left eigenstate of $H$. Moreover, this shows that 
we can guarantee that one eigenvalue of $H$ will be zero by choosing $\alpha = 3\delta$. We anticipate that the right eigenvector associated with this eigenvalue will be the ground state, and that this eigenvector will represent the thermal Boltzmann distribution at late times.

With these choices, the instantaneous ground state $\ket{0}_R = \ket{0(A,D,\delta,I_a,I_b)}_R$, and we can write
\begin{equation}
    \partial_\tau \ket{0}_R = \left(A\alpha \frac{\partial }{\partial A} - D\delta \frac{\partial }{\partial D} + \frac{\partial \delta}{\partial \tau} \frac{\partial }{\partial \delta} + \frac{\partial I_a}{\partial \tau} \frac{\partial }{\partial I_a} + \frac{\partial I_b}{\partial \tau} \frac{\partial }{\partial I_b} \right) \ket{0}_R\,,
    \label{eq:partial-tau-phi0}
\end{equation}
where because of the dilute limit we in fact have $\frac{\partial I_a}{\partial \tau} = 0$. Then we can seek to minimize the LHS of the adiabaticity condition \eqref{eq:adiabaticityn0}, namely $\delta_A^{(0)}$, by using our ability to choose the one remaining free rescaling parameter $D(\tau)$. We do so by choosing $\frac{\partial \delta}{\partial \tau}$ to minimize $||\partial_\tau \ket{0_R}||^2\equiv\Bra{0}_R\overleftarrow{\partial_\tau}\partial_\tau\Ket{0}_R$. From \eqref{eq:partial-tau-phi0}, this means that we choose
\begin{equation}
    \frac{\partial \delta}{\partial \tau} = -\frac{\text{Re} \Bra{0}_R \overleftarrow{\frac{\partial }{\partial \delta}} \left( A\alpha \frac{\partial }{\partial A} - D\delta \frac{\partial }{\partial D} + \frac{\partial I_b}{\partial \tau} \frac{\partial }{\partial I_b} \right) \Ket{0}_R }{\Bra{0}_R\overleftarrow{\frac{\partial }{\partial \delta}}\frac{\partial }{\partial \delta}\Ket{0}_R}.
\end{equation}
Additionally, for any state
\begin{equation}
    w(\chi,\tau) = \sum_i w_i(\tau) \psi_i(\chi),
\end{equation}
the coefficients evolve according to
\begin{equation}
    \partial_\tau w_i = -\sum_j H_{ij} w_j.
\end{equation}
Using this framework, we can numerically solve for the evolution of the system (through its basis coefficients) and the maximally adiabatic frame $A(\tau),D(\tau)$ by explicitly writing the numerical values of the truncated effective Hamiltonian matrix at each time. Some additional specifics of the numerical implementation of the system evolution are detailed in Appendix \ref{sec:earlynumerical}. In this manner we can solve for an optimally adiabatic choice of time-dependent rescalings $A(\tau)$ and $D(\tau)$ and simultaneously solve for the evolution of the distribution function. 

\begin{figure}
    \centering
    \includegraphics[width=\textwidth]{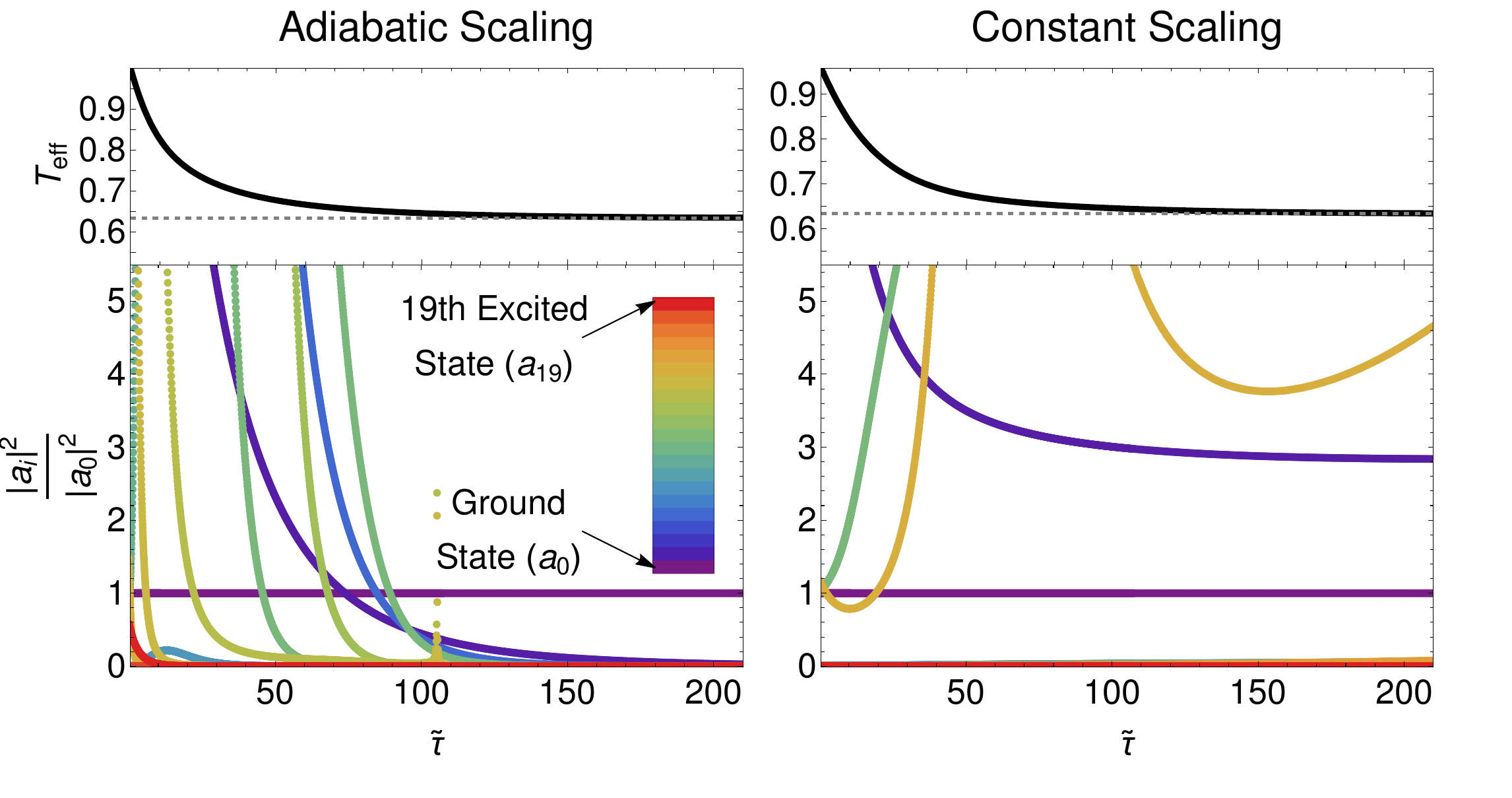}
    \caption{Using 20 basis states, the resulting eigenstate coefficients $a_i$ (bottom panels) ordered by color, and effective temperature $T_{\rm eff} = I_a/I_b$ (top panels) as a function of rescaled time $\tilde{\tau} = \lambda_0 \ell_{\rm Cb} \tau$. In the left panels, the adiabaticity-maximizing choice of scaling $D(\tau)$ described in the text is used, while in the right panels, the scaling is chosen to be constant. 
    Note that the coefficients $a_i$ are normalized relative to the ground state coefficient $a_0$. Therefore, by definition, the ground state occupation appears as a straight line at 1 in both lower panels.
    Both choices for $D(\tau)$ reproduce the same physical dynamics, as exemplified by the identical effective temperatures  in the two top panels. We can see in the lower-left panel that in the adiabatic frame, the ground state of the rescaled distribution function becomes dominant on roughly the same time-scale as the $T_{\rm eff}$ levels off and the system thermalizes, showing that adiabatic evolution provides a reasonable physical interpretation for the rescaled system, even though in this case the initial condition is very far from the instantaneous ground state. Furthermore, the decay of excited state coefficients is somewhat ordered, with the longest-lived excited mode being the first excited state. In the time-independent frame, evolution is clearly non-adiabatic, and the ground state does not become dominant, showing the importance of the choice of frame for understanding the evolution of the system. } \label{fig:moneyplot}
\end{figure}

An example result of this procedure is shown in Figure \ref{fig:moneyplot}. We use the initial condition
\begin{equation}
    f=\frac{1}{24 \pi} e^{-4 \chi} (3 + 3 \sqrt{3} - 12 \sqrt{3} \chi + 
   8 \sqrt{3} \chi^2)\,,
\end{equation}
that we have chosen somewhat arbitrarily and that is intended only as an example in which the evolution begins  far from the time-dependent ground state, not as a representation of some physical system. Without dynamical rescaling it is not clear that the AH interpretation holds, but using the adiabaticity-maximizing rescaling, it becomes clear that the ground state dominates on approximately the same time scale as the thermalization of the system in the rescaled picture. The instantaneous ground state that we can identify from the state at the initial time using the maximally adiabatic scaling frame (which we identify without using prior knowledge of the equilibrium solution beyond the choice of basis) proves to be the attractor solution for this system, as we hypothesized in the introduction to this Section. Importantly, unlike in the case considered by BSY~\cite{Brewer:2022vkq}, the rescaled eigenstates are still time-dependent and although we have found an adiabatic rescaling meaning that the adiabaticity condition {\it is} satisfied it is not satisfied trivially via its LHS vanishing as in BSY.  
This is a promising initial demonstration of the broad applicability of the adiabatic framework.

\subsection{Example 2: Longitudinally expanding gluon gas} \label{sec:expanding}

We now turn to considering an anisotropic gas of gluons which is expanding only longitudinally, for which we will make no assumptions about whether $f$ is over- or under-occupied. 
We shall assume that the collision kernel takes
the small-angle scattering form~\eqref{eq:small-angle-kernel}.
With these assumptions, Eq.~\eqref{eq:Boltzmann-Milne} 
takes the form
\begin{equation}
    \frac{\partial f}{\partial y} - p_z \frac{\partial f}{\partial p_z} = \tau \lambda_0 \ell_{\rm Cb} \left[ I_a \nabla_{\bm{p}}^2 f + I_b \nabla_{\bm{p}} \cdot \left( \frac{\bm{p}}{p} f (1+f) \right) \right] \,.
\end{equation}
We will assume that the gluon distribution function has azimuthal symmetry. In the next two subsections, we 
shall first describe the early, pre-hydrodynamic, attractor for this kinetic theory and then describe the late-time attractor that can describe hydrodynamization. In both cases we shall use the AH framework, but we shall find it convenient to use different bases 
for the early and late time attractors.

\subsubsection{Early-time, pre-hydrodynamic, attractor} \label{sec:pre-hydro}

At early times, we cast $f$ into the rescaled form
\begin{equation}
    f(p_\perp,p_z,y) = A(y) w \left(\frac{p_\perp}{B(y)},\frac{p_z}{C(y)},y\right).
\end{equation}
From here, we can use the effective Hamiltonian in~\eqref{eq:Heff-smallanglescatt-BC} to generate the time evolution of $w$ according to $\partial_y w = - H w$. Furthermore, we make the approximation $p \approx B \zeta$ in the both the denominator of the terms in \eqref{eq:Heff-smallanglescatt-BC} proportional to $I_b$ and the evolution of $I_b$ itself, which is consistent with the fact that at early times the hierarchy $p_z^2 \ll p_\perp^2$ holds. To simplify the analysis, we will also drop 
the $I_b f^2$ term in the collision kernel which corresponds to dropping
the terms in the effective Hamiltonian \eqref{eq:Heff-smallanglescatt-BC} that are explicitly dependent on $w$. We discuss the limitations of this approximation in Appendix~\ref{app:f2}. With these approximations, Eq.~\eqref{eq:Heff-smallanglescatt-BC} becomes
\begin{align} \label{eq:expandingh}
    H =&\alpha + \beta \zeta \partial_\zeta + (\gamma-1) \xi \partial_\xi - \tau \lambda_0 \ell_{\rm Cb} I_a \left[ \frac{1}{B^2} \left( \frac{1}{\zeta} \partial_\zeta + \partial_\zeta^2 \right) + \frac{1}{C^2} \partial_\xi^2 \right] \\ \nonumber
&- \frac{\tau \lambda_0 \ell_{\rm Cb} I_b}{B \zeta} \left[ 2 + \zeta \partial_\zeta + \xi \partial_\xi \right] \,. 
\end{align}
We will use the basis
\begin{equation}
\psi_{ij}^{(R)} = N_{ij} {\rm L}_i^1(\zeta) {\rm He}_{2j}(\xi) \exp{\lbrace-\left(\xi^2/2+\zeta\right)\rbrace}\,, \;\; \psi_{ij}^{(L)} = N_{ij} {\rm L}_i^1(\zeta) {\rm He}_{2j}(\xi)\,,
\end{equation}
where the ${\rm L}_i^1(\zeta)$ are associated Laguerre polynomials, ${\rm He}_{2j}(\xi)$ are probabilist's Hermite polynomials~\cite{abramowitz1965handbook}, and $N_{ij}$ is a normalization constant chosen such that the basis satisfies the orthonormality condition
\begin{align}
    \frac{1}{(2\pi)^2} \int_{-\infty}^\infty d\xi \int_0^\infty d\zeta \; \zeta \; \psi_{ij}^{(L)} \psi_{kl}^{(R)} &= \delta_{ik} \delta_{jl} \,.
\end{align}

\begin{figure}
\centering
\begin{subfigure}{0.65\textwidth}
    \includegraphics[width=\textwidth]{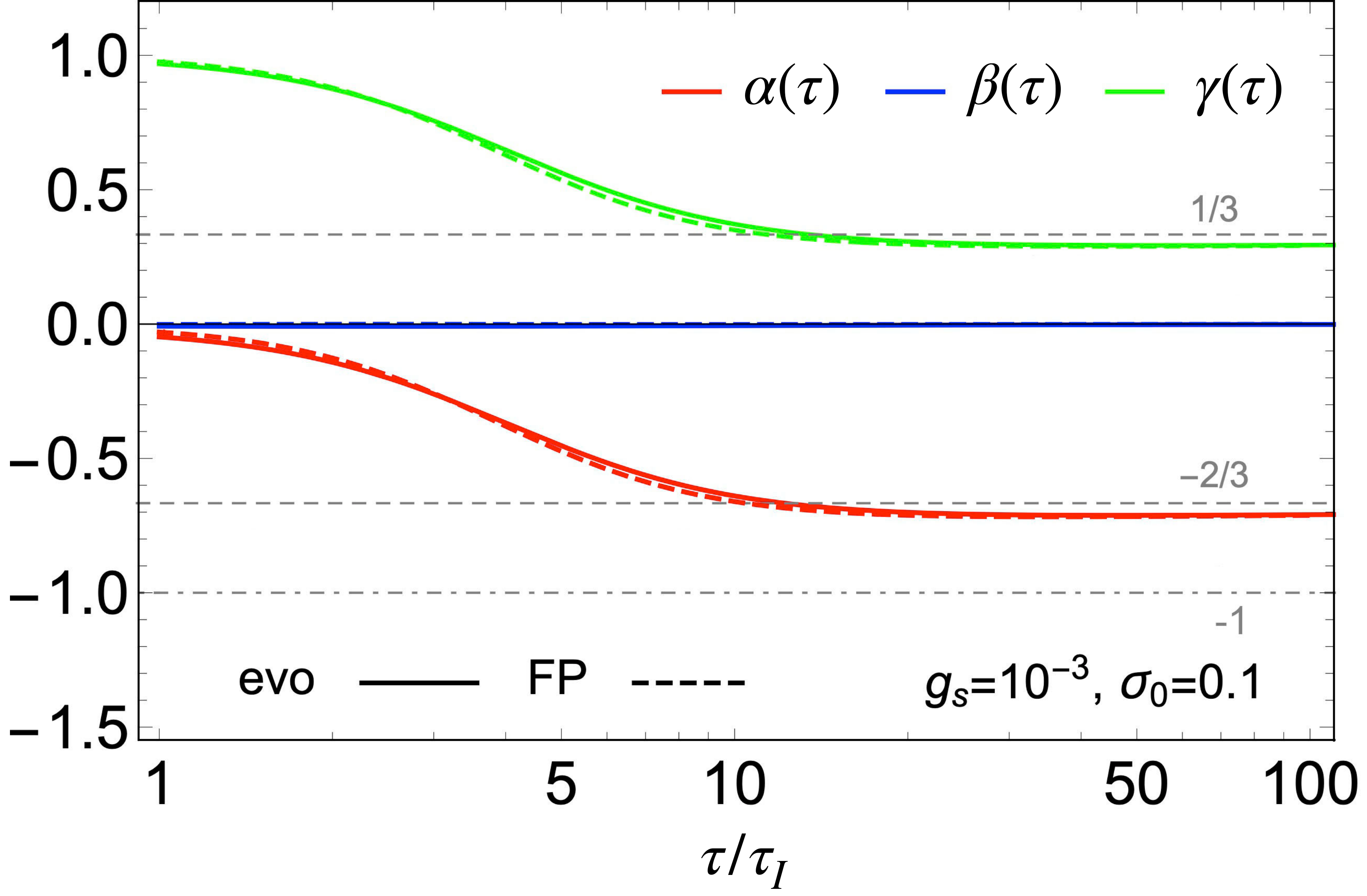}
\end{subfigure}
\hfill
\begin{subfigure}{0.7\textwidth}
    \includegraphics[width=\textwidth]{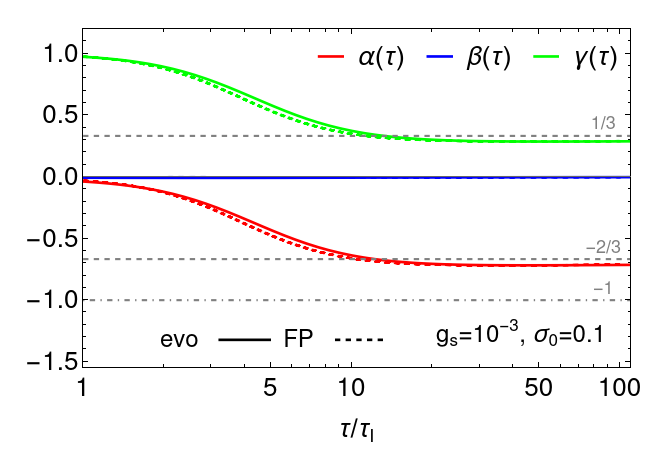}
\end{subfigure}
\caption{A comparison of the time evolution of the extracted scaling exponents $\alpha$, $\beta$ and $\gamma$ in kinetic theory with the weak coupling $g=10^{-3}$ for an initial condition~\eqref{eq:BSY-initial-conditions} with $\sigma_0=0.1$ which approaches the BMSS fixed point with $\alpha=-2/3$, $\beta=0$ and $\gamma=1/3$ (indicated by the gray dashed lines;  for later reference, we also include a dot-dashed line describing the dilute fixed point $\alpha = -1$). The top panel is taken from BSY~\cite{Brewer:2022vkq}; the solid curves (labeled ``evo'') represent scaling exponents found analytically by BSY by applying the adiabatic hydrodynamization framework to the kinetic theory with a simpler collision kernel than the one we employ.
The colored dashed curves (labeled ``FP'') depict the scaling exponents extracted
from a numerical analysis of
the early-time dynamics of the kinetic theory with the full small-angle scattering collision kernel~\cite{Brewer:2022vkq}. 
In the bottom panel, the dotted curves are the same numerical calculation as in the top panel, while the solid curves depict the scaling exponents that we have found in this work using the full small-angle scattering collision kernel and adiabaticity-maximizing scalings chosen at each time. Specifically, we used a basis consisting of 40 basis states made up of products of 10 transverse polynomials ${\rm L}_i^1(\zeta)$ with 4 longitudinal polynomials ${\rm He}_{2j}(\xi)$. We see a satisfying agreement across the three methods. The deviation of all three from the BMSS scaling values at late times is due to the time dependence of the Coulomb logarithm, and is discussed at length in 
Ref.~\cite{Brewer:2022vkq}
}
\label{fig:bsycomparison1}
\end{figure}

\begin{figure} 
\centering
\begin{subfigure}{0.65\textwidth}
    \includegraphics[width=\textwidth]{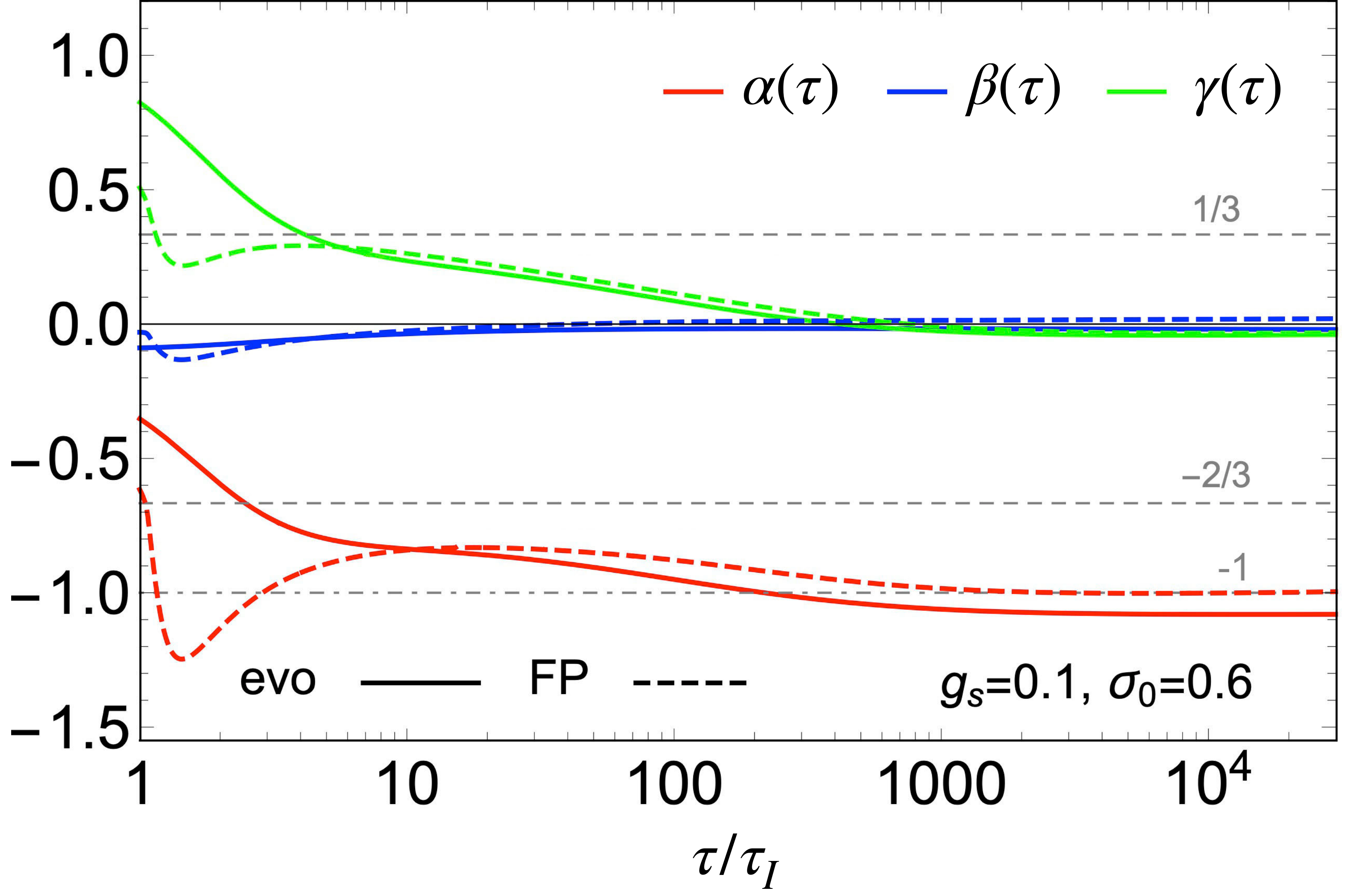}
\end{subfigure}
\hfill
\begin{subfigure}{0.7\textwidth}
    \includegraphics[width=\textwidth]{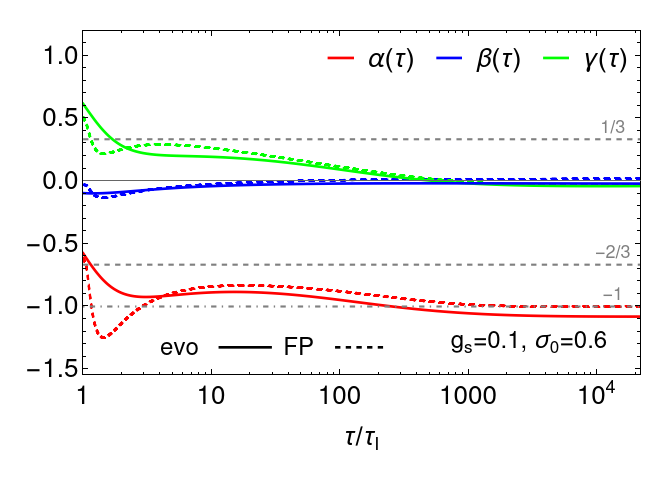}
\end{subfigure}
\caption{A comparison of scaling exponents as in Fig.~\ref{fig:bsycomparison1}, but for a more strongly coupled, highly occupied initial condition with $g_s=0.1$ and $\sigma_0=0.6$ which initially seems to approach the BMSS fixed point (as happens definitively at smaller coupling) but then evolves to  the ``dilute fixed point'' with $\alpha=-1$, $\beta=\gamma=0$ (indicated with gray dash-dotted lines). As in Fig.~\ref{fig:bsycomparison1}, the colored dashed curves in both panels (labeled ``FP'') represent numerically extracted scaling exponents~\cite{Brewer:2022vkq}, while the solid curves (labeled ``evo'') in the top panel were found using the simplified analytic solution of BSY~\cite{Brewer:2022vkq}. The solid curves in the bottom panel are from this work and were found using the adiabaticity-maximizing method described in the text, employing the same 40 basis states described in Fig. \ref{fig:bsycomparison1}. All three are in good agreement. At early and intermediate times, our new adiabatic hydrodynamization results perhaps agree somewhat better with the scaling exponents extracted numerically 
than the BSY results in the top panel do.
}
\label{fig:bsycomparison2}
\end{figure}

\begin{figure}
\centering
\begin{subfigure}{0.7\textwidth}
    \includegraphics[width=\textwidth]{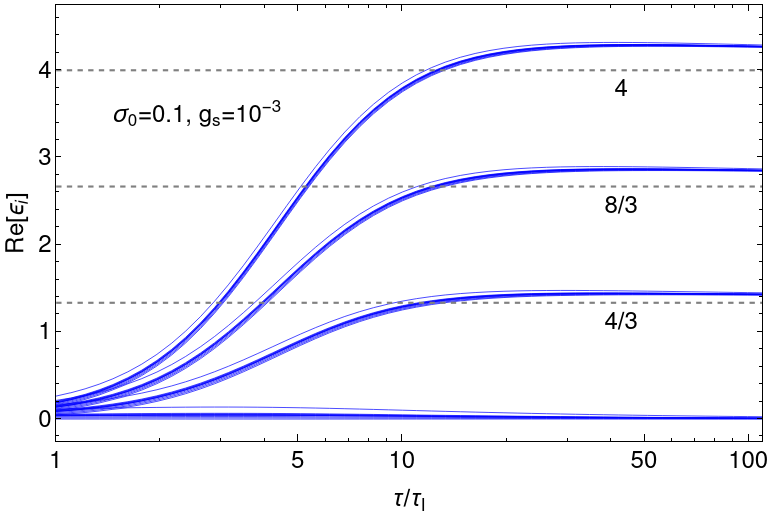}
    \label{fig:bsy1energy}
\end{subfigure}
\hfill
\begin{subfigure}{0.7\textwidth}
    \includegraphics[width=\textwidth]{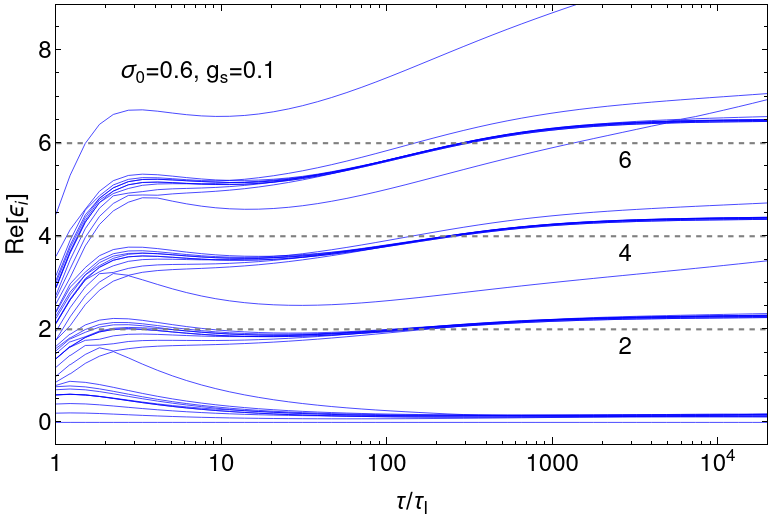}
    \label{fig:bsy2energy}
\end{subfigure}
\caption{Eigenvalues corresponding to the time-dependent eigenstates of the effective Hamiltonian $H$ for the two initial conditions presented in Figs.~\ref{fig:bsycomparison1} and \ref{fig:bsycomparison2}. For both initial conditions, many of these effective energy levels are 
clustered into 
degenerate or near-degenerate groups for much of the system's evolution, with large energy gaps between these groups at late times. We can associate each cluster of eigenvalues with a longitudinal mode, and any splitting within a cluster comes from the small effects of differing transverse modes, as discussed in the text.  Dotted lines show the energy levels expected from BSY (see Eq.~\eqref{eq:BSYenergies}) using 
$\alpha=-2/3$, $\beta=0$, $\gamma=1/3$ for the top panel 
and
$\alpha=-1$, $\beta=\gamma=0$ for the bottom panel 
to reflect the approximate late-time values of these scaling exponents, as seen in Figs. \ref{fig:bsycomparison1} and \ref{fig:bsycomparison2} respectively.}
\label{fig:energies}
\end{figure}

\begin{figure} 
\centering
\begin{subfigure}{.75\textwidth}
    \includegraphics[width=\textwidth]{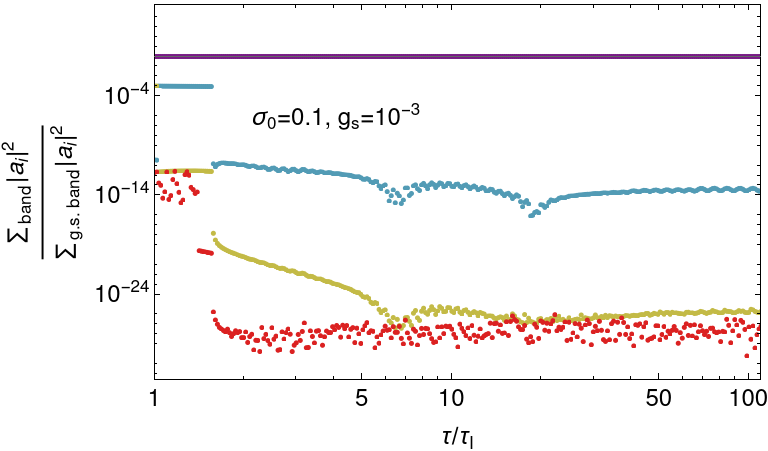}
\end{subfigure}
\hfill
\begin{subfigure}{.75\textwidth}
    \includegraphics[width=\textwidth]{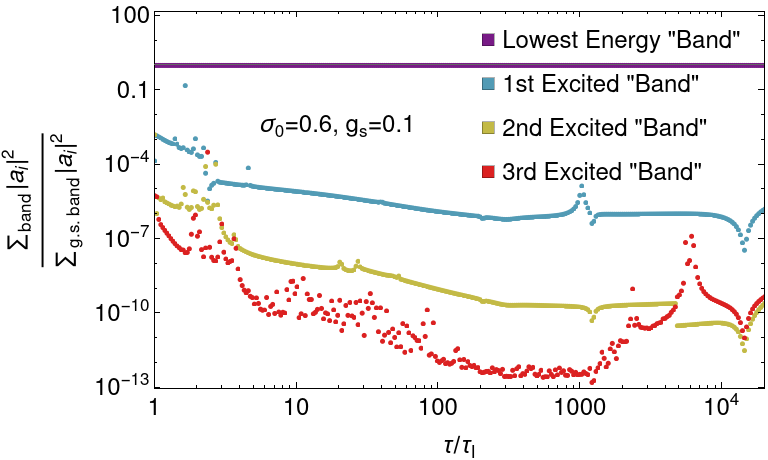}
\end{subfigure}
\caption{Sums over each energy ``band'' of eigenstate coefficients $a_i$ that tell us about the occupation of each band of eigenstates in the distribution function as a function of time for the two initial conditions presented in Figs.~\ref{fig:bsycomparison1} and \ref{fig:bsycomparison2}. $N_\perp=10$ and $N_z=4$ are the number of transverse and longitudinal basis states, respectively. Then each of the $N_z$ ``bands'' contains $N_\perp$ states, ordered by energy. That is, the first ``band" contains the states with the lowest $N_\perp$ energies, the second contains the states with the next-lowest $N_\perp$ energies, etc. The large downward jump in the excited bands at an early time ($\tau/\tau_I\approx 1.6$) in the more weakly coupled example (top panel) is due to a level crossing in which one state moves from the second band to the first band. We report the coefficients in this way because (as seen in Fig. \ref{fig:energies}) the energies of the eigenstates are clustered into well-separated groups at late times. Therefore rather than comparing eigenstate occupation to occupation of a single ground state, it is more meaningful to compare occupation of the higher-energy bands of eigenstates to the lowest-energy band. We can see that for both of these initial conditions, the state of the system is initially very close to its instantaneous ground state, and then remains very close to its ground state as the ground state evolves. This is consistent with the notion of adiabatic evolution.}
\label{fig:coefficients}
\end{figure}

As in the previous example, we will choose $\alpha$ in such a way as to guarantee a zero eigenvalue. Calculating matrix elements according to
\begin{align}
    H_{ijkl} \equiv \frac{1}{(2\pi)^2} \int_{-\infty}^\infty d\xi \int_0^\infty d\zeta \; \zeta \; \psi_{ij}^{(L)} H \psi_{kl}^{(R)} \,,
\end{align}
we note that the first row of the effective Hamiltonian with our choice of basis is
\begin{equation}
    H_{1j1l}=\delta_{1j}\delta_{1l}(\alpha-2\beta-\gamma+1)\ .
\end{equation}
We therefore choose
\begin{equation}
    \alpha = \gamma + 2\beta - 1 
\end{equation}
to ensure that the left basis state which projects onto the particle number has zero eigenvalue.
We will then choose $\gamma(y)$ and $\beta(y)$ according to the adiabatic solutions for these scaling exponents found previously by BSY~\cite{Brewer:2022vkq} upon analyzing the kinetic theory with a simpler version of the collision kernel (omitting the term proportional to $I_b$) than we are considering here. That is, we choose $\gamma(y)$ and $\beta(y)$ according to Eq.~\eqref{eq:bsyscalings}. In the BSY case, they found that it is possible to choose scalings $\gamma(y)$ and $\beta(y)$ such that their simplified effective Hamiltonian is completely time independent, and therefore perfectly adiabatic. We adopt the same scalings that BSY employed in the hope that the evolution of the rescaled system will still be adiabatic, even though the inclusion of $I_b$ terms in the collision kernel means that our effective Hamiltonian will have some time dependence.
For the sake of comparison with BSY, we use the same initial conditions they do: 
\begin{equation}
    f(p_\perp,p_z;\tau_I) = \frac{\sigma_0}{g_s^2}\exp \left( - \frac{p_\perp^2+\xi_0^2 p_z^2}{Q_s^2}\right)
    \label{eq:BSY-initial-conditions}
\end{equation}
with initial anisotropy $\xi_0=2$ and initial time $\tau_I Q_s=70$, and the values of the coupling $g_s$ and overoccupancy of hard gluons $\sigma_0$ in Figs.~\ref{fig:bsycomparison1} and \ref{fig:bsycomparison2} chosen as in BSY. The scaling exponents found for these initial conditions in the way we have described above are plotted in 
Figs.~\ref{fig:bsycomparison1} and~\ref{fig:bsycomparison2} and, as expected, the scaling exponents we extract match both BSY and numerically calculated scaling exponents very well. 
Although this prescription for the scaling exponents 
does not yield a time-independent effective Hamiltonian with perfectly adiabatic time evolution, 
we can see that the eigenvalues of the extracted time-dependent rescaled eigenstates, shown in Fig.~\ref{fig:energies}, are similar to the eigenvalues calculated by BSY:
\begin{equation} \label{eq:BSYenergies}
    \epsilon_{nm} = 2n(\gamma-1)-2m\beta \;\;\;\;\; n,m=0,1,2,\dots\,.
\end{equation}
As can be seen both in Fig.~\ref{fig:energies} and in this analytic expression (with the understanding that $\beta$ is small relative to $\gamma$), there is a large separation between the energies of states which have differing longitudinal modes, while within each longitudinal mode the transverse modes are nearly degenerate. That is, unlike in the simple model presented in Sec.~\ref{sec:isotropic} where there was a single ground state separated from all other states by a gap, here we have a subset of states which act collectively as a ``ground state" in our adiabaticity picture, with these states nearly degenerate with each other and separated from all other states by a gap. 
As such, in Fig.~\ref{fig:coefficients}, we compare the relative occupation of the clustered subsets of states rather than occupation of individual eigenstates. We see that for the initial conditions considered, which begin with the system primarily occupying the lowest energy set of states, the system remains very near this ``ground" set of states as the system evolves. In that sense, we might call the system's evolution ``quasi-adiabatic'', since despite the lack of a single adiabatic ground state, we still have a picture in which we have a restricted set of preferred degrees of freedom for the system.
We shall see in Sect.~\ref{sec:non-scaling}  that at later times as the kinetic theory hydrodynamizes, the energies of all but one of this set of ground states rises and the system evolves into a single isolated adiabatically evolving state.

Unfortunately, the choice of basis that we have used here to describe the early pre-hydrodynamic attractor is limited
in its utility later, during hydrodynamization, because
we have employed the approximation $p \approx p_\perp$ in the effective Hamiltonian and in $I_b[f]$. Once the system begins to isotropize, a different basis will be necessary in order to continue to evolve $f$ to hydrodynamization.

\subsubsection{Late-time, hydrodynamizing attractor} \label{sec:hydro}

The approximation $p \approx p_\perp$ made in the previous 
Section to calculate the matrix elements was motivated by the fact that at very early times the longitudinal expansion drives $p_z$ downward, making the momentum distribution anisotropic with $p_z\ll p_\perp$.
This motivated us to choose basis states that are products of functions of $p_z$ and $p_\perp$. 
In this Section, we seek to describe hydrodynamization within the AH framework.
In the late stages of hydrodynamization, the distribution function must approach local thermal equilibrium, meaning that the momentum distribution must approach isotropy.
To describe hydrodynamiztion, therefore, it behooves us to choose a basis in which an isotropic
thermal distribution
is well described by the set of states that the (truncated) basis spans. A natural choice of coordinates via which to accomplish this is $p = \sqrt{p_\perp^2 + p_z^2}$ and $u = p_z/p$. 
As before, we introduce a rescaling in the $p$ coordinate such that $p = D(y) \chi$ to find an adiabatic frame. However, because $u$ is a coordinate that takes values in a bounded interval $-1 \leq u \leq 1$,  rescaling $u$ to a different variable $\tilde{u} = r(y) u $ would induce a more complicated inner product in terms of $\tilde{u}$ because the limits of integration would become time-dependent. For this reason, we shall not introduce any rescaling of $u$ in the present discussion. (However, in the next Section we shall need to introduce the parameter $r(y)$ as a way of recovering adiabaticity beyond the regime where scaling solutions exist.)

With the motivations above, in this Section we cast $f$ in the form
\begin{equation}
    f(p,u,y) = A(y) w \left( \frac{p}{D(y)} , u, y \right) \, ,
\end{equation}
where $w = w(\chi, u , y)$. As in the previous Section and as discussed in Appendix~\ref{app:f2}, we will drop the $I_b f^2$ terms in the collision kernel, which corresponds to dropping the explicitly $w$-dependent terms in the effective Hamiltonian. With this, the effective Hamiltonian that generates the time evolution of $w$ is given by
\begin{align} 
    H &= \alpha + \delta \chi \partial_\chi - u^2 \chi \partial_\chi - u(1-u^2) \partial_u - \tau \lambda_0 \ell_{\rm Cb} \frac{I_b}{D} \left( \frac{2}{\chi} + \partial_\chi \right) \nonumber \\
    & \quad   - \tau \lambda_0 \ell_{\rm Cb} \frac{I_a}{D^2} \left[ \frac{2}{\chi} \partial_\chi + \partial_\chi^2 + \frac{1}{\chi^2} \partial_u \left( (1 - u^2) \partial_u f \right) \right] \, . \label{eq:Heff-thermal-noBose}
\end{align}

Next, we choose the basis 
\begin{align}
    \psi_{nl}^{(R)} = N_{nl} e^{-\chi} L_{n-1}^{(2)}(\chi) P_l(u) \, , & & \psi_{nl}^{(L)} = N_{nl} L_{n-1}^{(2)}(\chi) P_l(u) \, ,
\end{align}
where $L_{n-1}^{(2)}(\chi)$ are associated Laguerre polynomials, $P_l(u)$ are Legendre polynomials, and the normalization coefficients $N_{nl}$ are such that
\begin{align}
    \frac{1}{4\pi^2} \int_{-1}^1 du \int_0^\infty d\chi \, \chi^2 \psi_{mk}^{(L)} \psi_{nl}^{(R)} = \delta_{kl} \delta_{mn} \, .
\end{align}
Since the thermal state of the system described by Eq~\eqref{eq:Heff-thermal-noBose} is a Boltzmann distribution (and not a Bose-Einstein distribution because we dropped the $I_b f^2$ term in the collision kernel), this basis is well-equipped to describe it.

As before, the eigenvalue associated with the mode that carries the particle number can be calculated explicitly. In this case, it is $\alpha - 3\delta + 1$ (where here $\delta\equiv -\partial_y D/D$), and so we set $\alpha=3\delta -1$ so as to fix this eigenvalue to zero. 
Since the ground state is the state that will carry the particle number at late times, we are thus guaranteeing that the late-time thermal distribution has eigenvalue zero.

We then have to choose how we evolve $D$ or, equivalently, $\delta$. In this case, we find that a good description is achieved by taking $D$ to follow the scale that determines the effective temperature at late times, namely
\begin{equation} \label{eq:D-evol}
    \delta=-\frac{\partial_y D}{D} = - \rho \left( 1 - D \left\langle \frac{2}{p} \right\rangle  \right) \, ,
\end{equation}
where $\rho$ is a dimensionless parameter which we can choose so as to control the speed with which $D(y)$ follows the typical momentum scale of gluons $\langle \frac{2}{p}\rangle^{-1}$. In practice, we choose $\rho = 10$.
When thermalization is approached, $D \to 1/\langle \frac{2}{p}\rangle \sim I_a/I_b = T_{\rm eff}$ where $\langle\frac{2}{p}\rangle \sim I_b/I_a$. Note that $\langle\frac{2}{p}\rangle = I_b/I_a$
only for a dilute system with $f \ll 1$.
Nonetheless, because the final state of the system in the examples we consider is in fact dilute, in practice we will often ignore the distinction.

We choose the initial condition to be
\begin{equation} \label{eq:init-cond-iso}
    f({\bf p}, \tau = \tau_I) = \frac{\sigma_0}{2 g_s^2} e^{- \sqrt{2} p/Q_s}
\end{equation}
and initialize the system at $\ln (\tau_I Q_s ) = 4$, with $\sigma_0 = 10^{-2}$, and scan over couplings $g_s \in \{1, 1.5, 2, 2.6\}$. 
{{In order to ensure that we initialize the evolution in this Section close enough to the hydrodynamization stage of the bottom-up scenario of 
BMSS~\cite{Baier:2000sb}, which occurs after the pre-hydrodynamic stages that we described via AH in Sect.~\ref{sec:pre-hydro},
we could initialize the distribution function at a later time $\tau_I$ with a lower initial occupancy $\sigma_0$ than we employed in that section.
What we have done instead is to choose a smaller $\sigma_0$ with a comparable $\tau_I$, relative to the choices we made in Sect.~\ref{sec:pre-hydro},
but to employ substantially
larger values of the coupling $g_s$.
The stronger coupling ensures that the dynamics of the system will have been more rapid in the previous stages, making these choices appropriate.}} Our goal in Sect.~\ref{sec:non-scaling} will be to find a single formulation that describes the early time dynamics that we 
analyzed in Sect.~\ref{sec:pre-hydro} followed smoothly and adiabatically by the hydrodynamization that we describe here.  Since here we are treating the two regimes separately, we must initialize our analysis of the late-time, hydrodynamizing, attractor here in a way that resembles the state of the system at the end of our AH analysis of the pre-hydrodynamic scaling regime in Sect.~\ref{sec:pre-hydro}.

\begin{figure}
    \centering
    \includegraphics[width=0.49\textwidth]{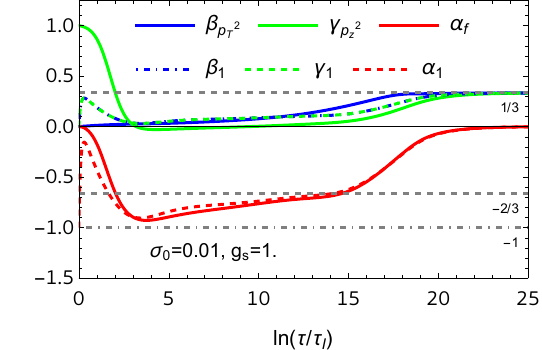}
    \includegraphics[width=0.49\textwidth]{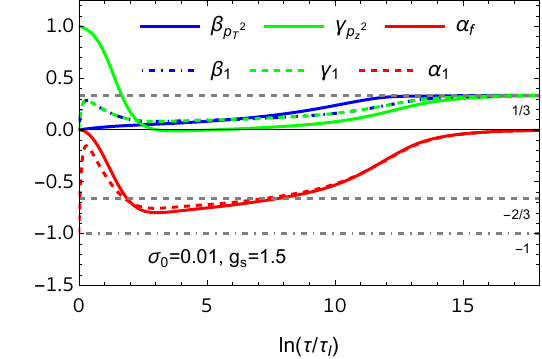}
    \includegraphics[width=0.49\textwidth]{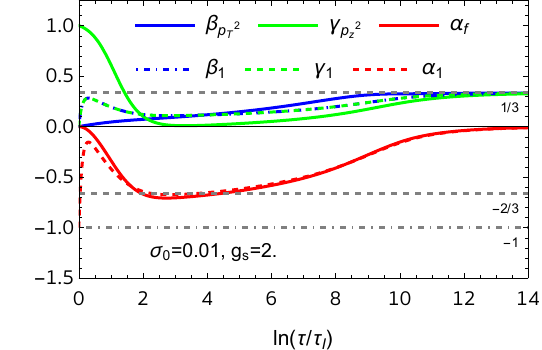}
    \includegraphics[width=0.49\textwidth]{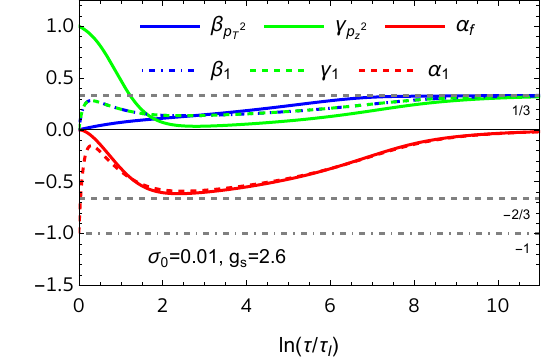}
    \caption{Evolution of the typical momentum scales encoded in the scaling exponents $\alpha, \beta, \gamma$ for the initial condition specified in Eq.~\eqref{eq:init-cond-iso}. We used $N_{\rm states} = 36$. From left to right and top to bottom, $g_s = 1, 1.5, 2, 2.6$. In order to test how well-adapted the basis is to the dynamics of the longitudinally expanding gluon gas as it hydrodynamizes, we plot two sets of scaling exponents: the solid lines describe the scaling exponents as calculated from the moments $\langle p_z^2 \rangle$ and $\langle p_\perp^2 \rangle$ computed using the full distribution, and the dashed lines represent the evolution of those scales as would be prescribed only by a single basis state, the one which carries the particle number and which, we shall see below, becomes the lowest energy eigenstate of the effective Hamiltonian at late times.
    For all four values of the coupling, we see the system hydrodynamize: it follows an attractor solution that brings it to $\alpha=0$, $\beta=\gamma=1/3$ which, as described at the beginning of Section~\ref{sec:scaling}, are the values of the exponents that characterize the kinetic theory of a boost-invariant longitudinally expanding hydrodynamic fluid in local thermal equilibrium. Furthermore, the agreement between the dashed and solid curves indicates that the evolution of the typical momentum scales $\langle p_z^2 \rangle_1$ and $\langle p_\perp^2\rangle_1$ described only by the single basis state that carries the particle number are quite similar to those computed from the full distribution function. This agreement at intermediate as well as at late times suggests that, as we shall indeed confirm below, the evolution of the state of the system as it hydrodynamizes is described well by the adiabatic evolution of a single, lowest energy, state of the effective Hamiltonian. }
    \label{fig:scalings-D}
\end{figure}

To test how well our evolution equation~\eqref{eq:D-evol} performs, Figure~\ref{fig:scalings-D} shows the evolution of $\alpha, \beta, \gamma$ {introduced at the beginning of this Section~\ref{sec:scaling}, evaluated as in \eqref{eq:scaling-exponents-2} via the characteristic momenta and occupancies of the distribution function computed via the full evolution equation, namely}
\begin{align}
    \beta_{\langle p_T^2\rangle} &= - \frac{1}{2} \partial_y \ln \langle p_\perp^2 \rangle \, ,  \\ 
    \gamma_{\langle p_z^2\rangle} &= -\frac{1}{2} \partial_y \ln \langle p_z^2 \rangle \, , \\
    \alpha_{\langle f \rangle} &= \partial_y \ln \langle f \rangle 
\label{eq:scaling-exponents-3}
\end{align}
and as calculated from the single basis state $\psi_{10}^{(R)}$ (which carries the particle number) alone, namely 
\begin{align}
    \beta_1 &\equiv -\frac12 \partial_y \ln \langle p_\perp^2 \rangle_1 = - \partial_y \ln D  \, , \\
    \gamma_1 &\equiv -\frac12 \partial_y \ln \langle p_z^2 \rangle_1 = - \partial_y \ln D \, , \\
    \alpha_1 &\equiv \partial_y \ln \langle f_1 \rangle_1 = - 1 - 3 \partial_y \ln D \, ,
\end{align}
where the average $\langle \cdot \rangle_1$ denotes
\begin{equation}
    \langle X \rangle_1 \equiv \frac{\int_{\bf p} X f_1 }{\int_{\bf p} f_1} \, ,
    \label{eq:single-state-average}
\end{equation}
with $f_1$ being the 
particle-number-carrying basis state
\begin{equation}
    f_1({\bf p}, y) = A(y) \psi_{10}^{(R)}(\chi = p/D(y), u = p_z/p ) \, ,
\end{equation}
and where $A$ is chosen to be proportional to $e^{-y} D^{-3}$, such that $\alpha = 3\delta - 1$. As we can see via the comparison between dashed and solid curves in Fig.~\ref{fig:scalings-D}, the information about how the typical longitudinal and transverse scales evolve in time is described well at late times and reasonably well at intermediate times, provided the separation between the rate of change of the longitudinal and transverse scales is not large.

\begin{figure}
    \centering
    \includegraphics[width=0.7\textwidth]{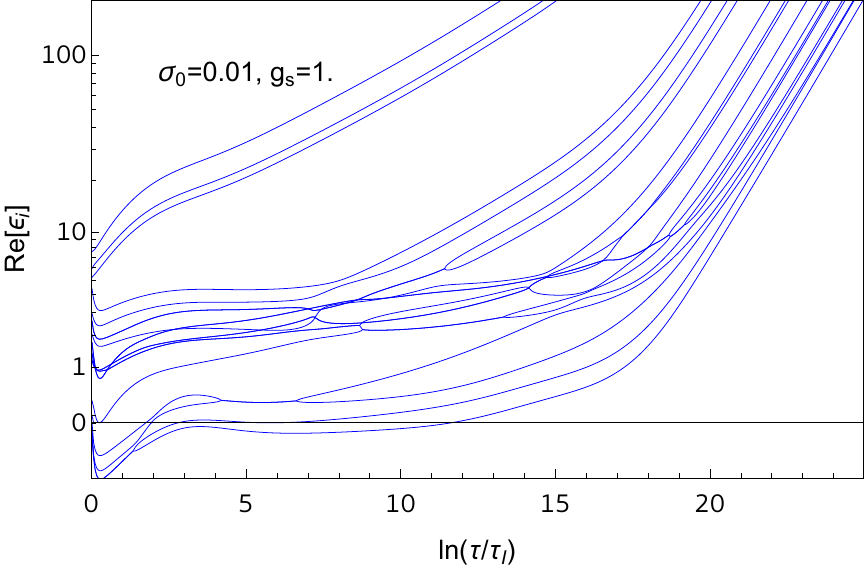}
    \caption{Plot of the eigenvalues of the effective Hamiltonian~\eqref{eq:Heff-thermal-noBose} for $g_s = 1$ and the initial condition given in Eq.~\eqref{eq:init-cond-iso}. We used $N_{\rm states} = 36$; here we display the 20 states of even $u \to -u$ parity, which are the ones with nonzero occupancy. We see a cluster of low energy eigenstates (and clusters of higher energy eigenstates also) at early times. Before $\ln(\tau/\tau_I)=15$, all states except one separate from the single ground state of the effective Hamiltonian $H$ that has eigenvalue zero. Comparing to the top-left ($g_s=1$) panel of Fig.~\ref{fig:scalings-D}, we see that the hydrodynamization phenomenon in that Figure is described by the adiabatic evolution of the isolated ground state of the effective Hamiltonian seen here.
    }
    \label{fig:eigenvals-lateTime}
\end{figure}

\begin{figure}
    \centering
    \includegraphics[width=0.8\textwidth]{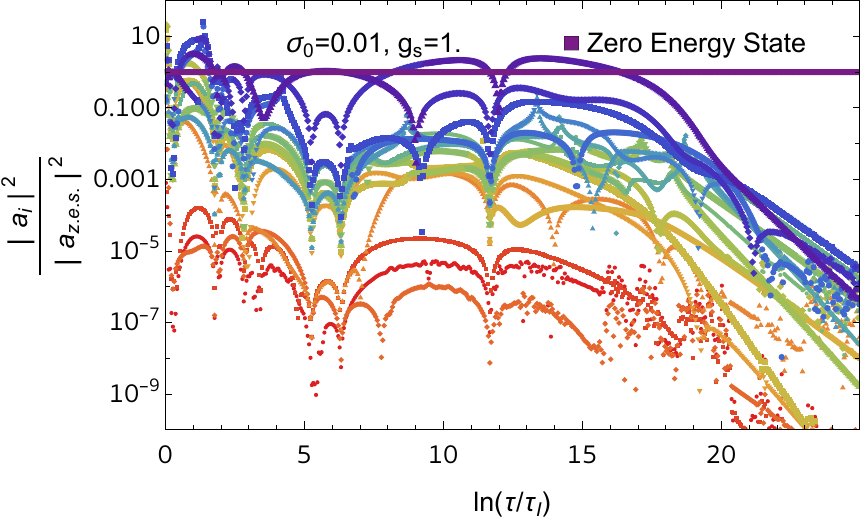}
    \caption{Plot of the coefficients in the $H$ eigenstate decomposition of the distribution function for $g_s = 1$ and the initial condition given in Eq.~\eqref{eq:init-cond-iso}. We used $N_{\rm states} = 36$; here we display the 20 states of even $u \to -u$ parity, which are the ones with nonzero occupancy. The coefficients are normalized relative to the occupation of the zero-energy-eigenvalue state that carries the particle number. At late times, the distribution function is composed almost entirely of this state whose adiabatic evolution describes the hydrodynamization of this kinetic theory.
    }
    \label{fig:coefficients-lateTime}
\end{figure}

\begin{figure}
    \centering
    \includegraphics[width=0.7\textwidth]{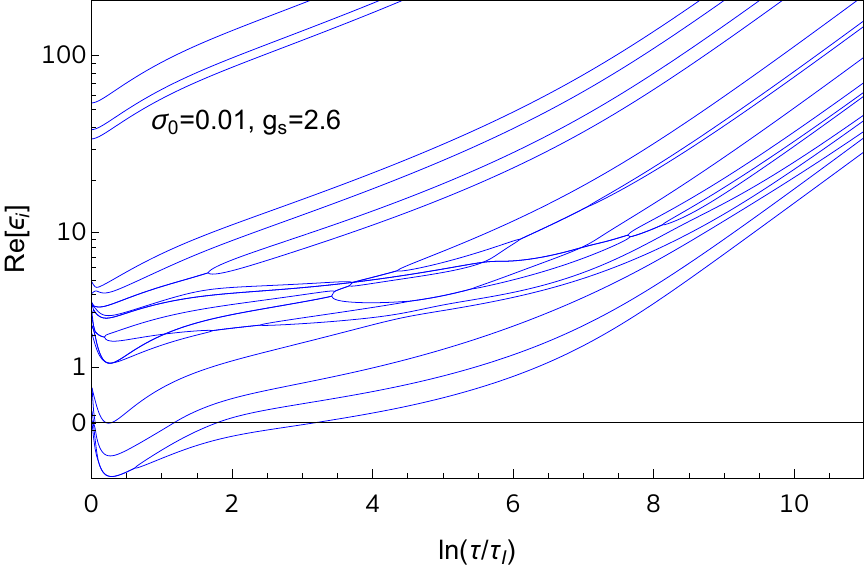}
    \caption{Plot of the eigenvalues of the effective Hamiltonian~\eqref{eq:Heff-thermal-noBose} for $g_s = 2.6$ and the initial condition given in Eq.~\eqref{eq:init-cond-iso}. We used $N_{\rm states} = 36$; here we display the 20 states of even $u \to -u$ parity, which are the ones with nonzero occupancy. Before $\ln(\tau/\tau_I)=5$, all states except one separate from the single ground state of the effective Hamiltonian $H$ that has eigenvalue zero. Comparing to the bottom-right ($g_s=2.6$) panel of Fig.~\ref{fig:scalings-D}, we see that the hydrodynamization phenomenon in that Figure is described by the adiabatic evolution of the isolated ground state of the effective Hamiltonian seen here.
    }
    \label{fig:eigenvals-lateTime-2}
\end{figure}

\begin{figure}
    \centering
    \includegraphics[width=0.8\textwidth]{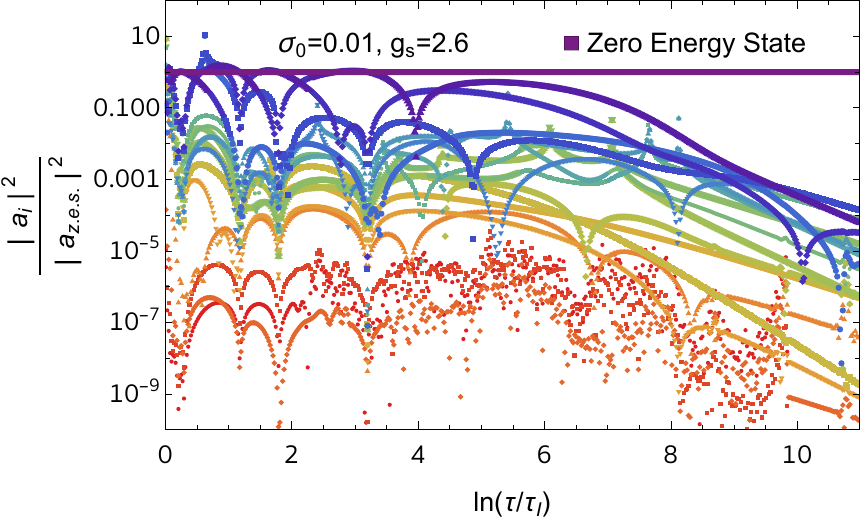}
    \caption{Plot of the coefficients in the $H$ eigenstate decomposition of the distribution function for $g_s = 2.6$ and the initial condition given in Eq.~\eqref{eq:init-cond-iso}. We used $N_{\rm states} = 36$; here we display the 20 states of even $u \to -u$ parity, which are the ones with nonzero occupancy. The coefficients are normalized relative to the occupation of the 
    zero-energy-eigenvalue state that carries the particle number. At late times, the distribution function is composed almost entirely of this state whose adiabatic evolution describes the hydrodynamization of this kinetic theory.
    }
    \label{fig:coefficients-lateTime-2}
\end{figure}

We see {from the final values of the scaling exponents in Fig.~\ref{fig:scalings-D} that all of the solutions we consider here hydrodynamize, reaching the scaling form with the values of the scaling exponents that describe a boost-invariant longitudinally expanding kinetic theory fluid in} local thermal equilibrium at late times. It was not possible to see this with the choice of basis and scalings employed in the previous Section. 
Moreover, as we can see from Figures~\ref{fig:eigenvals-lateTime} and~\ref{fig:eigenvals-lateTime-2}, hydrodynamization follows after an energy gap opens up between a single isolated ground state and all the excited states, with the dynamics from then on being governed by the adiabatic evolution of the ground state of the effective Hamiltonian, with the evolution of this state bringing the system to a thermal distribution at late times.
The dominance of this state, which by construction has energy eigenvalue ${\rm Re}\, \epsilon_n = 0$, is apparent from the late time behavior of 
Figs.~\ref{fig:coefficients-lateTime} and~\ref{fig:coefficients-lateTime-2}.  At late times, during hydrodynamization, the distribution function is composed almost entirely of the adiabatically evolving ground state. The occupation of any of the  higher energy states is small and declining.
Therefore, with the basis and scaling we have introduced here, AH describes an attractor that hydrodynamizes.

In addition to seeing how
pre-hydrodynamic scaling and its attractor (Sect.~\ref{sec:pre-hydro}) and the attractor that brings this kinetic theory through the hydrodynamization process (this Section) are each described naturally via AH, the other important conclusion that we can draw from these results is simply that a longitudinally expanding gas of gluons can reach thermal equilibrium even when the collision kernel only includes small-angle scattering, provided that the coupling is large enough. We shall see in the next Section that if the coupling is too weak (in this kinetic theory with only small-angle scattering)  the system dilutes away before it can hydrodynamize.

\section{Adiabaticity beyond scaling}
\label{sec:non-scaling}

In the previous Section, we described the hydrodynamization of a longitudinally expanding gas of gluons in two stages: first, a pre-hydrodynamic attractor; then, an attractor that describes hydrodynamization. For each stage separately, we found a basis and a scaling such that the dynamics are described naturally via the adiabatic evolution of a few (prehydrodynamic attractor) or one (hydrodynamization) lowest energy eigenstate(s) of the (different) effective Hamiltonians that we were able to construct explicitly in each case.  This is a pleasing validation of the power of the AH framework, a confirmation that in two quite different contexts it provides an intuitive description of the loss of memory of earlier stages and the scaling evolution of the attractor.
What would be even better, though, would be a single seamless AH description of the entire dynamics, including the prehydrodynamic attractor and the subsequent hydrodynamization. 
What we have done so far does not tell us how to connect the early time dynamics with the hydrodynamic regime. If we are to capture both of these different scaling regimes that within a single description, that description cannot be described by a single scaling form throughout. This means that to achieve this goal we must
generalize the AH framework beyond the scaling assumption. This is what we do in this Section.

So far, we have discussed how to seek and find adiabatically evolving ground states of theories where the distribution function acquires a scaling form, and all of the time dependence of this state can be encoded in a few time-dependent variables. The main simplification in the dynamics is that by a suitable choice of these time-dependent variables the ground state of the effective Hamiltonian becomes as slowly varying as possible, and therefore evolves adiabatically. From the point of view of the decomposition of the distribution function in basis states, this is implemented by having a basis that is ``co-moving'' with the typical width of the distribution as a function of momentum.

However, a more general picture is possible following the same logic. The essential ingredient of the above is that the basis is well-adapted to the physical dynamics of the problem. When the system undergoes dynamics that drive it to a universal scaling form, what we have discussed so far in the previous sections is ideal. However, when the system transitions between two such phases, the time-dependent parameters of the basis should be adapted in such a way that they can adequately describe the intermediate dynamics.  If we can find such parameters so that the energy levels of the Hamiltonian are gapped and the ground state is slowly evolving in time as before, then we can succeed in connecting the different phases with a simple, unified, adiabatic description.

This is precisely the situation in the kinetic theory for a longitudinally expanding gluon gas with only small-angle scattering in the collision kernel that we have investigated in Sec.~\ref{sec:expanding}. There, we had one scaling form long before hydrodynamics and another as the system hydrodynamizizes. Our goal now is to smoothly connect them. We anticipate that the construction that follows can be extended to more general kinetic theories as well. With this in mind, in Sect.~\ref{sec:non-scaling-H} we will first describe the differences between the approach of this Section and that of Sect.~\ref{sec:scaling-H} in general terms. Then, in 
Sect.~\ref{sec:connectstages} we proceed to describe the boost-invariant kinetic theory of gluons undergoing small-angle scatterings from early to late times, quantitatively.

\subsection{Effective Hamiltonians outside the scaling regime} \label{sec:non-scaling-H}

Because of the preceding discussion, we are led to consider distribution functions that are not scaling, but nonetheless have explicit time dependence through a parameter that we introduce by hand. For definiteness, consider the case where we write
\begin{equation}
    f({\bf p},y) = A(y) w \left( \frac{p}{D(y)}, u, r(y), y \right) \, ,
\end{equation}
{{where $u=p_z/p$ as in Sect.~\ref{sec:hydro} and where here we have introduced}} $r(y)$, a parameter that will be encoded \textit{in the definition} of the basis we will expand $w$ on {and that we will use to important ends below. Note now that when we write $\partial_y w$ in what follows, we shall always understand this derivative to be defined as acting only on the last argument of $w$, not on the $y$-dependence inside $r$ and $D$. The idea of this definition is to absorb the ``fast'' evolution of the scales in the distribution function into the parameter $r(y)$ and the scaling function $D(y)$ and, by doing so, make the evolution of the basis state coefficients we choose to set up the description ``slow,'' which in turn enables the eigenstates to be slowly-varying functions of time, as required for AH to take place.}

As before, we assume we have a boost-invariant kinetic theory with a specified collision kernel. The effective Hamiltonian that implements $\partial_y w = - {H}_{\rm eff} w$ is given by
\begin{equation}
    H_{\rm eff} = \frac{\partial_y A}{A} + (\partial_y r) \partial_r - \frac{\partial_y D}{D} \chi \partial_\chi - u^2 \chi \partial_\chi - u(1-u^2) \partial_u  - \tau \tilde{\mathcal{C}}[f = A w(\chi, u, r, y)]_{\chi = p/D(y)} \,\, ,
\end{equation}
where $r(y)$ is a function we have to specify based on the same considerations as before, from which it will follow that the evolution of the distribution function is dominated by the ground state(s) of this effective Hamiltonian.

Upon making the small-angle scattering approximation, this effective Hamiltonian becomes
\begin{align}
    H_{\rm eff} &= \alpha + (\partial_y r) \partial_r + \delta \chi \partial_\chi - u^2 \chi \partial_\chi - u(1-u^2) \partial_u \nonumber \\  
    & \quad - \tau \lambda_0 \ell_{\rm Cb} \frac{I_a}{D^2}  \left[ \frac{2}{\chi} \partial_\chi + \partial_\chi^2 + \frac{1}{\chi^2} \frac{\partial}{\partial u} \left( (1 - u^2) \frac{\partial f}{\partial u} \right) \right] \nonumber \\ 
    & \quad - \tau \lambda_0 \ell_{\rm Cb} \frac{I_b}{D} \left[ \frac{2(1+Aw)}{\chi} + (1+2Aw) \partial_\chi \right] \, , \label{eq:Heff-smallanglescatt-Dr}
\end{align}
where we have reintroduced $\delta \equiv - \partial_y D/D$ and $\alpha \equiv \partial_y A/A$. As before, in what follows we shall drop the $I_b f^2$ terms in the collision kernel, which corresponds to dropping the explicit $w$ terms in the last line of $H_{\rm eff}$.

Other than the fact that there is a new operator in $H_{\rm eff}$ due to the new time-dependent parameter $r(y)$, the next logical steps follow those of Section~\ref{sec:scaling-H} almost identically. If the criterion $\delta_A^{(n)} \ll 1$ is satisfied in tandem with an energy gap between the lowest energy state(s) and all higher energy states, then the ground state will rapidly come to dominate the evolution of the system. When the evolution of the system comes to be governed by the adiabatic evolution of the instantaneous ground state of $H_{\rm eff}$, this provides an intuitive path to understanding how the kinetic theory loses almost all memory of its initial state  as an attractor emerges and to 
identifying what degrees of freedom do get transported from the initial to the final state. Furthermore, here we shall be able to do all of this in a way that follows the evolution of an instantaneous ground state that connects the early-time pre-hydrodynamic evolution in a heavy ion collision smoothly through to
hydrodynamization, and explain the emergence of the small set of low-energy degrees of freedom describing hydrodynamics.

\subsection{Example: Longitudinally expanding gluon gas from free-streaming until hydrodynamics} \label{sec:connectstages}

Having discussed in general terms how to find an adiabatically evolving ground state that captures the dynamics of a weakly coupled gluon gas without assuming a single scaling form throughout the evolution, we now discuss the concrete setup that we will use to describe the longitudinally expanding gluon gas from free-streaming through prehydrodynamic attractor to hydrodynamization and hydrodynamics, explicitly and quantitatively.

The first ingredient for this is the choice of basis on which to expand the rescaled distribution function $w(\chi,u,r,y)$. The main requirement that we need in our basis is that it must be able to accurately describe both of the self-similar solutions we have observed in the previous Section: the early-time prehydrodynamic attractor described in Section~\ref{sec:pre-hydro} and the late-time hydrodynamizing attractor described in Section~\ref{sec:hydro}. To this end, we choose
\begin{align}
    \psi_{nl}^{(R)} = N_{nl} e^{-\chi} e^{-u^2 r^2/2} L_{n-1}^{(2)}(\chi) Q_l^{(R)}(u;r) \, , & & \psi_{nl}^{(L)} = N_{nl} L_{n-1}^{(2)}(\chi) Q_l^{(L)}(u;r) \, ,
\end{align}
where the polynomials $Q_l^{(R)}(u;r)$, $Q_l^{(L)}(u;r)$ are polynomials on $u$ of degree $l$, constructed such that
\begin{equation}
    \int_{-1}^1 du \, e^{- u^2 r^2/2} Q_l^{(L)}(u;r) Q_k^{(R)}(u;r) = 2 \delta_{lk} \, .
\end{equation}
That is to say, they are constructed such that they are orthogonal with respect to the measure $e^{- u^2 r^2/2}$ on the $(-1,1)$ interval, and are normalized by setting $Q_0^{(L)} = 1$, $Q_0^{(R)} = 2/J_0(r) $, and $Q_k^{(L)} = J_0(r) Q_k^{(R)}/2$ for $k \geq 1$, where $J_0(r) = \int_{-1}^1 du \, e^{-u^2 r^2/2}$, consistent with the definition~\eqref{eq:Jn-moments} that we will introduce later. In fact, in the $r\to 0$ limit, they are equivalent to Legendre polynomials, whereas in the $r \to \infty$ limit they approach Hermite polynomials in shape (after an appropriate rescaling of the $u$ coordinate by $r$).

As before, the normalization coefficients $N_{nl}$ are chosen such that
\begin{align}
    \frac{1}{4\pi^2} \int_{-1}^1 du \int_0^\infty d\chi \, \chi^2 \psi_{mk}^{(L)} \psi_{nl}^{(R)} = \delta_{kl} \, .
\end{align}
In essence, what this basis does is that it reproduces the late-time basis of 
Sect.~\ref{sec:hydro} if $r = 0$, and resembles the early-time basis of Sect.~\ref{sec:pre-hydro} when $r \gg 1$, with the identification $\xi \sim r u$, $\zeta \sim \chi$. In particular, we expect that at early times, when $\langle p_z^2 \rangle \ll \langle p_\perp^2 \rangle$, $r$ will assume the role of $C$ and encode the longitudinal expansion of the system. In this way, it provides us with a parameter that makes the basis flexible enough so that the physical state of the kinetic theory, i.e.~its distribution function, can always be well described by a small set of basis states including both at early and at late times.  This flexibility that this choice of basis incorporates means that the dynamics we describe need not, and will not, be characterized by a single scaling form throughout. We shall see, though, that at sufficiently weak coupling the time evolution dictated by the effective Hamiltonian ensures that at early times the system follows the prehydrodynamic scaling of Sect.~\ref{sec:pre-hydro} and at later times it hydrodynamizes as in Sect.~\ref{sec:hydro}, with the basis chosen so as to yield an efficient description throughout.

In principle, we could choose $r$ by maximizing the degree to which the evolution of the system is adiabatic. However, we know from the previous sections that, at early times, the maximally adiabatic basis choice implies that there is a group of quasi-degenerate ``ground states'' that will drive the evolution. In contrast to this, at later times the ground state is unique and evolves to become the thermal distribution. At early times, even though the gaps among these low-energy states are extremely small, the time derivative of the ground states themselves is also small. As the system approaches thermalization, a gap must open up between an isolated lowest energy state and the other state(s) that were previously almost degenerate, and all these eigenstates of the effective Hamiltonian have to rearrange themselves into different functional forms. Hence, maximizing adiabaticity during this stage of the evolution might be too strict of a condition for what is actually needed, which is that the physical state of the system is dominated by a (set of) ground state(s) throughout. 

Instead of maximizing adiabaticity, we shall choose $r$ here based upon what we already know about the physical behavior of the gluon distribution at early and late times from Sects.~\ref{sec:pre-hydro} and \ref{sec:hydro}, by matching each to an approximate evolution of the distribution function. We motivate our choice of $r$ in the following way: if the distribution function were dependent on $p$ and $u$ in a factorized form as
\begin{equation}
    w \sim e^{-y} \frac{D_0^3}{D^3} e^{-\chi} \frac{e^{-u^2r^2/2}}{\int_{-1}^1 dv \, e^{-v^2 r^2/2} } \, , \label{eq:w-r-motivation}
\end{equation}
then we can choose $r$ such that the evolution equation for the $\langle u^2 \rangle$ moment of the distribution function is exactly satisfied (for this specific functional form of $w$). That is to say, such that $\int_{\bf p} u^2 \partial_y f = \int_{\bf p} u^2 \big( p_z \partial_{p_z} f  - C[f] \big) $. The idea is to maximize the degree to which the evolution can be approximately described by the first basis state of our construction.
We then introduce the following integral moments
\begin{equation} \label{eq:Jn-moments}
    J_n(r) = \int_{-1}^1 du \, u^n e^{-u^2 r^2/2} \, ,
\end{equation}
which allow us to derive the following equation:
\begin{equation} \label{eq:r-evol}
    \partial_y r = - \frac{1}{r} \frac{J_0}{J_4 J_0 - J_2^2} \left[ -2 ( J_2 - J_4) + \frac{\tau \lambda_0 \ell_{\rm Cb} I_a }{D^2} (J_0 - 3 J_2) \right] \, ,
\end{equation}
which we use to fix the evolution of $r$ in the analysis that follows. It can be seen from Eq.~\eqref{eq:w-r-motivation} that the parameter $r$ describes how anisotropic the basis states with which we describe the distribution function are. The case when $r = 0$ corresponds to a fully isotropic basis. And indeed, the time evolution described by Eq.~\eqref{eq:r-evol} approaches $r\rightarrow 0$ (isotropizes) at late times in a carefully balanced way: as $\tau$ grows large, the value of $r$ will be driven to a configuration where $J_0 = 3 J_2$
which (by inspection of the expressions for these integral moments)
is only satisfied with $r=0$. (Note that at $r=0$, $J_0=2$, $J_2=2/3$, and $J_4=2/5$.) 
In more detail, the approach to isotropy $r \to 0$ happens in such a way that
the quantity in square brackets on the right-hand side of \eqref{eq:r-evol} evolves toward zero (meaning that $r$ becomes constant) even while
the time-dependent quantity $\tau \lambda_0 \ell_{\rm Cb} I_a/D^2$ grows with time without bound.
That is, $J_0-3J_2$ evolves toward zero inversely proportional to this quantity
with the proportionality constant being
$2(J_2-J_4)$.  This means that 
$r$ evolves toward 0, as that is where $J_0=3J_2$, and this corresponds to isotropization.
With this, we see that the time at which the basis states become isotropic is controlled by the dynamical quantity $\tau \lambda_0 \ell_{\rm Cb} I_a/D^2$.

As in \eqref{eq:D-evol}, we let the $D$ scaling evolve as
\begin{equation} \label{eq:Dr-evol}
    \frac{\partial_y D}{D} = 10 \left( 1 - D \left\langle \frac{2}{p} \right\rangle  \right) \, ,
\end{equation}
with the simple idea that $D$ should follow the typical momentum scale of the temperature, and match the (inverse) effective temperature of the system at late times. As before, we choose $A$ such that the eigenstate that carries particle number has a vanishing eigenvalue, i.e., $\alpha = 3\delta - 1$, with $\delta = - \partial_y D / D$.

\begin{figure}
    \centering
    \includegraphics[width=0.49\textwidth]{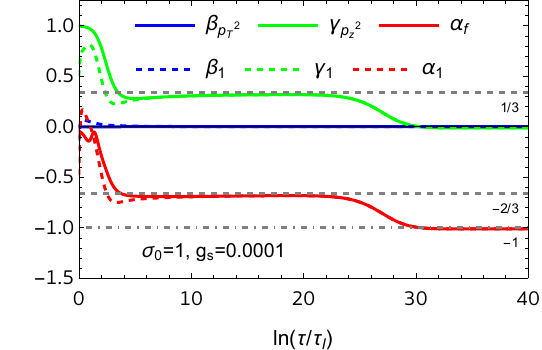}
    \includegraphics[width=0.49\textwidth]{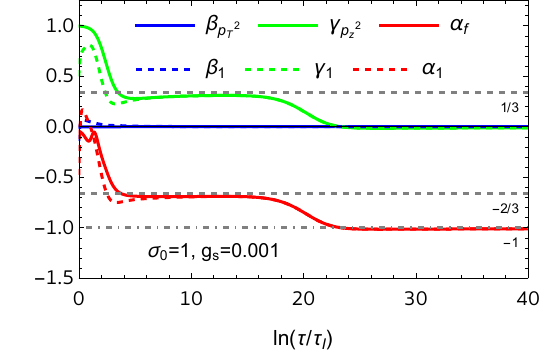}
    \includegraphics[width=0.49\textwidth]{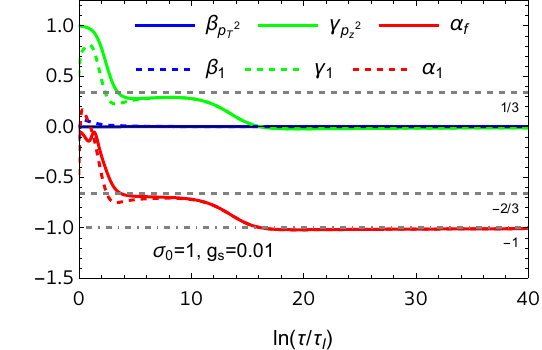}
    \includegraphics[width=0.49\textwidth]{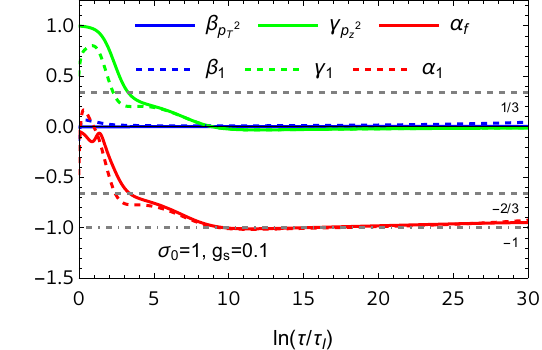}
    \caption{Evolution of the typical momentum scales encoded in the scaling exponents $\alpha, \beta, \gamma$ for weakly coupled kinetic theories with the initial condition specified in Eq.~\eqref{eq:init-cond}. From left to right and top to bottom, $g_s = 10^{-4}, 10^{-3}, 10^{-2}, 10^{-1}$.  We used $N_{\rm states} = 15$. In order to test how well-adapted the basis is to the dynamics of the gluon gas, we plot two sets of scaling exponents: the solid lines describe the scaling exponents as calculated from the moments $\langle p_z^2 \rangle$ and $\langle p_\perp^2 \rangle$, and the dashed lines represent the evolution of those scaling exponents as described only by the basis state $\psi^{(R)}_{10}$ that carries the particle number.
    }
    \label{fig:scalings-Dr-weak}
\end{figure}

\begin{figure}
    \centering
    \includegraphics[width=0.49\textwidth]{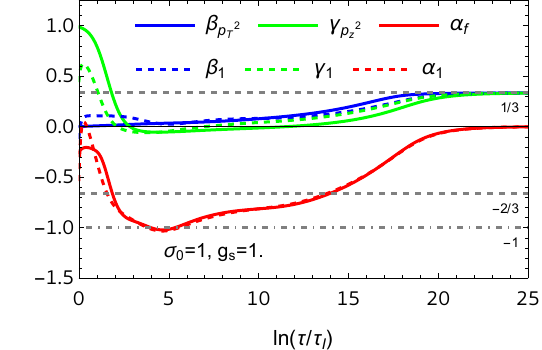}
    \includegraphics[width=0.49\textwidth]{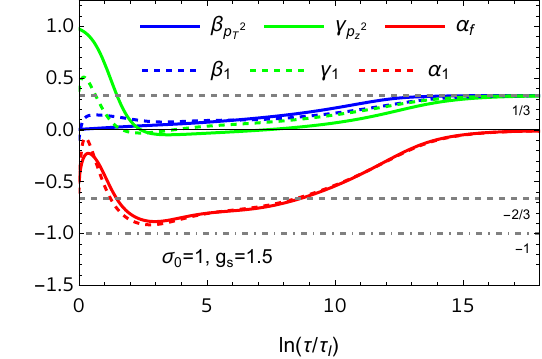}
    \includegraphics[width=0.49\textwidth]{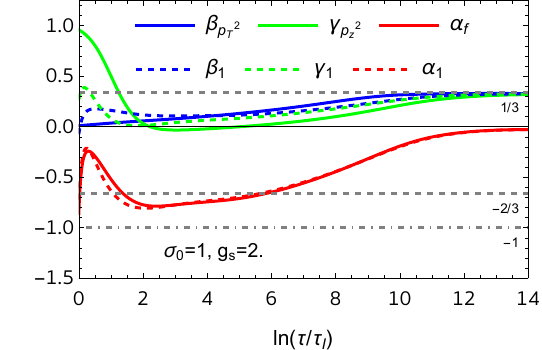}
    \includegraphics[width=0.49\textwidth]{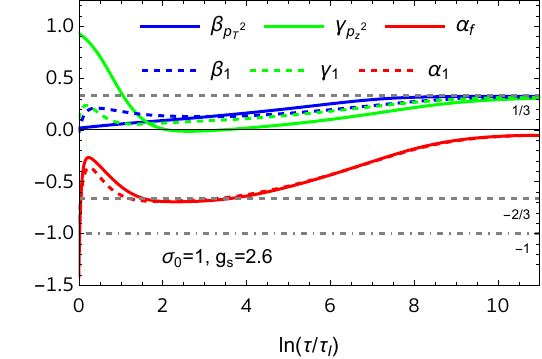}
    \caption{Evolution of the typical momentum scales encoded in the scaling exponents $\alpha, \beta, \gamma$ for more strongly coupled kinetic theories with the initial condition specified by Eq.~\eqref{eq:init-cond} as in Fig.~\ref{fig:scalings-Dr-weak}. From left to right and top to bottom, $g_s = 1, 1.5, 2, 2.6$. We used $N_{\rm states} = 21$. As before, the solid lines describe the scaling exponents as calculated from the moments $\langle p_z^2 \rangle$ and $\langle p_\perp^2 \rangle$, and the dashed lines represent the evolution of those scaling exponents as described only by the basis state $\psi^{(R)}_{10}$ that carries the particle number.
    }
    \label{fig:scalings-Dr-strong}
\end{figure}

At the very early stages, $\tau \to 0$ suppresses the effect of collisions and $r$ becomes large as $r \sim e^y$. When $r$ is large, its evolution equation reduces to that of $C$ in the work of BSY~\cite{Brewer:2022vkq} with the identification $C = D/r$, where $D$ stays essentially constant because $\langle 1/p \rangle$ is dominated by the $p_\perp$ dependence of the distribution during the BMSS ($\gamma \approx 1/3$ and $\alpha \approx -2/3$) and dilute ($\gamma \approx 0$ and $\alpha \approx -1$) stages
that we introduced at the beginning of Section~\ref{sec:scaling}, and described in Sect.~\ref{sec:pre-hydro} (see Fig.~\ref{fig:bsycomparison2}). 
Throughout this second stage, $r$ slowly decreases (in a non-exponential way with $y$). As we explained just before, as time progresses the term explicitly proportional to $\tau$ in the evolution equation \eqref{eq:r-evol} for $r$ dominates and the system is driven to a state where $J_2/J_0 = 1/3$,  which corresponds to $ r = 0$, where the basis functions $Q_l$ are Legendre polynomials by construction and thus correspond to spherical harmonics. By itself, this does not mean that the distribution function is isotropic, because this would require that all of the excited states with nontrivial profiles along $u$ decay away, which is something we will verify by showing that an energy gap opens up dynamically as the system hydrodynamizes. Nonetheless, because the hydrodynamic state corresponds to an isotropic distribution, and this state will be dynamically approached from a highly anisotropic state, we expect that having a basis that becomes isotropic at the same time as the system hydrodynamizes will provide a simpler description of the dynamics than an isotropic basis. In this way, this setup is well-equipped to describe isotropization of the gluon distribution and hydrodynamization, as we will demonstrate explicitly in what follows.

In Figs.~\ref{fig:scalings-Dr-weak} and~\ref{fig:scalings-Dr-strong} we see a comparison of the evolution of the typical scales of the problem in terms of the longitudinal and transverse scaling exponents $\gamma$ and $\beta$, respectively, for eight choices of the coupling $g_s$. For these solutions, we have chosen the initial condition to be
\begin{equation} \label{eq:init-cond}
    f({\bf p}, \tau = \tau_I) = \frac{\sigma_0}{g_s^2} e^{- \sqrt{2} p/Q_s} e^{- r_I^2 u^2 /2 } Q_0^{(R)}(u;r_I)
\end{equation}
with $r_I = \sqrt{3}$, $\tau_I Q_s = 1$, and $\sigma_0 = 1$.
Our initial conditions are inspired by previous works~\cite{Kurkela:2015qoa,Mazeliauskas:2018yef,Boguslavski:2023jvg} so as to match them in qualitative terms. Furthermore, they are suitably chosen to study the bottom-up thermalization scenario~\cite{Baier:2000sb} with $f \sim 1/g_s^2$ at the earliest times $\tau Q_s \sim 1$ of a heavy-ion collision when the typical momentum scale characterizing $f$ is $Q_s$. In practice, we choose $\langle p \rangle = 3\, Q_s / \sqrt{2} \approx 2.12 \,Q_s$, which is close to the initial typical momentum in the simulations of Ref.~\cite{Boguslavski:2023jvg}. With all of 
these choices in hand, the only parameter we vary in Figs.~\ref{fig:scalings-Dr-weak} and~\ref{fig:scalings-Dr-strong} is the coupling constant $g_s$.

We stress that our initial condition resembles that of Ref.~\cite{Mazeliauskas:2018yef} more than that of Refs.~\cite{Kurkela:2015qoa,Boguslavski:2023jvg}, which suffices for our purposes as we will not attempt to precisely describe the IR physics of the distribution function. 
  Indeed, such a goal would require us to keep, at the very least, the $I_b f^2$ terms we have neglected in the kinetic equation, and ultimately the $1 \leftrightarrow 2$ processes we have omitted throughout.

To test how well our choice of evolution equations~\eqref{eq:r-evol} and~\eqref{eq:Dr-evol} perform, Figures~\ref{fig:scalings-Dr-weak} and~\ref{fig:scalings-Dr-strong} show the evolution of $\alpha, \beta, \gamma$ as calculated from the characteristic occupancy and momenta of the full distribution via \eqref{eq:scaling-exponents-2}, namely \eqref{eq:scaling-exponents-3},
and as calculated from only the evolution of the single basis state that carries the particle number, which is governed by the evolution of $D$ and $r$ {{in such a way that here the resulting scaling exponents are}
\begin{align}
    \beta_1 &\equiv -\frac12 \partial_y \ln \langle p_\perp^2 \rangle_1 = -\frac{1}{2} \partial_y \ln  \left( D^2 - \frac{D^2 J_2(r)}{J_0(r)} \right) \, , \\
    \gamma_1 &\equiv -\frac12 \partial_y \ln \langle p_z^2 \rangle_1 = -\frac{1}{2} \partial_y \ln  \frac{D^2 J_2(r)}{J_0(r)} \, , \\
    \alpha_1 &\equiv \partial_y \ln \langle f_1 \rangle_1 = - 1 - \partial_y \ln \frac{D^3 J_0(r)^2}{J_0(\sqrt{2} r)} \, .
\end{align}
where the average $\langle \cdot \rangle_1$ is as defined previously in \eqref{eq:single-state-average},
where $f_1$ is given by the particle number-carrying state in the basis that we use throughout this Section
\begin{equation}
    f_1({\bf p}, y) = A(y) \psi_{10}^{(R)}(\chi = p/D(y), u = p_z/p ; r(y)) \, ,
\end{equation}
and, as previously mentioned, $A$ is chosen to be proportional to $e^{-y} D^{-3}$, such that $\alpha = 3\delta - 1$.}

As we can see via the (successful) comparison between the dashed and solid curves in Figs.~\ref{fig:scalings-Dr-weak} and~\ref{fig:scalings-Dr-strong}, the information about how the typical longitudinal and transverse scales evolve in time (solid curves) is already well-captured by 
our choices of $D(y)$ and $r(y)$
that determine the dashed curves
for all values of the coupling constant that we have considered. We describe the physical regimes corresponding to the stages seen in the evolution of the scaling exponents in Figs.~\ref{fig:scalings-Dr-weak} and~\ref{fig:scalings-Dr-strong} below, after looking at the behavior of the pressure anisotropy during the same dynamical evolution.

\begin{figure}
    \centering
    \includegraphics[width=0.48\textwidth]{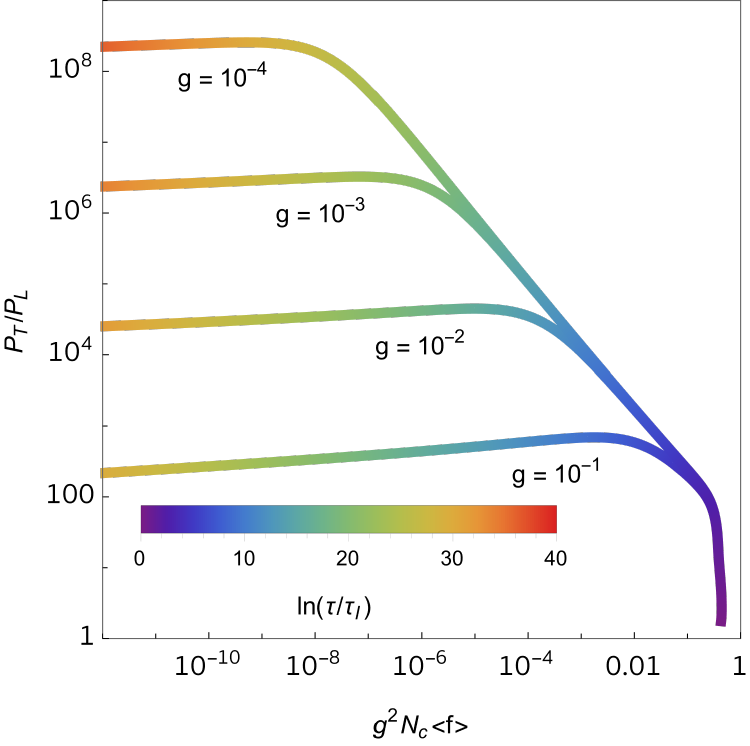}
    \includegraphics[width=0.48\textwidth]{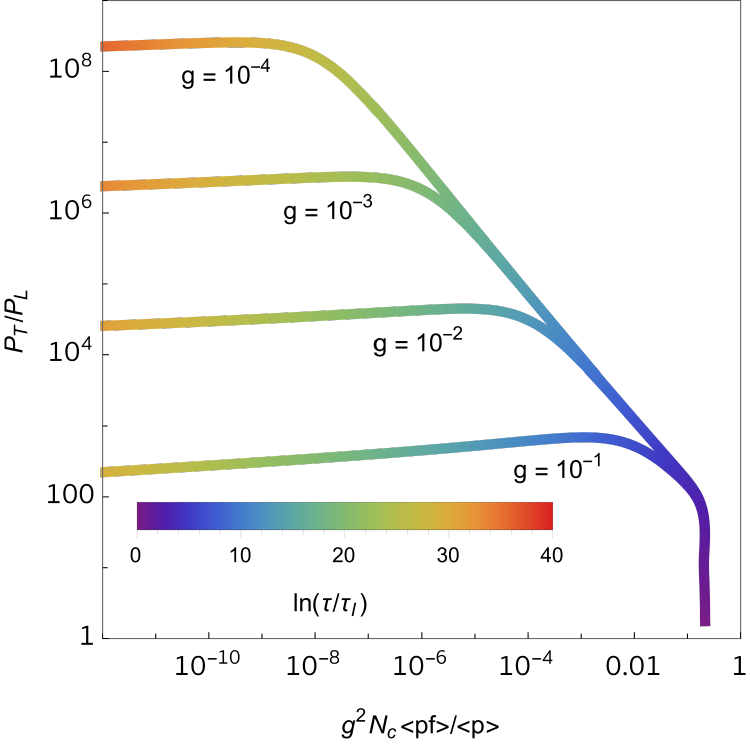}
    \caption{Left: evolution of the pressure anisotropy as a function of the occupancy $g_s^2 N_c \langle f \rangle$ of the distribution function, with the evolution time along each curve depicted by the coloring. Right: evolution of the pressure anisotropy as a function of the energy-weighted occupancy $g_s^2 N_c \langle p f \rangle/\langle p \rangle$ of the distribution function. Both plots were obtained from weakly coupled kinetic theories with $g_s = 10^{-1}, 10^{-2}, 10^{-3}, 10^{-4}$, with initial conditions specified by Eq.~\eqref{eq:init-cond}.  Here we used $N_{\rm states} = 15$.
    }
    \label{fig:occupancies-anisotropies-weak}
\end{figure}

\begin{figure}
    \centering
    \includegraphics[width=0.48\textwidth]{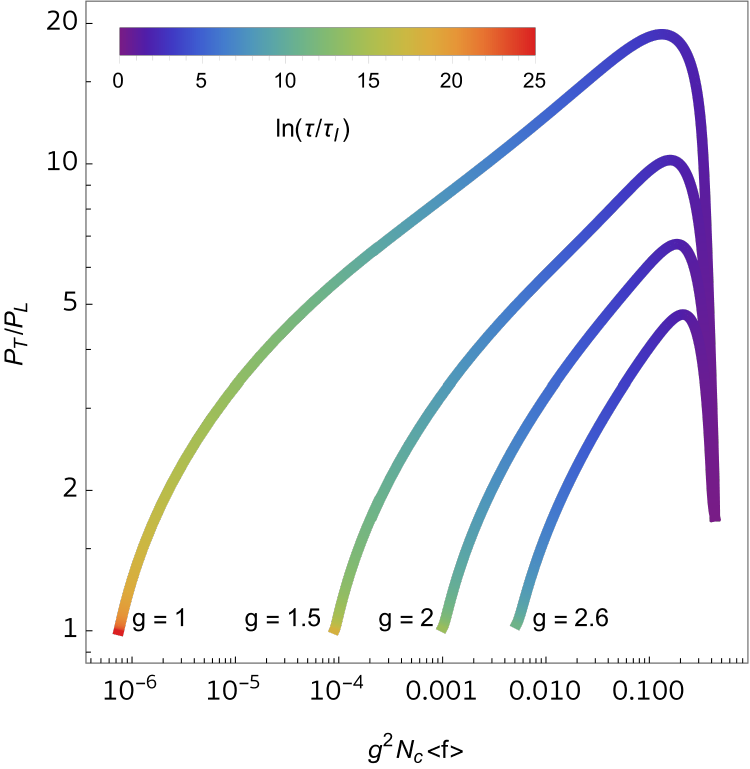}
    \includegraphics[width=0.48\textwidth]{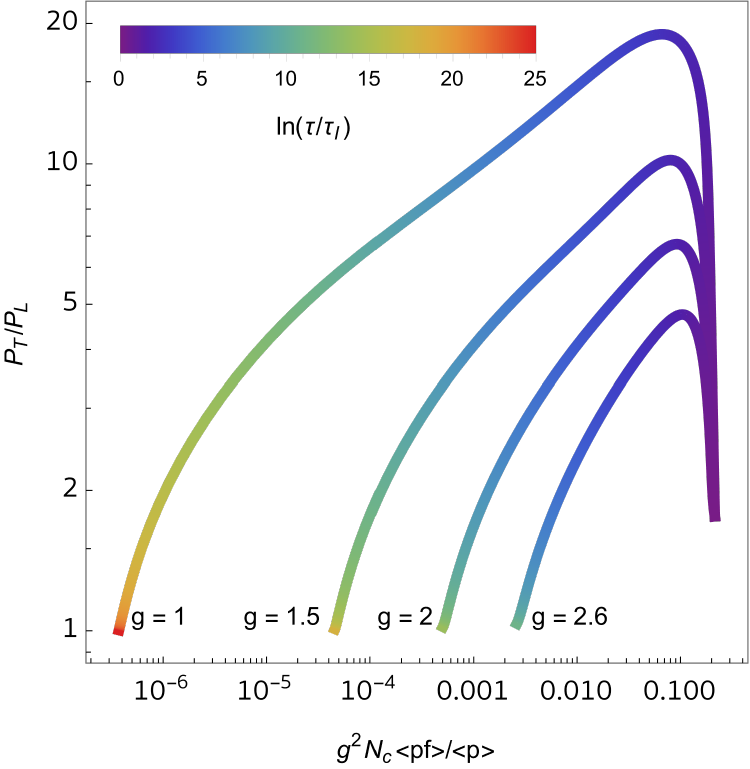}
    \caption{Left: evolution of the pressure anisotropy as a function of the occupancy $g_s^2 N_c \langle f \rangle$ of the distribution function, with the evolution time represented by color. Right: evolution of the pressure anisotropy as a function of the energy-weighted occupancy $g_s^2 N_c \langle p f \rangle/\langle p \rangle$ of the distribution function. Both plots were obtained from more strongly coupled kinetic theories with $g_s = 1, 1.5, 2, 2.6$, with initial conditions specified by Eq.~\eqref{eq:init-cond}. Here we used $N_{\rm states} = 21$.
    }
    \label{fig:occupancies-anisotropies-strong}
\end{figure}

Let us investigate how the pressure anisotropy and the degree of occupation of the distribution function evolve together as a function of time. We quantify the occupation by calculating $g_s^2 N_c\langle f \rangle$, for the same choice of initial condition and the same values of the coupling constant as in 
Figs.~\ref{fig:scalings-Dr-weak} and~\ref{fig:scalings-Dr-strong}. We follow this quantity together with the pressure anisotropy in the left panel and together with $g_s^2 N_c \langle pf \rangle / \langle p \rangle$ (as introduced in~\cite{Kurkela:2015qoa}) in the right panel of 
Figs.~\ref{fig:occupancies-anisotropies-weak} and~\ref{fig:occupancies-anisotropies-strong}, which show, respectively, weakly coupled kinetic theories ($g_s \in \{10^{-1}, 10^{-2}, 10^{-3}, 10^{-4}\}$) and more strongly coupled kinetic theories ($g_s \in \{1, 1.5, 2, 2.6\}$).

At weak coupling, in Figs.~\ref{fig:scalings-Dr-weak} and \ref{fig:occupancies-anisotropies-weak} one can see three distinct stages in the evolution of the gluon gas:
\begin{enumerate}
    \item There is first a stage of free streaming, where the occupancy stays fixed and the pressure anisotropy grows, driven by a depletion of the distribution at large $p_z$ due to the boost-invariant longitudinal expansion. During this very early free-streaming stage, $\alpha$ and $\beta$ begin near zero and $\gamma$ begins near 1.
    \item The BMSS fixed point~\cite{Baier:2000sb} dominates starting at $y \sim 4$ (that is, $\tau/\tau_I \sim 50$), independently of the coupling constant, and drives the dynamics with its characteristic scalings ($\gamma \approx 1/3, \beta \approx 0, \alpha \approx -2/3$).
    \item The maximal value of the anisotropy scales approximately with $g_s^{-2}$. After this maximal value is reached, which takes longer and longer times at weaker and weaker couplings, the system enters the dilute regime with $\alpha=-1$ and $\beta=\gamma=0$ found in Ref.~\cite{Brewer:2022vkq}, and (very) slowly becomes more isotropic as the occupancy drops rapidly.
\end{enumerate} 
However, at such weak couplings, the system does not reach local thermal equilibrium within a time of at least $\tau/\tau_I \sim 10^{13}$. We do not evolve the system to even larger (arbitrarily large) times because the system will be too dilute for any possible hydrodynamization to be of physical interest to us.

At stronger couplings, we see in Figs.~\ref{fig:scalings-Dr-strong} and \ref{fig:occupancies-anisotropies-strong} that the second stage of the weakly coupled scenario disappears essentially completely, with the system evolving directly from the free-streaming regime to a regime where the scaling exponents are close to the values that, at weak coupling, correspond to the dilute fixed point. There is then, however, a last stage of the evolution, namely hydrodynamization, whose analysis is our goal in this Section. 
The three stages in the evolution at stronger couplings are:
\begin{enumerate}
    \item Growth of the pressure anisotropy without changing the typical occupancy until $y \sim 3$. As at weak coupling, during this very early free-streaming stage, $\alpha$ and $\beta$ begin near zero and $\gamma$ begins near 1.
    \item 
    The system evolves directly from the free-streaming stage to a regime in which $\beta$ and $\gamma$ are close to zero and $\alpha$ is close to -1. At these stronger values of the coupling, the scaling exponents are not constant during this epoch, {{but at least for $g_s$ around 1 to 1.5  we see a regime that resembles the dilute scaling found at weak coupling.}} During this epoch, both the pressure anisotropy and the occupation drop together. During both the free-streaming stage and this stage, the dynamical evolution is approximately described via a scaling function with a pre-hydrodynamic form as in Sect.~\ref{sec:pre-hydro}, which is to say by a pre-hydrodynamic attractor. 
    Once we take $g_s$ as large as 2.6, though, it is no longer clear whether there is a distinct dilute-like regime as the system quickly evolves onward to$\ldots$

    \item Hydrodynamization. Isotropization and approach to local thermal equilibrium, with a thermalization time that  ranges between $\tau_{\rm th}/\tau_I \sim 10^{4-5}$ at $g_s = 2.6$ and $\tau_{\rm th}/\tau_I \sim 10^{8-10}$ at $g_s = 1$, suggesting that $\tau_{\rm th}/\tau_I$ scales exponentially with an inverse power of the coupling constant, as expected in kinetic theories with only small angle scatterings, 
    whose thermalization times have been estimated to scale parametically as $\tau_{\rm th}/\tau_I 
    \sim \exp(-1/\alpha_s^{1/2})$~\cite{Mueller:1999fp,Bjoraker:2000cf,Tanji:2017suk}. Due to the absence of gluon splittings, the thermalization time of this kinetic theory is much longer than that of QCD EKT~\cite{Kurkela:2015qoa,Kurkela:2014tea}. Note that hydrodynamization, namely the approach to isotropization and thermalization which is signified in Fig.~\ref{fig:scalings-Dr-strong} by the three scaling exponents rising away from their dilute values toward their hydrodynamic values $\alpha=0$, $\beta=\gamma=1/3$, begins at $\tau/\tau_I\sim 10^{5-6}$ for $g_s=1$, {{and at earlier and earlier times for larger values of the coupling.}} Furthermore, at the largest value of the coupling that we have investigated, namely $g_s=2.6$ which corresponds to $\alpha_s\sim0.5$ and $\lambda_{\rm 't~Hooft}\sim 20$, as noted above we see that the evolution proceeds essentially directly from the early free-streaming phase to hydrodynamization, with the intermediate phase 2.~so brief as to be not distinguishable. During hydrodynamization and during the subsequent hydrodynamic evolution, the dynamical evolution is described via a scaling function as in Sect.~\ref{sec:hydro}, which is to say by a hydrodynamizing attractor.
\end{enumerate}

It is important to note that the distribution function takes on different scaling forms in the pre-hydrodynamic stage (stage 2 above; scaling as in Sect.~\ref{sec:pre-hydro}) and during hydrodynamization (stage 3 above; scaling as in Sect.~\ref{sec:hydro}). We have achieved a unified and continuous description of both stages and the transition from one to the other even though the distribution function does not take on a scaling form during that transition.

We see that for values of $g_s$ ranging from 1 (which is much weaker than appropriate for the description of hydrodynamization in a heavy ion collision) to 2 (reasonable) to 2.6 (a little on the large side) we have been able to obtain a complete description of hydrodynamization, beginning from free-streaming, with the form of the scaling function then changing smoothly from its pre-hydrodynamic form (as in Sect.~\ref{sec:pre-hydro}) to the form needed to describe hydrodynamization and ultimately isotropization and thermalization that we first introduced in Sect.~\ref{sec:hydro}. Our description is adiabatic throughout, but the dynamics governed by the adiabatic evolution of the ground state(s) of the effective Hamiltonian changes from that of the pre-hydrodynamic attractor to that of the hydrodynamizing attractor. We describe the transition from one attractor to the next further below.

In the remainder of this Section, we confirm that AH is working as described via careful inspection of how the eigenvalues of the effective Hamiltonian, and the occupation of the associated eigenstates, evolve with time during the dynamics illustrated in the Figs.~\ref{fig:scalings-Dr-weak}-\ref{fig:occupancies-anisotropies-strong} above. Can we confirm that the system is indeed dominated by an adiabatically evolving band of instantaneous ground states and ultimately by an adiabatically evolving isolated instantaneous ground state? If so, these state(s) encode the information about the initial state that survives each stage in the process of hydrodynamization.

\begin{figure}
    \centering
    \includegraphics[width=0.7\textwidth]{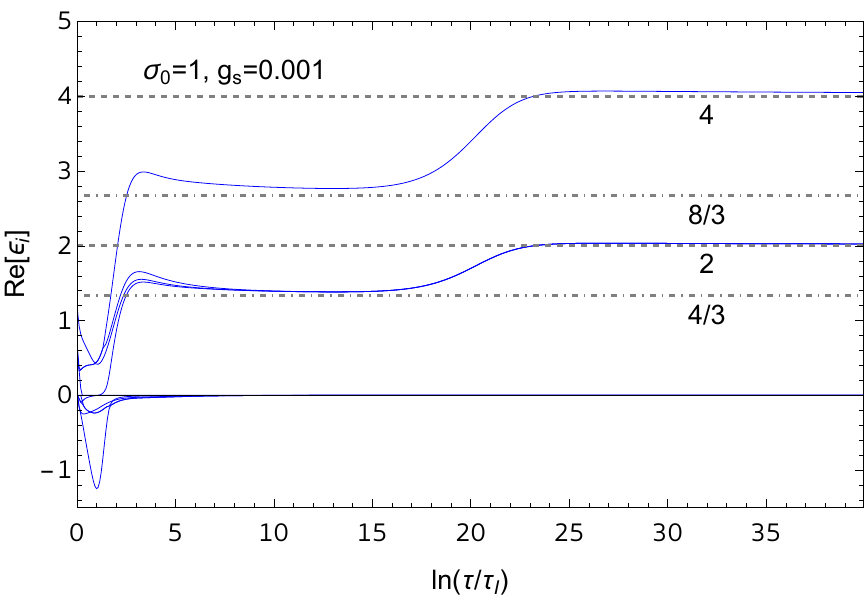}
    \caption{Plot of the instantaneous eigenvalues of the effective Hamiltonian~\eqref{eq:Heff-smallanglescatt-Dr} for $g_s = 10^{-3}$ and the initial condition given in Eq.~\eqref{eq:init-cond}.  We used $N_{\rm states} = 15$; here we display the 9 states of even $u \to -u$ parity, which are the ones with nonzero occupancy. As first analyzed by BSY~\cite{Brewer:2022vkq}, the system rapidly reaches the pre-hydrodynamic attractor which first describes evolution to and at the BMSS fixed point and subsequently describes the approach to and evolution at the dilute fixed point. 
    }
    \label{fig:eigenvals-Dr-gs10-3}
\end{figure}

\begin{figure}
    \centering
    \includegraphics[width=0.8\textwidth]{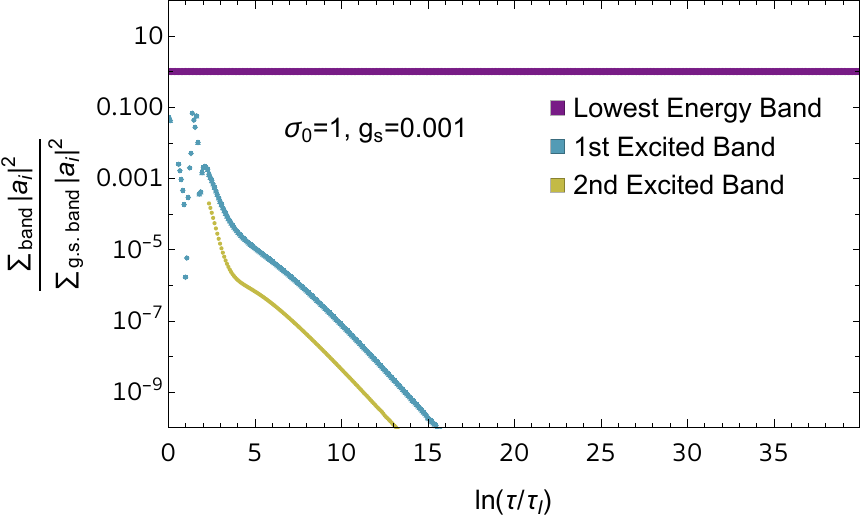}
    \caption{Plot of the coefficients in the $H_{\rm eff}$~\eqref{eq:Heff-smallanglescatt-Dr} eigenstate decomposition of the distribution function for $g_s = 10^{-3}$ and the initial condition given in Eq.~\eqref{eq:init-cond}, grouped by bands of nearly degenerate eigenvalues (see Fig.~\ref{fig:eigenvals-Dr-gs10-3}). We used $N_{\rm states} = 15$; here we display the 9 states of even $u \to -u$ parity, which are the ones with nonzero occupancy. The coefficients are normalized relative to the occupation of the lowest energy band that carries the particle number. We see that the occupation of each of the higher bands decay steeply and exponentially starting already during the approach to the BMSS regime. Only the lowest energy band is relevant to describing the evolution of this kinetic theory from then on. 
    }
    \label{fig:coefficients-Dr-gs10-3}
\end{figure}

We start at very weak coupling with $g_s=10^{-3}$, where the dynamics is that which we first explored in Sect.~\ref{sec:pre-hydro}. As we have seen in Figs.~\ref{fig:scalings-Dr-weak} and \ref{fig:occupancies-anisotropies-weak} and discussed above,
the system does not reach hydrodynamization. As we can see from Figures~\ref{fig:eigenvals-Dr-gs10-3} and~\ref{fig:coefficients-Dr-gs10-3}, a band of lowest energy eigenstates of the effective Hamiltonian
quickly becomes dominant, starting from the time at which the BMSS fixed point is reached and continuing through the approach to, and evolution at, the dilute fixed point. In the BMSS regime, the occupation of the excited states decays exponentially, and these states stay irrelevant at later times, even as the instantaneous energy levels change their values during the transition from the BMSS regime to the dilute regime. This is in precise agreement with the picture first introduced by BSY~\cite{Brewer:2022vkq}, where the adiabatic evolution of a set of states with the same, lowest, instantaneous energy eigenvalue dominates the pre-hydrodynamic dynamics.

\begin{figure}
    \centering
    \includegraphics[width=0.7\textwidth]{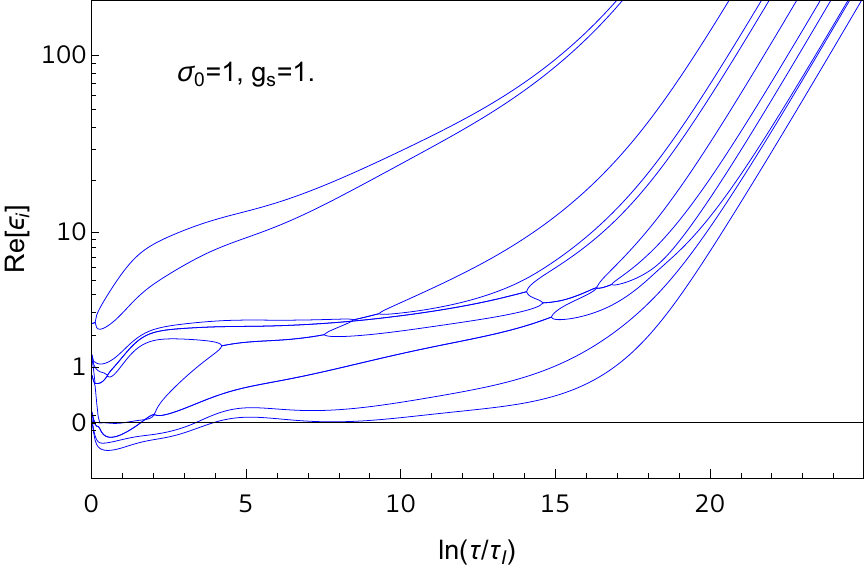}
    \caption{Plot of the instantaneous eigenvalues of the effective Hamiltonian~\eqref{eq:Heff-smallanglescatt-Dr} for an intermediate coupling $g_s = 1$ and the initial condition given in Eq.~\eqref{eq:init-cond}. We used $N_{\rm states} = 21$; here we display the 12 states of even $u \to -u$ parity, which are the ones with nonzero occupancy. We see that a gap opens up within the band of nearly degenerate low-lying eigenvalues at $\ln(\tau/\tau_I)\sim 12$, which corresponds to the time in the top-left panel of Fig.~\ref{fig:scalings-Dr-strong} when the scaling exponents begin to evolve away from the dilute regime in the direction of their hydrodynamic values. After this gap opens, the evolution is governed by an isolated instantaneous ground state, corresponding to the isolated  lowest eigenvalue. The adiabatic evolution of this state describes hydrodynamization. 
    }  
    \label{fig:eigenvals-Dr-gs1}
\end{figure}

\begin{figure}
    \centering
    \includegraphics[width=0.8\textwidth]{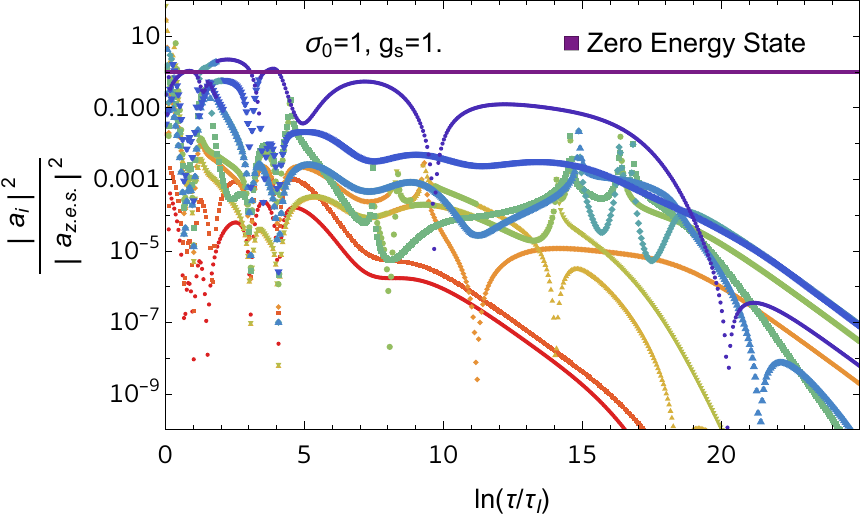}
    \caption{Plot of the coefficients in the $H_{\rm eff}$~\eqref{eq:Heff-smallanglescatt-Dr} eigenstate decomposition of the distribution function for $g_s = 1$ and the initial condition given in Eq.~\eqref{eq:init-cond}. We used $N_{\rm states} = 21$; here we display the 12 states of even $u \to -u$ parity, which are the ones with nonzero occupancy. The coefficients are normalized relative to the occupation of the zero-energy-eigenvalue state that carries the particle number, which at late times is the instantaneous ground state with the lowest eigenvalue. We see that after the gap between this eigenvalue and the others opens up (see Fig.~\ref{fig:eigenvals-Dr-gs1}), the occupation of each of the higher energy states falls away.
    }
    \label{fig:coefficients-Dr-gs1}
\end{figure}

\begin{figure}
    \centering
    \includegraphics[width=0.48\linewidth]{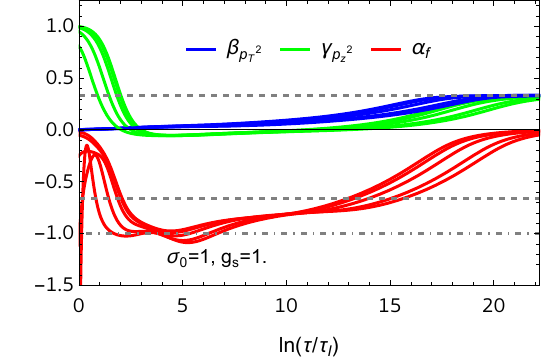}
    \includegraphics[width=0.48\linewidth]{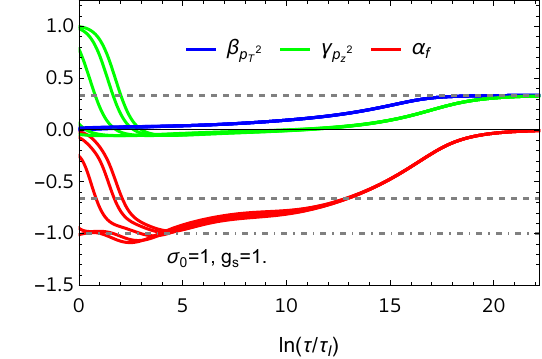}
    \caption{Instantaneous rates of change of the typical scales $\langle p_\perp^2 \rangle$, $\langle p_z^2 \rangle$ and occupancy $\langle f \rangle$, for $g_s = 1$ and $N_{\rm states} = 21$, for initial conditions specified by $r_I \in \{0.5, 1, \sqrt{3}, 3, 5\}$. Left: curves as obtained directly from the solution to the dynamics of the system. Right: curves shifted in time (horizontally) so as to match with the $r_I = 0.5$ curve at the final time. We see that for all these initial conditions the dynamics of the system collapses onto the same curve after a time of $y \sim 5$. Comparing with Figure~\ref{fig:eigenvals-Dr-gs1}, we see that this happens while there is still a band of low energy levels around $\epsilon = 0$, but after a gap has emerged relative to the higher excited states. The rest of the evolution is clearly governed by an attractor that may be understood as the adiabatic evolution of the ground state (band), as we discuss in the text.}
    \label{fig:varying-r}
\end{figure}

This picture changes qualitatively at late times when we consider the intermediate value of the coupling $g_s = 1$. From Figs.~\ref{fig:eigenvals-Dr-gs1} and~\ref{fig:coefficients-Dr-gs1}, we see that at early times there are bands of energy eigenvalues with the lowest such band remaining close to degenerate until $y \equiv \ln(\tau/\tau_I) \sim 12$, at which time a gap starts opening up, consistent with the beginning of hydrodynamization, i.e.~the beginning of the departure from the dilute regime and the approach to isotropization and local thermal equilibrium seen in the top-left panel of 
Fig.~\ref{fig:scalings-Dr-strong} and discussed above. Meanwhile, we see in Fig.~\ref{fig:coefficients-Dr-gs1} that between $y \sim 2$ and $y \sim 10$ the occupation of a small group of low-lying states is significant while above $y\sim 12$ and certainly for $y > 15$ the occupation of all states other than the one with the isolated lowest eigenvalue drop away and only the instantaneous ground state of the effective Hamiltonian is needed to describe hydrodynamization in this kinetic theory.
This confirms that the AH scenario is realized sequentially in this kinetic theory, with different attractors (first pre-hydrodynamic; later hydrodynamizing) describing different out-of-equilibrium regimes corresponding to different stages in the loss of memory of the initial condition, with each collapsing the state onto a smaller set of degrees of freedom than in the previous stage, until local thermal equilibrium is reached and only the hydrodynamic evolution of the state with a thermal distribution remains. To further substantiate this, in Figure~\ref{fig:varying-r} we show that the instantaneous rates of change of $\langle p_\perp^2 \rangle$, $\langle p_z^2 \rangle$ and $\langle f \rangle$ collapse onto a single curve after a short time for initial conditions which differ in their initial value of the anisotropy parameter $r$.

With these results in hand, we have a unified and continuous description of the transition from one attractor to the next.
The pre-hydrodynamic attractor is described via the adiabatic evolution of a band of lowest-energy eigenstates, with the system having previously been attracted into some superposition of the eigenstates in this band. The mix within this superposition may evolve as the  eigenstates in this band evolve adiabatically but there is  no mixing with higher energy states. 
These few lowest-energy eigenstates, and their time-evolution, encode whatever information about its initial conditions the system ``remembers'' during the epoch when its evolution is described via the pre-hydrodynamic attractor.
The transition from this pre-hydrodynamic attractor to hydrodynamization is precipitated by the energy of all the basis states but one in this band rising, leaving a single isolated ground state of the evolving effective Hamiltonian. As this happens, the occupancy of all 
the other states from what used to be the low-lying band decay rapidly to zero.  That is to say, the system ``forgets'' all aspects of its initial state except those encoded in a single instantaneous eigenstate and its time-evolution.
From then on, during
hydrodynamization the system is described by the adiabatic evolution of a single isolated lowest energy eigenstate of the evolving effective Hamiltonian.

\begin{figure}
    \centering
    \includegraphics[width=0.7\textwidth]{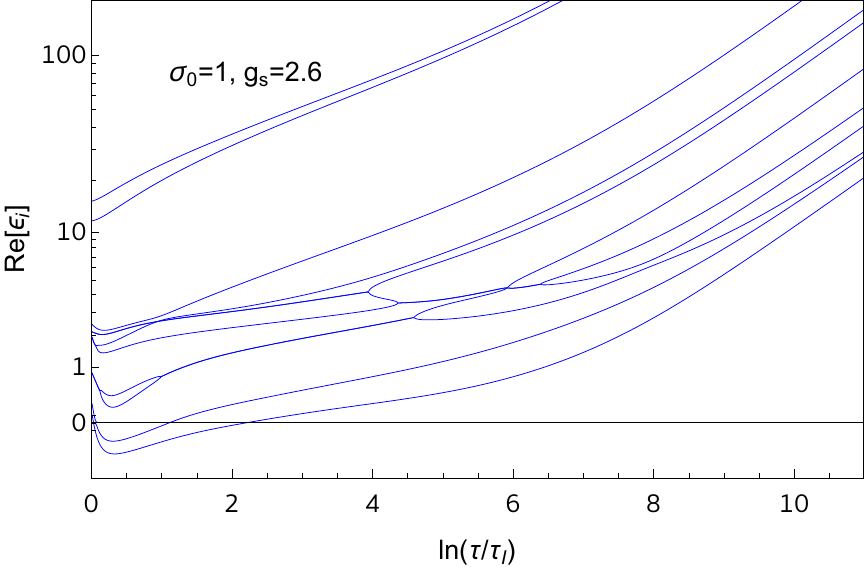}
    \caption{Plot of the eigenvalues of the effective Hamiltonian~\eqref{eq:Heff-smallanglescatt-Dr} for $g_s = 2.6$ and the initial condition given in Eq.~\eqref{eq:init-cond}. We used $N_{\rm states} = 21$; here we display the 12 states of even $u \to -u$ parity, which are the ones with nonzero occupancy. At this large coupling, it is hard to say whether there is a regime at early times where a band of low eigenvalues dominates, since a gap between a single lowest eigenvalue and all the others opens up so early. This corresponds to the rapid onset of hydrodynamization that we have seen in the bottom-right panel of Fig.~\ref{fig:scalings-Dr-strong}. 
    }
    \label{fig:eigenvals-Dr-gs26}
\end{figure}

\begin{figure}
    \centering
    \includegraphics[width=0.8\textwidth]{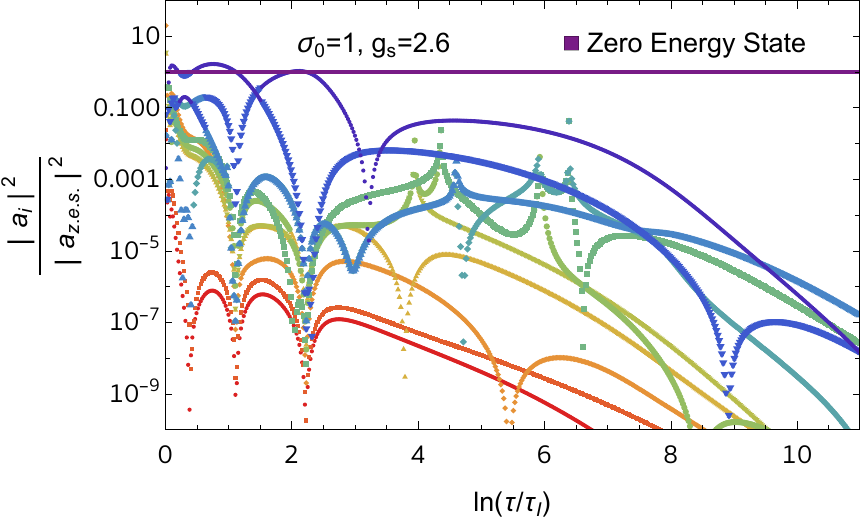}
    \caption{Plot of the coefficients in the $H_{\rm eff}$~\eqref{eq:Heff-smallanglescatt-Dr} eigenstate decomposition of the distribution function for $g_s = 2.6$ and the initial condition given in Eq.~\eqref{eq:init-cond}. We used $N_{\rm states} = 21$; here we display the 12 states of even $u \to -u$ parity, which are the ones with nonzero occupancy. The coefficients are normalized relative to the occupation of the zero-energy-eigenvalue state that carries the particle number, which at late time is the instantaneous ground state of $H_{\rm eff}$. We see that after the (here quite early time at which the) gap between this eigenvalue and the others opens up (see Fig.~\ref{fig:eigenvals-Dr-gs1}), the occupation of each of the higher energy states falls away. 
    }
    \label{fig:coefficients-Dr-gs26}
\end{figure}

\begin{figure}
    \centering
    \includegraphics[width=0.48\linewidth]{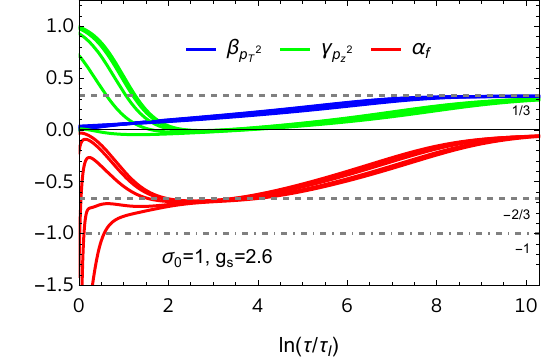}
    \includegraphics[width=0.48\linewidth]{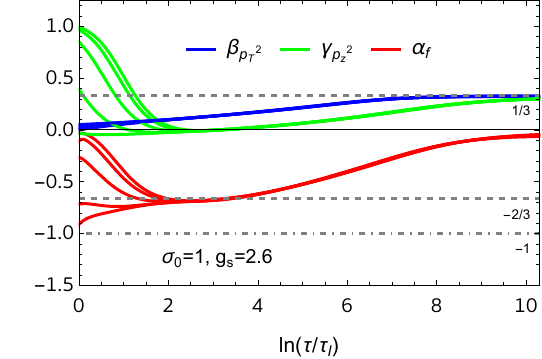}
    \caption{Instantaneous rates of change of the typical scales $\langle p_\perp^2 \rangle$, $\langle p_z^2 \rangle$ and occupancy $\langle f \rangle$, for $g_s = 2.6$ and $N_{\rm states} = 21$, for initial conditions specified by $r_I \in \{0.5, 1, \sqrt{3}, 3, 5\}$. Left: curves as obtained directly from the solution to the dynamics of the system. Right: curves shifted in time (horizontally) so as to match with the $r_I = 0.5$ curve at the final time. We see that all curves collapse onto the same curve after a time of $y \sim 3$, or $\tau \sim 20 \, \tau_I$.}
    \label{fig:varying-r-g26}
\end{figure}

When we analyze the eigenvalues and eigenstate occupations at the still larger value of the coupling $g_s=2.6$ in Figs.~\ref{fig:eigenvals-Dr-gs26} and \ref{fig:coefficients-Dr-gs26}, the middle stages of the sequential process that we have analyzed have become so short as to be indistinguishable: the system proceeds relatively directly from the earliest pre-hydrodynamic stage to hydrodynamization, which brings it to isotropization and local thermal equilibrium in a shorter time. This evolution is described just as well via the AH scenario as the dynamics with more distinct epochs found at intermediate and weak coupling, above. As we can see from Fig.~\ref{fig:varying-r-g26}, attractor behavior in the scaling exponents is observed as early as $\tau=20\,\tau_I$.
We see that
a gap opens up between the instantaneous ground state of the effective Hamiltonian with an isolated lowest eigenvalue and all higher energy states, with this happening 
before $\tau=20\,\tau_I$ (exactly around the onset of the attractor behavior displayed in Fig.~\ref{fig:varying-r-g26}), which is to say at a much earlier time here with $g_s=2.6$ than what we found with $g_s=1$ in Figs.~\ref{fig:eigenvals-Dr-gs1} and \ref{fig:coefficients-Dr-gs1}.
After this happens, hydrodynamization and the approach to isotropization and equilibrium is duly realized by the rapid decay of excited states, with the subsequent evolution of the full distribution function being described by the adiabatic evolution of the instantaneous ground state. 

In our analyses of this kinetic theory at $g_s=1$ and $g_s=2.6$, we have thus achieved all of the goals that we set out to achieve, seeing in concrete terms how all the physical processes and regimes involved in the early pre-hydrodynamic dynamics and the subsequent hydrodynamization and hydrodynamic evolution in this kinetic theory are described via Adiabatic Hydrodynamization in a fashion that yields understanding and intuition for what is happening when, and why and how it happens.

\

\section{Outlook} \label{sec:outlook}

We have demonstrated via concrete examples that attractors in kinetic theory can be described by the low-energy instantaneous eigenstates of the operator that generates the time evolution of the system (i.e., the effective Hamiltonian) in an appropriate ``adiabatic'' frame, defined by the requirement that the instantaneous ground state of the effective Hamiltonian evolves in time as slowly as possible. In this way, we have demonstrated that the AH framework is ideally equipped to describe the process of hydrodynamization of kinetic theory, and in particular should greatly simplify the analysis of the QCD EKT description of the initial stages of heavy-ion collisions.
Furthermore, we have formulated a prescription to find the optimal adiabatic frame, even though in practice we showed that the AH scenario can be realized without finding the strictly optimal solution for this frame.

We showed explicitly via our analysis of the kinetic theory for a longitudinally expanding gluon gas obtained by making the small-angle elastic scattering approximation to the QCD EKT collision kernel that the 
processes by which the system loses memory of its initial state proceed in stages -- stages that are distinct if the kinetic theory is sufficiently weakly coupled.  Each out-of-equilibrium attractor stage encodes only a  subset of the information in the initial condition, characterized precisely by the adiabtically evolving low-energy instantaneous eigenstate(s) of the effective Hamiltonian in the adiabatic 
frame. The adiabatic description of the earliest prehydrodynamic era is governed by a band consisting of  degenerate (e.g.~in the analysis of BSY~\cite{Brewer:2022vkq} or Sect.~\ref{sec:pre-hydro}) or almost degenerate (e.g.~in the analysis of Sect.~\ref{sec:non-scaling})  lowest-energy instantaneous eigenstates. Subsequently, further gaps open up and hydrodynamization begins as the system evolves to have a single isolated lowest-energy eigenvalue, with the corresponding adiabatically evolving instantaneous ground state of the effective Hamiltonian describing hydrodynamization and the subsequent hydrodynamic evolution of a distribution in local thermal equilibrium.
As the number of ground states in the adiabatic description of the dynamics
becomes progressively smaller as the system passes from one attractor stage to the next, more information about the initial state of the kinetic theory is forgotten, until only a single state remains. The adiabatic evolution of this state describes hydrodynamization and at late times, as hydrodynamization concludes, this state corresponds to the thermal distribution.

At this point, we comment on our expectations for what will happen when this framework is applied to QCD EKT in full. While it is very likely that for realistic values of the coupling in full QCD EKT 
it may be hard to identify distinct attractor regimes, we anticipate that at sufficiently weak coupling we will see a variation on the themes that we have explored concretely in Sect.~\ref{sec:non-scaling}, where the evolution is described adiabatically throughout but not via a single scaling form, with the system initially following pre-hydrodynamic attractor(s) and subsequently hydrodynamizing, as described by a different attractor. For realistic values of the coupling these processes may become less distinct. But we expect that at larger values of the coupling, as in our analysis of the kinetic theory we have used in Sect.~\ref{sec:non-scaling},  
the entire evolution will be described via the adiabatic evolution of a small number of instantaneous ground states, with that number decreasing to one as gaps in the low-lying spectrum of the effective Hamiltonian open up.
Most of the loss of memory of the initial conditions for the kinetic theory will occur at the very beginning as all eigenstates of the effective Hamiltonian except those in a low-lying band 
rapidly become irrelevant. Further loss of memory of what came before will then correspond to the opening up of further gaps
in the instantaneous spectrum of the effective Hamiltonian, and the last-remaining, lowest energy, instantaneous ground state will adiabatically evolve so as to become the state that describes a thermal distribution function as the system hydrodynamizes. We leave the explicit implementation and verification of this picture in QCD EKT to future work.

There are at least two substantial advances that need to be realized in order to realize the vision above. The first is to restore the full collision kernel of QCD EKT in the kinetic theory and to add quarks and antiquarks to the gluons that we have focused on throughout this work. Likely the most important aspect of this will be to add number-nonconserving processes in the collision kernel, starting with $1 \leftrightarrow 2$ processes, as these are not included in the small-angle scattering collision kernel that we have employed throughout this work.
Restoring $1 \leftrightarrow 2$ processes will result in hydrodynamization happening more quickly at weak (and likely at any given value of the) coupling.
It will be very interesting to watch the Adiabatic Hydrodynamization scenario 
in action fully explicitly in that case, as we have done here.
We anticipate that the massive simplification of the 
problem we have achieved here by demonstrating that the system quickly becomes dominated by the low-energy eigenstates of the 
time-evolution operator 
will, when realized in QCD EKT, be of practical importance in addition to being of value for the understanding and physical intuition that it yields. A full numerical simulation of QCD EKT carries along a lot of unnecessary information corresponding to the description of the evolution of all the higher energy states of the time-evolution operator.  Realizing its description in the language of adiabatic hydrodynamization should therefore make it more practicable to make advances in the second important direction, namely introducing initial, prehydrodynamic, hydrodynamizing and hydrodynamic states with geometries that are more similar to those of heavy ion collisions (HICs) than the longitudinally-boost invariant transversely-infinite geometry that we have employed throughout this work.
Adding transverse expansion into the kinetic equation is of foremost importance, so as to be able to describe how a (finite) droplet of QCD matter produced in a HIC expands and eventually falls apart, especially in situations where the QCD matter does not fully hydrodynamize. 
Understanding the pre-hydrodynamic dynamics in small collision systems in which hydrodynamization may not be complete is of considerable interest as it is one of the few ways in which experimental measurements may shed light on QCD matter in the act of hydrodynamizing. Applying Adiabatic Hydrodynamization in such a setting will require applying it in geometries where the transverse extent of the matter described by kinetic theory is finite and spatial gradients and transverse expansion are important.  This presents considerable challenges in general; the simplification introduced via understanding and employing the adiabatic hydrodynamization approach could help to overcome these challenges.
Another feature to include in future work is spacetime rapidity dependence in the kinetic equation, which would in principle allow one to make more quantitative statements about observables as one moves away from the mid-rapidity region in a HIC.

Our results pave the way for an intuitive understanding of the dynamics of hydrodynamization in QCD EKT, and more generally of the emergence of scaling 
phenomena during the approach to equilibrium in kinetic theory in other contexts. Presumably after some modification, the approach that we have presented in explicit detail in this work could also be suitable to provide the same kind of understanding and intuition for the physics of thermalization during and after the reheating epoch in the early Universe, which has already been described via kinetic theory 
(see, e.g.,~\cite{Harigaya:2013vwa,Mukaida:2022bbo,Mukaida:2024jiz}). It may also be of interest in the context of the preheating epoch of some inflationary models. And, it almost goes without saying that we expect that the understanding of the dynamics of out-of-equilibrium QCD matter via the approach we have laid out here could be particularly valuable in Bayesian analyses of HICs (e.g., JETSCAPE~\cite{Kauder:2018cdt} or \textit{Trajectum}~\cite{Nijs:2020roc}) due to the simplicity gained by only needing to describe the pre-hydrodynamic and hydrodynamizing distribution function in terms of the instantaneous low-energy eigenstates of the time-evolution operator. Bayesian analyses like these, simplified and empowered via employing adiabatic hydrodynamization in QCD EKT, offer a path to 
making direct connections between experimental data from HICs and the QCD description of their initial stages.

\acknowledgments

We thank J\"urgen Berges, Jasmine Brewer, Aleksi Kurkela, Aleksas Mazeliauskas, Govert Nijs and Yi Yin for useful discussions. This work is supported by the U.S.~Department of Energy, Office of Science, Office of Nuclear Physics under grant Contract Number DE-SC0011090\@.

\appendix

\section{Numerical implementation} \label{app:numerics}

In this Appendix, we discuss the concrete numerical setup we used to obtain the results described in the main text of the paper. Section~\ref{sec:earlynumerical} describes the general setup we used to solve for the evolution of the distribution function when the time-dependent parameters we introduced to find an adiabatic frame corresponded to coordinate rescalings only, i.e., where we take the basis to be time-independent when written as a function of the rescaled coordinates, as in Sections \ref{sec:isotropic} and \ref{sec:expanding}. Section~\ref{sec:r-numerical} describes the additions needed for the case when the basis is not time-independent, as required in the description of Section~\ref{sec:connectstages}.

\subsection{Hamiltonian evolution for scaling regimes} \label{sec:earlynumerical}

Here we will detail the numerical implementation of the scaling solutions described in 
Sects.~\ref{sec:isotropic} and \ref{sec:pre-hydro}, starting with an explanation of the details of the implementation for Sect.~\ref{sec:isotropic}, then the details for the similar case of \ref{sec:pre-hydro}. As discussed in the main text, we choose a basis \eqref{eq:boltzmannbasis}, onto which we project a distribution function and effective Hamiltonian operator \eqref{eq:isotropich}. Matrix elements of the effective Hamiltonian can be expressed in terms of time-independent integrals over the basis functions; using a projection as in Eq.~\eqref{eq:hprojection} the matrix elements can be written in the form
\begin{align}
    H_{kk'} = \alpha+ \delta h^{(1)}_{kk'} - \lambda_0 \ell_{\rm Cb} \frac{I_a}{ D^2} h^{(2)}_{kk'}- \lambda_0 \ell_{\rm Cb} \frac{I_b}{D} h^{(3)}_{kk'}\,,
\end{align}
where
\begin{align}
    h^{(1)}_{kk'} &= \int d^3\chi \psi_k^{(L)} \chi \partial_\chi \psi_{k'}^{(R)}\\
    h^{(2)}_{kk'} &= \int d^3\chi \psi_k^{(L)} \left(\frac{2}{\chi} \partial_\chi + \partial_\chi^2\right) \psi_{k'}^{(R)}\\
    h^{(3)}_{kk'} &= \int d^3\chi \psi_k^{(L)} \left( \frac{2}{\chi} + \partial_\chi \right) \psi_{k'}^{(R)}\,.
\end{align}

The Hamiltonian also in general depends on the non-linear functionals $I_a,I_b,$ and $\ell_{\rm Cb}$. As noted in the text, we write the Coulomb logarithm $\ell_{\rm Cb}$ as
\begin{align}
     \ell_{\rm Cb}[f] =\ln \left( \frac{\sqrt{\langle p^2 \rangle}}{m_D} \right)\, ,
\end{align}
where $m_D$ is the Debye mass $m_D=2N_cg_s^2I_b$. 
While in principle $\ell_{\rm Cb}$ could become negative if $g_s$ is sufficiently large, we don't encounter this issue in any of our solutions. Since $f_I \propto 1/g_s^2$, such problem cannot possibly appear at the initial time for our choice of initial conditions, and we have not encountered a situation where the dynamics modifies this later on.

The functionals $I_a,I_b$ we will treat as independent variables, which we will evolve by writing their time-derivatives in terms of our scaling parameters and time-independent integrals, as we do for the Hamiltonian. For 
Sect.~\ref{sec:isotropic}, this is
\begin{align}
    \partial_\tau I_b &= I_b(\alpha - 2 \delta) + AD^2 \lambda_{k} H_{kk'} w_{k'} \\
    \partial_\tau I_a &= I_a(\alpha-3\delta) +AD^3 q_k H_{kk'} w_{k'}
\end{align}
where
\begin{align}
    \lambda_k &= \int \frac{d^3\chi}{(2\pi)^3}\, \psi_k^{(L)} \frac{2}{\chi} \psi_{k'}^{(R)} \\
    q_k &= \int \frac{d^3\chi}{(2\pi)^3} \psi_k^{(L)}\, \psi_{k'}^{(R)}\,,
\end{align}
although in the specific case of Sect.~\ref{sec:isotropic}, $\partial_\tau I_a=0$ because we have taken the dilute limit in that Section and have no expansion, so $I_a=K_N$ is conserved. 

In Sect.~\ref{sec:pre-hydro}, we express the Hamiltonian \eqref{eq:expandingh} as
\begin{equation}
\begin{split}
H_{ijkl} = \alpha \delta_{ijkl} + \beta h^{(1)}_{ijkl} + (\gamma-1) h^{(2)}_{ijkl} - \tau \lambda_0 \ell_{\rm Cb}  \frac{I_a}{B^2} h^{(3)}_{ijkl} \\- \tau \lambda_0 \ell_{\rm Cb} \frac{I_a}{C^2} h^{(4)}_{ijkl} - \tau \lambda_0 \ell_{\rm Cb}\frac{I_b}{B} h^{(5)}_{ijkl}
\end{split}
\end{equation}
where 
\begin{align}
h^{(1)} &\equiv \zeta \partial_\zeta, \\
h^{(2)} &\equiv \xi \partial_\xi, \\
h^{(3)} &\equiv \frac{1}{\zeta} \partial_\zeta + \partial_\zeta^2, \\
h^{(4)} &\equiv \partial_\xi^2, \;\;\text{and}\\
h^{(5)} &\equiv \frac{2}{\zeta} + \partial_\zeta + \frac{\xi}{\zeta} \partial_\xi
\,.
\end{align}
The evolution equations of the functionals $I_a,I_b$ in Sect.~\ref{sec:pre-hydro} are
\begin{align}
    \partial_y I_b &= (\alpha - \beta - \gamma) I_b - ABC \lambda_{ij} H_{ijkl} w_{kl} \\
    \partial_y (\tau I_a) &= (1+2\alpha - 2\beta - \gamma) (\tau I_a-K_{\tilde{N}}) - 2\tau A^2B^2C w_{ij} q_{ijkl} H_{klmn} w_{mn}\,,
\end{align}
where
\begin{align}
    \lambda_{ij} &= \frac{1}{(2\pi)^2} \int_{-\infty}^\infty d\xi \int_0^\infty d\zeta \; \zeta \; \frac{2 \psi_R^{(ij)}}{\zeta} \\
    q_{ijkl} &= \frac{1}{(2\pi)^2} \int_{-\infty}^\infty d\xi \int_0^\infty d\zeta \; \zeta \; \psi_R^{(ij)} \psi_R^{(kl)}
\end{align}
and we have used the approximation $p\approx p_\perp=B\zeta$ in the expression for $\partial_y I_b$.

For convenience, we use the setup described in Appendix~\ref{sec:r-numerical} to calculate the solutions to the kinetic theory in Section~\ref{sec:hydro}. We describe this at the end of the end of that Appendix.

\subsection{Hamiltonian evolution beyond the scaling regime} \label{sec:r-numerical}

Here we describe the setup needed to evolve the system in the time-dependent basis of Section~\ref{sec:connectstages}.
As discussed in the main text, we choose to expand the rescaled distribution function on a basis defined by
\begin{align}
    \psi_{nl}^{(R)} = N_{nl} e^{-\chi} e^{-u^2 r^2/2} L_{n-1}^{(2)}(\chi) Q_l^{(R)}(u;r) \, , & & \psi_{nl}^{(L)} = N_{nl} L_{n-1}^{(2)}(\chi) Q_l^{(L)}(u;r) \, ,
\end{align}
where the polynomials $Q_l^{(R)}(u;r)$, $Q_l^{(L)}(u;r)$ are polynomials on $u$ of degree $l$, constructed such that
\begin{equation}
    \int_{-1}^1 du \, e^{- u^2 r^2/2} Q_l^{(L)}(u;r) Q_k^{(R)}(u;r) = 2 \delta_{lk} \, ,
\end{equation}
which are normalized by setting $Q_0^{(L)} = 1$, $Q_0^{(R)} = 2/J_0(r) $, and $Q_k^{(L)} = J_0(r) Q_k^{(R)}/2$ for $k \geq 1$, and the normalization coefficients $N_{nl}$ are chosen such that
\begin{align}
    \frac{1}{4\pi^2} \int_{-1}^1 du \int_0^\infty d\chi \, \chi^2 \psi_{mk}^{(L)} \psi_{nl}^{(R)} = \delta_{kl} \, .
\end{align}

The precise form of the distribution function we take is then
\begin{equation} \label{eq:f-app-r-explicit}
    f({\bf p},y) = \frac{e^{-y} D_0^3}{D^3(y)} \sum_{k=1}^{N_{\rm states}} c_k(y) \psi_{n(k) \, l(k)}^{(R)}(p/D, u; r) \, ,
\end{equation}
where we have chosen the time-dependent prefactor such that $\partial_y c_1 = 0$ due to the number-conserving property of the collision kernel. Letting $s(k) \equiv \sqrt{1/4 + 2k - 1} - 1/2$, we use
\begin{align}
    n(k) &= k - \frac{\lfloor s(k)  \rfloor \lfloor s(k) + 1  \rfloor }{2} \left\lfloor \frac{k}{\lfloor s(k) \rfloor \lfloor s(k) + 1  \rfloor / 2 + 1} \right\rfloor \\
    l(k) &= \lfloor s(k) + 1  \rfloor - n(k)
\end{align}
as indexing functions for the radial and angular basis functions. What these functions do is to enumerate the $(n,l)$ states as $(1,0), (1,1), (2,0), (1,2), (2,1), (3,0), \ldots$, where the enumeration is ordered by the value of $n+l$, and within each ``layer'' of fixed $n + l$, is goes from the smallest to the largest value of $n$. We do this so that a truncation done at a maximal value of $n+l$ is essentially a truncation on the number of ``nodes'' that the basis functions can accommodate. 

We define the matrix elements
\begin{align}
    a_{kk'} &= \frac{1}{4\pi^2} \int_{-1}^1 \! du \! \int_0^\infty \! d\chi \, \chi^2 \psi_{n(k')\, l(k') }^{(L)} \chi \partial_\chi \psi_{n(k)\, l(k) }^{(R)} \\
    b_{kk'} &= \frac{1}{4\pi^2} \int_{-1}^1 \! du \! \int_0^\infty \! d\chi \left[ \chi^2 e^{-\chi} \partial_\chi \psi_{n(k)\, l(k) }^{(L)}  \partial_\chi \! \left( e^\chi \psi_{n(k')\, l(k') }^{(R)} \right) \right. \nonumber \\ & \quad\quad\quad\quad\quad\quad\quad\quad\quad\quad \left. + (1 - u^2) \partial_u \psi_{n(k)\, l(k) }^{(L)} \partial_u \psi_{n(k')\, l(k') }^{(R)} \right] \\
    d_{kk'} &= \frac{1}{4\pi^2} \int_{-1}^1 \! du \! \int_0^\infty \! d\chi \, \chi^2 \partial_\chi \psi_{n(k')\, l(k') }^{(L)}  \psi_{n(k)\, l(k) }^{(R)} \\
    e_{kk'} &= \frac{1}{4\pi^2} \int_{-1}^1 \! du \! \int_0^\infty \! d\chi \, \chi^2 \psi_{n(k')\, l(k') }^{(L)} \left[ u (1-u^2) \partial_u + u^2 \chi \partial_\chi \right] \psi_{n(k)\, l(k) }^{(R)} \\
    f_{kk'} &= \frac{1}{4\pi^2} \int_{-1}^1 \! du \! \int_0^\infty \! d\chi \, \chi^2 \psi_{n(k')\, l(k') }^{(L)} \partial_r \psi_{n(k)\, l(k) }^{(R)}
\end{align}
and with them the evolution equation for the basis state coefficients is:
\begin{equation}
    \partial_y c_{k} = - [H_{\rm eff}]_{kk'} c_k' \, , \label{eq:app-c-evol}
\end{equation}
where
\begin{equation}
    [H_{\rm eff}]_{kk'} = -(1 - 3 \delta(y) ) \delta_{kk'} - e_{kk'} + \partial_y r(y) f_{kk'} + \delta(y) a_{kk'} + \frac{\tau \lambda_0 \ell_{\rm Cb} I_a}{D^2} \left[ b_{kk'} +  \left( \frac{I_b D}{I_a} - 1 \right) d_{kk'} \right] \, ,
\end{equation}
which we supplement with evolution equations for $D, r$ and $\tilde{\lambda} = I_b / \int_{\bf p} f$:
\begin{align}
    \partial_y D (y) &= 10 \big( D - D^2 \left\langle \frac{2}{p} \right\rangle  \big) \, , \label{eq:app-d-evol} \\
    \partial_y r &= - \frac{1}{r} \frac{J_0}{J_4 J_0 - J_2^2} \left[ -2 ( J_2 - J_4) + \frac{\tau \lambda_0 \ell_{\rm Cb} I_a }{D^2} (J_0 - 3 J_2) \right] \, , \label{eq:app-r-evol} \\
    \partial_y \tilde{\lambda} &= \tilde{\lambda} + \frac{ N_{10} }{c_1} \sum_{k, k'=1}^{N_{\rm states}} \frac{\delta_{0 \, l(k)}}{ N_{n(k)\,0}} \left( e_{k k'} - \partial_y r(y) f_{kk'} - \frac{\tau \lambda_0 \ell_{\rm Cb} I_a}{D^2} \left[ b_{kk'} + \left( \frac{I_b D}{I_a} - 1 \right) d_{kk'} \right] \right) c_{k'}  \, . \label{eq:app-lamb-evol}
\end{align}
The evolution equation for $\tilde{\lambda}$ is derived from the expression of $\tilde{\lambda}$ in terms of $D$ and the basis state coefficients:
\begin{equation}
    \tilde{\lambda} = \frac{  N_{10} }{D c_1} \sum_{k=1}^{N_{\rm states}} c_k \frac{\delta_{0 \, l(k)}}{ N_{n(k)\,0}} \, .
\end{equation}
We use this equation to eliminate $c_{N_{\rm states}}$ from the evolution equations, as this one is the most sensitive to truncation effects. Instead, whenever $c_{N_{\rm states}}$ appears we replace it by its expression in terms of the other basis state coefficients, $\tilde{\lambda}$ and $D$.

When solving the equations, whenever $I_a$ and $\ell_{\rm Cb}$ appear, they are evaluated in terms of the basis state coefficients and $r, D, \tilde{\lambda}$ through the expressions that define them~\eqref{eq:IaIb} and~\eqref{eq:lcb-def} by substituting~\eqref{eq:f-app-r-explicit} into these expressions.

Furthermore, in practice the matrix elements of $H_{\rm eff}$ require evaluating the integral moments $J_n$ introduced in Eq.~\eqref{eq:Jn-moments}. Especially when $r$ grows large (as in the weakly coupled cases we described in Figs.~\ref{fig:scalings-Dr-weak} and~\ref{fig:occupancies-anisotropies-weak}), it is more convenient to numerically solve for
\begin{equation}
    K_n = r^{n+1} \int_{-1}^1 \! du \, u^n e^{-u^2 r^2 /2 } \, .
\end{equation}
If $r$ does not grow large, as in the more strongly coupled systems we considered in Figs.~\ref{fig:scalings-Dr-strong} and~\ref{fig:occupancies-anisotropies-strong}, then it is sufficient to just evaluate these integrals numerically and have them stored for whenever they need to be evaluated. If $r$ grows arbitrarily large, then it is best to evolve $K_n$ as another variable to solve for in the system of ordinary differential equations comprised by Eqs.~\eqref{eq:app-c-evol},~\eqref{eq:app-d-evol},~\eqref{eq:app-r-evol} and~\eqref{eq:app-lamb-evol}. In particular, one can show that
\begin{equation}
    \partial_y K_n = 2 \partial_y r \, r^n \exp(-r^2/2) \, ,
\end{equation}
which we use to evaluate $K_n$ in the more weakly coupled examples of Section~\ref{sec:connectstages}.

For the solutions presented in Section~\ref{sec:hydro}, we used Eqs.~\eqref{eq:app-c-evol},~\eqref{eq:app-d-evol}, and~\eqref{eq:app-lamb-evol} with $r$ set to zero throughout.

\section{Remarks on the omitted $I_b f^2$ term} \label{app:f2}

In this Appendix, we return to the $I_b f^2$ term in the small-angle scattering collision kernel~\eqref{eq:small-angle-kernel} that, as we noted first in Sect.~\ref{sec:AH}, we have omitted throughout this paper. Here we provide a further discussion of the limitations of this approximation, as well as what challenges would need to be faced in order to reinstate the $I_b f^2$ term and treat it within the framework discussed herein.

Quantitatively, BSY~\cite{Brewer:2022vkq} showed that there is no large discrepancy in the dynamics of the hard sector at early times when this term is included/omitted, which we have further verified in Section~\ref{sec:pre-hydro} by comparing to the numerical results that they obtained using the complete small-angle scattering collision kernel, including the $I_b f^2$ term. As such, this approximation does not seem to alter the dynamics significantly at early times. Furthermore, as time goes by, the distribution function becomes more and more dilute, and by the time that the system begins to approach hydrodynamics the condition $f \ll 1$ is certainly satisfied, and therefore dropping the $f^2$ term seems to be a robust approximation in solving the kinetic equation that the small-angle scattering approximation defines.

However, upon closer inspection, one finds that this can only be strictly true for the hard sector, because the equilibrium distribution of the collision kernel is actually the Bose-Einstein distribution if the $I_b f^2$ term is included. 
At small $p$, the Bose-Einstein distribution behaves as $T_{\rm eff}/p$. (Recall that $T_{\rm eff} = I_a/I_b$.) Furthermore, if the condition
\begin{equation} \label{eq:f2-condition}
    I_a \lim_{p \to 0} \int_{-1}^1 du \, p^2 \frac{\partial f}{\partial p} + I_b \lim_{p \to 0} \int_{-1}^1 du \, p^2 f^2 = 0 \, 
\end{equation}
is not satisfied, then the RHS of the kinetic equation specified by the collision kernel~\eqref{eq:small-angle-kernel} generates a term proportional to the delta function $\delta^{(3)}({\bf p})$ due to the action of $\nabla^2_{\bf p}$ on $1/p$ (or equivalently $\nabla_{\bf p} \cdot (\hat{p}/p^2)$), which in turn sources divergent contributions to $f$ and, see \eqref{eq:IaIb}, hence to $I_a$ and thus makes the kinetic equation itself ill-defined. This condition is the same as the one given in 
Ref.~\cite{Blaizot:2014jna} for this collision kernel in a non-expanding geometry. One way to avoid this problem altogether, as discussed in 
Refs.~\cite{Blaizot:2016iir,BarreraCabodevila:2022jhi} for the non-expanding case, is to include $1 \leftrightarrow 2$ number-changing processes to the collision kernel from the full QCD EKT description, which then ensures that the above condition is satisfied. However, this alternative is beyond the scope of the present work. As we note in Section~\ref{sec:outlook}, introducing $1 \leftrightarrow 2$ processes is an important goal for future work.

The preceding argument would seem to indicate that restoring the $I_b f^2$ term in the small-angle scattering collision kernel would introduce pathologies for some initial conditions, but not all of them: note that a Bose-Einstein distribution with temperature given by $T_{\rm eff}$ satisfies~\eqref{eq:f2-condition}.
On the other hand, though, the dynamical evolution that starts from a distribution function that satisfies $\lim_{p \to 0} p f = 0$ (which implies both $\lim_{p \to 0} p^2 f^2 = 0$ and $\lim_{p \to 0} p^2 \partial f/ \partial p = 0$) will not encounter this issue at early times\footnote{We note, however, that in the case of a non-expanding plasma it has been shown that this issue will appear after some finite time if the initial condition is overpopulated~\cite{Blaizot:2011xf,Blaizot:2014jna}, and has been interpreted as the formation of a Bose-Einstein condensate. The formation of such a condensate is forced as the system thermalizes by the simultaneous conservation of number density and energy density. However, it is unclear whether one can engineer initial conditions that would do the same for a longitudinally expanding plasma, as the energy density in this case is not conserved and thus a Bose-Einstein condensate is not guaranteed to form. If a condensate were to form, then the discussion surrounding this boundary condition would have to be modified to allow for the exchange of particles with the condensate.} because no term proportional to a delta function will be generated on the RHS of the kinetic equation as~\eqref{eq:f2-condition} trivially vanishes. Then, at later times it will approach the shape of a Bose-Einstein distribution and must ultimately become arbitrarily close to it. The evolution toward a Bose-Einstein distribution has to satisfy the condition~\eqref{eq:f2-condition} at all times in order to avoid the emergence of (unphysical) pathologies, and as such, the numerical method one chooses to solve this equation must be equipped to handle this. In particular, if a term proportional to $1/p$ at small $p$ is generated, its coefficient must be such that~\eqref{eq:f2-condition} is satisfied. If the distribution is isotropic, then the integrals over $u$ in~\eqref{eq:f2-condition} are trivial, and
the term that arises is 
$T_{\rm eff}/p$. If, on the other hand, the asymptotic behavior of $f$ as $p \to 0$ is not isotropic, but instead $f \approx g(u)/p$, then Eq.~\eqref{eq:f2-condition} implies that $g(u)$ must satisfy $I_a \int_{-1}^1 g(u) du = I_b \int_{-1}^1 g^2(u) du $.

As a consequence of what we have just described, 
an additional restriction appears when we choose a basis to solve for the dynamics using the AH framework: if one wishes to have a basis state that permits the system to achieve full thermalization, then there must be at least one basis state that accommodates $T_{\rm eff}/p$ as the small $p$ asymptotic behavior. Furthermore, the coefficients in front of such basis states must be such that the physical distribution function satisfies the condition in Eq.~\eqref{eq:f2-condition} at all times, so as to not generate artificial singularities in the evolution of the system. This introduces an explicit additional scale into the basis, which need not be close to the typical hard scale of the distribution function (which one may infer, e.g., from $\langle 2/p \rangle^{-1}$ or $\langle p \rangle$) that we have used to find an approximation to the adiabatic frame in Sections~\ref{sec:hydro} and~\ref{sec:connectstages}. As such, the basis along the radial $p$ component of the distribution function needs to encode two scales, which means that they cannot be accommodated simply by a rescaling, and the situation becomes exactly analogous to the one we successfully dealt with in the angular $u$ direction in Sections~\ref{sec:non-scaling-H} and~\ref{sec:connectstages}. The problem is thus solvable, and whether this is done efficiently or not will crucially depend on whether the choice of basis permits an efficient calculation of the matrix elements of $H_{\rm eff}$. An alternative approach would be to not include any basis function that blows up as $p \to 0$, and simply approximate the Bose-Einstein distribution as best as possible with regular functions, in which case both terms in~\eqref{eq:f2-condition} vanish trivially, but at the cost of never describing fully accurately the IR behavior of the distribution function.

Last, we wish to point out an additional drawback of dropping the $I_b f^2$ term in the kinetic equation, which is that as a consequence the dynamics of $f$ does not satisfy strict energy conservation. This has the practical consequence that the late-time scaling regime with $\alpha = 0$ and $\gamma = \beta = 1/3$ is not guaranteed to last until arbitrarily long times: number conservation and isotropy only fix $\alpha = \gamma + 2\beta - 1$ and $\gamma = \beta$, whereas an energy-conserving collision kernel plus a late-time isotropic scaling regime guarantee that $\delta = \gamma = \beta = 1/3$ as well, because then the conservation properties imply
\begin{align}
    0 &= - \alpha \int_0^\infty d\chi \, \chi^2 w + (1/3 - \delta) \int_0^\infty  d\chi \, \chi^3 \partial_\chi w  \, , \label{eq:AppB-e-cons-1} \\
    0 &= - \alpha \int_0^\infty d\chi \, \chi^3 w + (1/3 - \delta) \int_0^\infty  d\chi \, \chi^4 \partial_\chi w  \, , \label{eq:AppB-e-cons-2}
\end{align}
which fix $\alpha = 0$, $\delta = 1/3$. Without the condition of energy conservation, the late-time scaling regime may take different values for the scaling exponents, and indeed, we observe that if we let our simulations run for arbitrarily long time, eventually a new scaling regime is reached, with $\delta = 1/5$ and $\alpha = -2/5$. This regime can be straightforwardly derived from a scaling analysis of the kinetic equation, by balancing the rate of change of $1/3 - \delta$ (coming from the expansion terms) with $A/D^2 \propto e^{-y}/D^5$ (the scaling behavior of the term that violates energy conservation in $\mathcal{C}[f]$). However, since this regime is clearly unphysical as it would not be present if the collision kernel conserved energy, we do not extend our simulations to such times. Restoring the $I_b f^2$ term would remove this last regime altogether, as energy conservation would then dictate $\alpha = 0$ and $\delta = \gamma = \beta = 1/3$, as prescribed by Eqs.~\eqref{eq:AppB-e-cons-1} and~\eqref{eq:AppB-e-cons-2}. 

\bibliography{main.bib}

\end{document}